\documentclass[a4paper,fleqn,usenatbib]{mnras}
\usepackage{newtxtext,newtxmath}
\usepackage[T1]{fontenc}
\usepackage{ae,aecompl}
\usepackage{graphicx}	% Including figure files
\usepackage{subfig}
\usepackage{amsmath}	% Advanced maths commands
\usepackage{amssymb}	% Extra maths symbols
\usepackage{color}

\title[Relativistic reconnection in merging flux tubes]{Relativistic resistive magnetohydrodynamic reconnection and plasmoid formation in merging flux tubes}
\author[B. Ripperda et al.]{
B. Ripperda$^{1,2}$\thanks{E-mail: ripperda@itp.uni-frankfurt.de},
O. Porth$^{2,3}$,
L. Sironi$^{4}$,
and R. Keppens$^{1}$
\\
$^{1}$Centre for mathematical Plasma Astrophysics, Department of Mathematics, KU Leuven, Celestijnenlaan 200B, B-3001 Leuven, Belgium\\
$^{2}$Institut f\"{u}r Theoretische Physik, Max-von-Laue-Str. 1, D-60438 Frankfurt, Germany\\
$^{3}$Astronomical Institute Anton Pannekoek, University of Amsterdam, Science Park 904, 1098 XH, Amsterdam, The Netherlands \\
$^{4}$Department of Astronomy, Columbia University, 550 W 120th St, New York, NY 10027, USA\\
}
\date{Accepted XXX. Received YYY; in original form ZZZ}
\pubyear{2018}
\begin{document}
\label{firstpage}
\pagerange{\pageref{firstpage}--\pageref{lastpage}}
\maketitle

\begin{abstract}
We apply the general relativistic resistive magnetohydrodynamics code {\tt BHAC} to perform a 2D study of the formation and evolution of a reconnection layer in between two merging magnetic flux tubes in Minkowski spacetime. 
Small-scale effects in the regime of low resistivity most relevant for dilute astrophysical plasmas are resolved with very high accuracy due to the extreme resolutions obtained with adaptive mesh refinement. Numerical convergence in the highly nonlinear plasmoid-dominated regime is confirmed for a sweep of resolutions. We employ both uniform resistivity and non-uniform resistivity based on the local, instantaneous current density. For uniform resistivity we find Sweet-Parker reconnection, from $\eta = 10^{-2}$ down to $\eta = 10^{-4}$, for a reference case of magnetisation $\sigma = 3.33$ and plasma-$\beta = 0.1$. {For uniform resistivity $\eta=5\times10^{-5}$ the tearing mode is recovered, resulting in the formation of secondary plasmoids. The plasmoid instability enhances the reconnection rate to $v_{\rm rec} \sim 0.03c$ compared to $v_{\rm rec} \sim 0.01c$ for $\eta=10^{-4}$.} For non-uniform resistivity with a base level $\eta_0 = 10^{-4}$ and an enhanced current-dependent resistivity in the current sheet, we find an increased reconnection rate of $v_{\rm rec} \sim 0.1c$.
{The influence of the magnetisation $\sigma$ and the plasma-$\beta$ is analysed for cases with uniform resistivity $\eta=5\times10^{-5}$ and $\eta=10^{-4}$ in a range $0.5 \leq \sigma \leq 10$ and $0.01 \leq \beta \leq 1$ in regimes that are applicable for black hole accretion disks and jets. The plasmoid instability is triggered for Lundquist numbers larger than a critical value of $S_{\rm c} \approx 8000$.}
\end{abstract}

%%%%%%%%%%%%%%%%%%%%%%%%%%%%%%%%%%%%%%%%%%%%%%%%%%

%%%%%%%%%%%%%%%%% BODY OF PAPER %%%%%%%%%%%%%%%%%%

\section{Introduction}

Astrophysical plasmas, where the dynamics is dominated by strong magnetic fields, are often a source of high-energy emission. Magnetic reconnection is generally acknowledged as the mechanism powering this emission, through the dissipation of magnetic energy. This process is conjectured to power high-energy emission in the form of flares from magnetars, pulsar wind nebulae and from black hole accretion disks and jets. 

{The magnetised corona above a turbulent accretion disk can be modelled as an ensemble of magnetic flux loops with footpoints tied to the disk. These flux tubes are hypothesised to interact with one another after which their energy is liberated in a reconnection event (\citealt{Tout1992}; \citealt{uzdensky2008}; \citealt{Goodman}; \citealt{Yuan2009}). The reconnecting flux tubes can drive the formation and ejection of energetic plasmoids. These plasmoids can orbit the black hole and power flaring emission from the inner accretion disk (see e.g. \citealt{Broderick2005}; \citealt{Noble2007}; \citealt{Marrone2008}; \citealt{Yuan2009}; \citealt{younsi_2015}). Such X-ray, infrared, and radio flares have been observed from the accretion disk region of Sgr A\mbox{*}, the supermassive black hole in the Galactic Centre  (\citealt{baganoff2001}; \citealt{genzel2003}; \citealt{eckart2006}; \citealt{Meyer2008}; \citealt{neilsen2013}; \citealt{Dexter2014}; \citealt{brinkerink2015}; \citealt{Gravity2018}).}

{The global dynamics of the plasma in the accretion disk is typically modelled within the fluid approximation of general relativistic magnetohydrodynamics (GRMHD). {Within the GRMHD description, reconnection of magnetic field lines occurs due to the resistivity of the plasma breaking Alfv\'{e}n's frozen in theorem (\citealt{Alfven1942}).} This resistivity can be purely numerically sourced, caused by insufficient grid resolution, or it can be explicitly modelled within the set of general relativistic resistive magnetohydrodynamics (GRRMHD) equations. It is however unclear how a realistic resistivity should be modelled and how the model affects the formation and structure of the reconnection layer, the efficiency of the energy conversion, and the production and energisation of plasmoids.} 

{The plasma in the aforementioned astrophysical environments is often magnetically dominated, such that the magnetic enthalpy density is larger than the thermal enthalpy density, i.e. such that the magnetisation}
\begin{equation}
\sigma = B^2 / (4 \pi h \rho c^2) \gtrsim 1,
\label{eq:sigmagg1}
\end{equation}
for a magnetic field with magnitude $B$ and a plasma with mass density $\rho$ and specific enthalpy $h$ and with $c$ the speed of light. 
%op: I found this discussion very incoherent.  Also it shouldn't really be here so high up in the intro.  In any case, I attempted to rectify the argumentation.  
{In a near stationary situation (\citealt{parker_1957}; \citealt{sweet1958}; \citealt{lyutikov_2003}; \citealt{lyubarsky2005}), the rate of reconnecting magnetic flux (the "reconnection rate") is proportional to the ratio of the current sheet inflow velocity to the sheet outflow velocity $v_{\rm in}/v_{\rm out}$. If the plasma can be assumed incompressible, it follows that $v_{\rm in}/v_{\rm out}=\delta/L$ where $\delta$ is the sheet thickness and $L$ its length along the large-scale reconnecting field.  The outflow speed can be estimated as $v_{\rm out} \sim v_{\rm A}$, where $v_{\rm A} = c \sqrt{{\sigma}/({\sigma + 1})}$ is the typical Alfv\'{e}n speed, which becomes close to the speed of light for highly magnetised plasma. 
According to the Sweet-Parker model, the width follows from continuity of the ideal- and resistive electric field across the sheet, giving $\delta=\eta/v_{\rm in}$, where $\eta$ is the resistivity, or the inverse conductivity, of the plasma. Hence the reconnection rate becomes $v_{\rm in}/v_{\rm out}=\eta/(L v_{\rm in})$. The Lundquist number of the plasma is defined as the ratio of the typical advective time scale, or Alfv\'{e}n time $\tau_{\rm A} \sim L/v_{\rm A}$, and the diffusion time scale $\tau_{\rm D} \sim L^2/\eta$:
\begin{equation}
S = \tau_{\rm D} / \tau_{\rm A} = \frac{L v_{\rm A}}{\eta} = \sqrt{\frac{\sigma}{\sigma+1}} \frac{cL}{\eta}.
\label{eq:lundquistnumber}
\end{equation}
The reconnection rate should then scale as $v_{\rm rec} \sim S^{-1/2}$.} For astrophysical plasmas, diffusion occurs on very small scales, resulting in a slow diffusion time scale compared to the Alfv\'{e}n time (or light crossing time $\tau_{\rm c} = L/c$ for $v_{\rm A} \approx c$). Therefore the Lundquist number is typically very large, resulting in Sweet-Parker reconnection rates (\citealt{komissarov2007b}) which are too small to explain the observed  high-energy emission. 

Resistivity in astrophysical systems is however generally considered to be non-uniform and strongly enhanced in the current sheet. The reconnection rate can be increased by a locally enhanced, non-uniform resistivity, that broadens the reconnection layer thickness $\delta$, but does not affect the ambient plasma. Non-uniform resistivity typically depends on plasma variables like the temperature, density, pressure and current density (\citealt{kulsrud}). {Recently, non-uniform resistivity models have been developed for collisional plasmas, based on non-relativistic kinetic modifications to the classical Spitzer resistivity (e.g. \citealt{hirvijoki2016}; \citealt{lingam2017}). In collisionless plasmas, the theory of anomalous resistivity arising from turbulence driven by kinetic instabilities has been developed by \cite{coroniti1977} and \cite{bychenkov1988}. In collisionless plasmas an appropriate model for resistivity can also be determined from Particle-in-Cell simulations (e.g. \citealt{che2017}) or by directly integrating Vlasov's kinetic equation (\citealt{Buchner2005}; \citealt{Buchner2006}). In non-relativistic plasmas the effects of anomalous resistivity have been studied extensively, concluding that localized resistivity can enable the fast \cite{petschek1964} reconnection mechanism (see e.g. \citealt{sato1979}; \citealt{scholer1989}; \citealt{ugai1992}; \citealt{erkaev2000}; \citealt{biskamp2001}; \citealt{kulsrud2001}; \citealt{uzdensky2003}).
This conclusion has been extended by e.g. \cite{zenitani2010}, showing that non-uniform, current-dependent resistivity can also enable fast reconnection in relativistic collisional plasmas. Anomalous resistivity models are extensively applied in magnetohydrodynamics (MHD) simulations of astrophysical plasmas, e.g. by \cite{Schumacher1997} for a setup of two coalescing flux tubes in non-relativistic plasma, or by \cite{Ohsuga2009} for global black hole accretion disk simulations in a pseudo-Newtonian gravitational potential.}

{From $v_{\rm rec} \sim \delta / L$, it is obvious that the reconnection rate also increases for a decreasing length $L$ of the layer. This can occur if the current sheet is broken up into smaller pieces due to the formation of plasmoids as a result of the tearing instability. A stationary reconnecting current sheet is unstable to a fast tearing instability, or plasmoid instability above a critical value of the Lundquist number that is typically quoted as $S \gtrsim S_{\rm c} = 10^4$ (\citealt{loureiro2007}), leading to a reconnection rate of order $10^{-2}v_{\rm A}$ (\citealt{samtaney2009}; \citealt{bhattacharjee2009}; \citealt{uzdensky2010}; \citealt{loureiro2012}; \citealt{huang2013}; \citealt{murphy2013}; \citealt{loureiro2016}; \citealt{uzdensky2016}; \citealt{comisso2016}; \citealt{comisso2017}; \citealt{tolman2018}). Special relativistic resistive magnetohydrodynamic (SRRMHD) simulations have confirmed this critical value of the Lundquist number in relativistic plasmas with a uniform resistivity (\citealt{komissarov2007b}; \citealt{zenitani2010}; \citealt{zanotti2011b}; \citealt{takamoto2013}; \citealt{baty2013}; \citealt{petri2014}; \citealt{delzanna2016}; \citealt{papini2018}). Plasmoids formed by the tearing instability can undergo mergers, bulk acceleration, and growth within the reconnection layer (\citealt{guo2014}; \citealt{sironi2014}; \citealt{sironi2016}; \citealt{werner2018}; \citealt{petropoulou2018}). These plasmoids are conjectured to power flares, accounting for intense variability (\citealt{komissarov2007b}; \citealt{giannios2009b}; \citealt{giannios2010b}; \citealt{giannios2013b}; \citealt{sironi2016}; \citealt{petropoulou2016}).}

{\cite{lazarian1999} proposed that externally driven turbulence can also enhance the reconnection rate in three-dimensional (3D) non-relativistic plasma models. \cite{loureiro2009} carried out two-dimensional (2D) simulations showing that turbulence can enhance reconnection and that the transition to the plasmoid-dominated regime is enabled at Lundquist numbers that are smaller than the classical critical value of $S_{\rm c} = 10^4$. \cite{beresnyak2017} has investigated reconnection in an initially laminar and thin current sheet with 3D non-relativistic resistive MHD, which spontaneously develops oblique tearing instabilities and becomes turbulent. \cite{striani2016} have shown that the kink mode, that is inaccessible in 2D, may loosen the restriction of the critical Lundquist number for plasmoid formation to $S \gtrsim 10^3$. Based on 3D resistive relativistic MHD simulations \cite{takamoto2018} recently claimed that turbulent motions due to a small perturbation in the magnetic field break up the current sheet before the plasmoid regime is reached, resulting in a smaller reconnection rate than in an equivalent 2D setup. There however, it is unclear whether numerical convergence is obtained in these demanding computational regimes, and numerical resistivity may play a role in the dynamics.}

{It remains a fundamental question how a semi-stationary current sheet can form from a stable plasma equilibrium without breaking up before the plasmoid regime is reached. For Sweet-Parker current layers in MHD theory, the growth rate of the plasmoid instability increases with the Lundquist number as $\propto S^{1/4}$ (\citealt{loureiro2007}; \citealt{samtaney2009}). However, since the thickness of a Sweet-Parker current sheet scales as $S^{-1/2}$, very long current sheets cannot realistically form for $S \gg10^4$. \cite{pucci2014} argued that for high, astrophysically relevant, Lundquist numbers, the maximum growth rate $\gamma_{\rm max}$ of the fastest growing mode of the tearing instability can become independent of the Lundquist number and it scales as  $\gamma_{\rm max} \simeq 0.6 \tau_{\rm A}^{-1}$.} 
%op: tau_A here?
This ``ideal'' tearing mode takes over and triggers plasmoid formation before the current sheet reaches a thickness for which the slow Sweet-Parker scaling is valid. The nonlinear evolution of the ideal tearing mode in a pre-existing current sheet has been shown to follow the asymptotic $\gamma_{\rm max} \simeq 0.6 \tau_{\rm A}^{-1}$ scaling by \cite{landi2015} with non-relativistic resistive MHD simulations. The ideal tearing mode in relativistic resistive MHD has recently been studied by \cite{delzanna2016} and \cite{papini2018}, confirming the growth rate of $\gamma_{\rm max} \simeq 0.6 \tau_{\rm A}^{-1}$, independent of $S$ for $S \gtrsim S_{\rm c} \approx 10^6$. The aforementioned 2D and 3D resistive MHD studies however, do not include the formation phase of the current sheet from a generic stable plasma equilibrium; The current sheet is already present at the initial time, which may affect the reconnection properties and the plasmoid formation threshold.
%The thickness of a Sweet-Parker current sheet scales as $\sim S^{1/2}$, whereas the ideal tearing mode becomes effective when a current sheet reaches a thickness of order $S^{-1/3}$. For the largest Lundquist numbers considered in this work, $S=10000$ and $S=20000$, the thickness for the ideal tearing mode to be effective $S^{-1/3}$, is approximately five times larger than $S^{-1/2}$. A forming and thinning current sheet can thus reach a critical thickness for which the tearing instability becomes so fast, that it can never remain in a stable Sweet-Parker regime and the plasmoid instability will determine the reconnection properties. 

To understand the complete process of reconnection it is required to model a developing current sheet starting from a non-reconnecting, stable plasma equilibrium {as has been pointed out by \cite{pucci2014} and \cite{uzdensky2016}}. If a stable plasma state is rearranged or compressed by a driving plasma flow, a current sheet can form that gradually becomes unstable to the tearing instability. 
In recent non-relativistic MHD simulations of two repelling flux tubes, we found that the interaction between such tubes can cause the formation of thin current sheets showing signs of plasmoid formation in both 2D and 3D (\citealt{Keppens}; \citealt{Ripperda}; \citealt{Ripperda2}). Multiple reconnection layers form in between and at the boundaries of the tilting flux tubes. A related and generic case of two stable, attracting flux tubes, with a net zero total current, has recently been studied with a combination of 2D relativistic force-free magnetodynamics, relativistic resistive MHD, and Particle-in-Cell (PiC) methods by \citealt{SironiPorth}. After the flux tubes are driven towards each other by a small velocity perturbation, a thin current sheet develops in between the coalescing structures. This stage is then followed by a merger of the flux tubes, resulting in fast reconnection. Due to the inclusion of the formation of the sheet in the model, the details of reconnection and plasmoid formation are not affected by the initial setup. Based on PiC simulations \cite{SironiPorth} find that macroscopic large-scale magnetic stresses lead to fast reconnection that is nearly an order of magnitude faster compared to plane-parallel cases where the tearing mode is triggered in a pre-existing current sheet. Yet, in their relativistic resistive MHD models the Lundquist number remains too small to trigger the plasmoid instability. PiC simulations have shown that relativistic 2D magnetostatic equilibria known as the `ABC' fields can as well lead to the formation of thin current sheets and fast reconnection (\citealt{east2015}; \citealt{yuan2016}; \citealt{nalewajko2016}; \citealt{Blandford2017}; \citealt{SironiPorth}).

{In this work we extend the force-free magnetodynamics simulations (i.e. in the limit of $\sigma \rightarrow \infty$) of \cite{SironiPorth} {to SRRMHD for high Lundquist numbers in 2D Minkowski spacetime with the {\it Black Hole Accretion Code} ({\tt BHAC}). {\tt BHAC} is a multidimensional framework that has been designed to solve the GRMHD equations in arbitrary spacetimes (\citealt{BHAC}) making use of constrained transport adaptive mesh refinement (AMR) (\citealt{olivares2018proc}). The framework has recently been extended to solve the GRRMHD equations (Ripperda et al., in prep.).} More details about the numerical methods applied in this work are given in Section \ref{sec:numerics}.}
%op: removing repetition
%In the setup explored here, no pre-existing current sheet is assumed, and instead we analyse the formation, the thinning and the break-up due to the plasmoid instability of a current sheet generated by the merger of two flux tubes initially in equilibrium.

We explore a range of Lundquist numbers by setting a uniform resistivity and compare reconnection properties and the plasmoid formation threshold for the forming current sheets. The influence of the magnetisation is explored by varying the initial pressure and density.
We also utilise a spatiotemporally dependent non-uniform resistivity model that depends on the current density, such that it increases in the current sheet. This prescription is compared to a fiducial case with uniform resistivity. 

The regime of high Lundquist numbers and low resistivity is very demanding for any resistive code and therefore serves as a restrictive test for the methods implemented in {\tt BHAC}. Exploring large Lundquist numbers proves to be very difficult, even in 2D simulations, since the reconnection layer becomes narrower for larger Lundquist numbers and the necessary time step and resolution to resolve the thinning current sheet become prohibitive. If the resolution is too low, numerical resistivity may be dominant over physically chosen resistivity. This is the same effect that causes reconnection and plasmoid formation in ideal MHD simulations. To study reconnection in high Lundquist number plasmas, it is essential to include physical resistivity combined with extreme resolutions, to assure that numerical resistivity is negligible. Diffusion due to resistivity is however only important on small scales, where large gradients of the electric and magnetic field may exist (\citealt{komissarov2007b}). Therefore, AMR can greatly enhance the efficiency of the simulation, by only resolving the small diffusion and reconnection regions in and around the current sheet with extremely high resolution. {Our AMR approach allows us to effectively resolve the plasmoid phase of the evolution with $16384^2$ cells. In previous numerical MHD studies of relativistic reconnection, the nonlinear phase at very late stages of the tearing instability for very high Lundquist numbers was never resolved with such high resolutions (\citealt{zenitani2010}; \citealt{zanotti2011b}; \citealt{baty2013}; \citealt{takamoto2013}; \citealt{petri2014}; \citealt{delzanna2016}).}

The paper is organised as follows: In Section \ref{sect:theory} the set of relativistic, resistive magnetohydrodynamics equations in Minkowski spacetime are given and in Section \ref{sec:numerics} we briefly describe the numerical methods that are used in this work. In Section \ref{sect:application} these methods are applied to the coalescence of two magnetic flux tubes causing magnetic reconnection, for a range of astrophysical conditions. We present our conclusions in Section \ref{sect:conclusions}.

\section{Relativistic resistive magnetohydrodynamics}
\label{sect:theory}
We solve the general relativistic resistive magnetohydrodynamics equations in Minkowski spacetime in $3+1$-split form (an extensive treatment of the implementation of GRRMHD in {\tt BHAC} will be provided in a future work). {We follow the derivation of the GRRMHD equations in \cite{Bucciantini}. For the remainder of this paper, we choose a $(-,+,+,+)$ signature for the spacetime metric. We adopt geometrised, dimensionless code units in which the vacuum permeability $\mu_0 = 1$, and vacuum permittivity $\epsilon_0 = 1$. Moreover we set the speed of light $c=1$ and we set all factors $4 \pi = 1$. In flat spacetime the evolution equations for a magnetised fluid are then given by}
\begin{equation}
\partial_t\mathbf{B} + \nabla \times \mathbf{E} = 0,
\label{eq:faraday4}
\end{equation}
\begin{equation}
\partial_t\mathbf{E} - \nabla \times \mathbf{B} = -\mathbf{J},
\label{eq:ampere4}
\end{equation}
\begin{equation}
\nabla \cdot \mathbf{B} = 0,
\label{eq:divB4}
\end{equation}
\begin{equation}
\nabla \cdot \mathbf{E} = q,
\label{eq:divE4}
\end{equation}
\begin{equation}
\partial_t D + \nabla \cdot \left(\rho \Gamma \mathbf{v}\right) = 0,
\label{eq:numberdensity}
\end{equation}
\begin{equation}
\partial_t\tau + \nabla \cdot \left(\mathbf{E}\times\mathbf{B} + h \Gamma^2 \mathbf{v}\right) = 0,
\label{eq:energy}
\end{equation}
\begin{equation}
\partial_t\mathbf{S} + \nabla \cdot \left(-\mathbf{E}\mathbf{E} - \mathbf{B}\mathbf{B} + h \Gamma^2 \mathbf{v}\mathbf{v} + \left[\frac{1}{2}\left(E^2+B^2\right) + p\right] \right) = 0,
\label{eq:momentum}
\end{equation}
{for electric field vector $\mathbf{E}$, magnetic field vector $\mathbf{B}$ and charge density $q$. The conserved quantities, as conserved by equations (\ref{eq:numberdensity})-(\ref{eq:momentum}), are the density $D$, the energy density $\tau$ and the energy flux density $\mathbf{S}$ given by
\begin{equation}
D = \Gamma \rho,
\label{eq:densityD}
\end{equation}
\begin{equation}
\tau = \frac{1}{2}(E^2 +B^2) + \rho h\Gamma^2 - p,
\label{eq:energytau}
\end{equation}
\begin{equation}
\mathbf{S} = \mathbf{E}\times\mathbf{B} + \rho h \Gamma^2 \mathbf{v},
\label{eq:momentumS}
\end{equation}}
{with the specific enthalpy $h = 1 + \frac{\hat{\gamma}}{\hat{\gamma}-1}p/\rho = 1 + 4p/\rho$ for a relativistic ideal gas with adiabatic index $\hat{\gamma} = 4/3$, $p$ the pressure, and $\rho$ the mass density as
measured in the frame co-moving with the fluid.} The fluid velocity as measured by an inertial observer is given by $\mathbf{v}$ and $\Gamma = (1-v^2)^{-1/2}$ is the Lorentz factor. The system is closed by the resistive Ohm's law
\begin{equation}
\mathbf{J} = q \mathbf{v} + \mathbf{J}_{\rm c} =  q \mathbf{v} + \frac{1}{\eta} \Gamma \left[\mathbf{E} + \mathbf{v}\times\mathbf{B} -\left(\mathbf{E}\cdot\mathbf{v}\right)\mathbf{v}\right],
\label{eq:ohmslaw}
\end{equation}
for the current density $\mathbf{J}$ and conduction current $\mathbf{J}_{\rm c}$, with resistivity $\eta(\mathbf{x},t)$ depending on spatial coordinates $\mathbf{x}$ and coordinate time $t$. {Other non-ideal transport effects like viscosity and thermal conduction are neglected in our model.
In {\tt BHAC} the general relativistic resistive magnetohydrodynamics equations are actually solved, however we adopt a flat spacetime, where the lapse function is set to unity, the shift vector is a three-vector equal to zero, and the spatial metric is the identity $3\times 3$ matrix (Ripperda et al., in prep.).}

\subsection{Non-uniform resistivity}
\label{sect:anomalousresistivity}
In its simplest form the resistivity is taken to be constant and homogeneous. However, a more realistic expectation would be a spatiotemporally varying description, depending on plasma parameters. In MHD, resistivity is considered as a macroscopic plasma property, whereas it should actually be calculated from kinetic physics. {Spatiotemporal resistivity can for example arise via a dependence on temperature (e.g. collisional Spitzer resistivity), through collisionless processes (anomalous resistivity) or through the emission or absorption of photons (radiative resistivity).} Non-uniform resistivity is generally thought of as a locally enhanced resistivity in a relatively small, confined area in the current sheet, compared to a smaller background value. It generally depends on the plasma temperature and the density but may also depend on other moments of the distribution function, like the current density (\citealt{kulsrud}). 
{In this work we assume a non-uniform resistivity model that depends purely on the current density $J$. The strength of the dependence is set by the constant parameter $\Delta_{\rm ei}$ as
\begin{equation}
\eta(\mathbf{x},t) = \eta_0 (1+\Delta_{\rm ei}^2 J)\,.
\label{eq:anomalousres_lingam}
\end{equation}
We parametrise microscopic effects through $\Delta_{\rm ei}$. In reality, $\Delta_{\rm ei}$ should depend on the underlying kinetic physics (e.g. on the electron pressure), and on the current density itself (\citealt{lingam2017}). For $\Delta_{\rm ei} \ll 1$ the non-uniformity of the resistivity is considered unimportant, but when $\Delta_{\rm ei} \sim \mathcal{O}(1)$ or even higher, current-dependent corrections in the resistivity become dominant. Such regimes are typically reached in reconnection zones with extremely strong and localized current density and length scales below the typical MHD cells. The non-uniform resistivity may strongly affect the behaviour of magnetic reconnection (\citealt{sato1979}; \citealt{scholer1989}; \citealt{ugai1992}; \citealt{erkaev2000}; \citealt{biskamp2001}; \citealt{kulsrud2001}; \citealt{uzdensky2003}; \citealt{zenitani2010}). The exact form of the non-uniform resistivity in collisionless plasmas can potentially be determined from kinetic simulations and yield sub-grid information for large-scale MHD simulations (see e.g. \citealt{Buchner2005}; \citealt{Buchner2006}; \citealt{che2017} for non-relativistic plasmas). For collisional plasmas, Spitzer resistivity (or further corrections such as  e.g. \citealt{hirvijoki2016}; \citealt{lingam2017} for non-relativistic plasmas) need to be used.}

We will numerically study effects of current-dependent, spatiotemporally varying resistivity in Section \ref{sect:application} by exploring a range of $\Delta_{\rm ei}$. For $J(\mathbf{x},t)$ we take the magnitude of the current density in relativistic plasmas as described by equation (\ref{eq:ohmslaw})  
\begin{equation}
J(\mathbf{x},t) = || q\mathbf{v} +  \Gamma \eta_0^{-1} \left[\mathbf{E} + \mathbf{v} \times \mathbf{B} - \left(\mathbf{E} \cdot \mathbf{v}\right)\mathbf{v}\right]||,
\label{eq:anomalousres_relativistic}
\end{equation}
such that the nonlinear part of the resistivity does not depend on $\eta_0$ but solely on the parameter $\Delta_{\rm ei}$ and the fluid variables $\mathbf{E}$, $\mathbf{B}$ and $\mathbf{v}$. {In future astrophysical applications, where a very small base resistivity $\eta_0$ is considered, the local resistivity $\eta(\mathbf{x},t)$ can be determined with an iterative method, instead of inserting $\eta_0$ directly into equation (\ref{eq:anomalousres_relativistic}), or the non-relativistic current density $\mathbf{J} = \nabla \times \mathbf{B}$ can be used to determine the local resistivity in a co-moving frame. These options have been compared here and show negligible differences for the chosen values of $\eta_0$ and $\Delta_{\rm ei}$.}

\section{Numerical methods}
\label{sec:numerics}
The hyperbolic equations with stiff source terms are solved with either an implicit-explicit (ImEx) scheme (\citealt{Palenzuela2}) or a Strang split scheme (\citealt{Komissarov}). 

For the ImEx scheme, equations (\ref{eq:faraday4}) and (\ref{eq:numberdensity})-(\ref{eq:momentum}), not containing any stiff source terms, are solved with an explicit second-order Runge-Kutta (RK) step. The left-hand side of equation (\ref{eq:ampere4}) is solved with the same second-order RK scheme to obtain the intermediate solution $\mathbf{E}\mbox{*}$. The stiff right-hand side of equation (\ref{eq:ampere4}) is added in an additional implicit step, solving for
\begin{equation}
\begin{split}
\mathbf{E} = & \frac{\eta}{\eta+\Gamma \Delta t}\left(\mathbf{E}{\mbox{*}}- q \mathbf{v} \Delta t\right) - \\
& \frac{\Gamma \Delta t}{\eta+\Gamma \Delta t}\left[\mathbf{v}\times\mathbf{B} - \Gamma \eta \frac{\left(\mathbf{E}{\mbox{*}}- q \mathbf{v} \Delta t\right) \cdot \mathbf{v}}{\eta \Gamma + \Delta t}\mathbf{v}\right],
\end{split}
\label{eq:E_imex_intermediate_comp_res}
\end{equation}
with the time step $\Delta t$ from the explicit RK step. This implicit inversion step is taken within a Newton-Raphson iteration to transform the conserved variables into primitive variables $\mathbf{v}$, $\rho$ and $p$. The iteration is considered to be converged if both the error on the primitive variables and on the electric field $\mathbf{E}$ are under a tolerance of $10^{-13}$.

In particular cases, the Strang split scheme is more robust and we choose to apply this method. Here, equations (\ref{eq:faraday4}) and (\ref{eq:numberdensity})-(\ref{eq:momentum}), not containing any stiff source terms, are solved with an explicit third-order Runge-Kutta (RK) step. The left-hand side of equation (\ref{eq:ampere4}) is solved with the same third-order RK scheme to obtain the intermediate solution $\mathbf{E}\mbox{*}$. The stiff part of equation (\ref{eq:ampere4}) is solved semi-analytically, by assuming a reduced, linear Amp\`{e}re's law $\partial_t (\mathbf{E}) = -\mathbf{J}$, such that
\begin{equation}
\partial_t (\mathbf{E}_{\|}) = -\frac{\Gamma}{\eta}\left[\mathbf{E}_{\|} - \left(\mathbf{v}\cdot\mathbf{E}\right)\mathbf{v}\right] - q\mathbf{v},
\label{eq:reducedamperepar}
\end{equation}
\begin{equation}
\partial_t (\mathbf{E}_{\perp}) = -\frac{\Gamma}{\eta}\left[\mathbf{E}_{\perp} + \mathbf{v}\times\mathbf{B}\right],
\label{eq:reducedampereperp}
\end{equation}
are solved to obtain the electric field at time $t + \Delta t$ 
\begin{equation}
\begin{split}
\mathbf{E} =  &  \mathbf{E}_{\|}\mbox{*} \exp\left[-\frac{\Delta t}{\Gamma \eta}\right] - \mathbf{v}\times\mathbf{B} + \\
 &  \left(\mathbf{E}_{\perp}\mbox{*} + \mathbf{v}\times\mathbf{B}\right)\exp\left[-\frac{\Gamma \Delta t}{\eta}\right] - q \mathbf{v} \Delta t,
\end{split}
\label{eq:Eparstrang}
\end{equation}
with $\mathbf{E}_{\|}$ the electric field vector parallel to the velocity field and $\mathbf{E}_{\perp}$ the electric field vector perpendicular to the velocity field, such that $\mathbf{E}\mbox{*} = \mathbf{E}_{\|}\mbox{*}+\mathbf{E}_{\perp}\mbox{*}$ is the electric field at the intermediate solution obtained from the explicit RK step. The velocity field $\mathbf{v}$ and the magnetic field $\mathbf{B}$ are obtained from the explicit RK step as well.

In both cases we use a Total Variation Diminishing Lax-Friedrichs scheme (TVDLF) and we employ a Cada reconstruction scheme (\citealt{Cada}) to compute the fluxes, and we use a Runge-Kutta integration with a Courant number of 0.4. The performance and accuracy of the schemes is briefly compared in Section \ref{sect:application}. The magnetic field is kept solenoidal, obeying equation (\ref{eq:divB4}) to roundoff-error by means of the staggered constrained transport scheme of \cite{balsara1999} (see \citealt{olivares2018proc} for details on the implementation in {\tt BHAC}).
The charge density (\ref{eq:divE4}) is obtained by numerically taking either the central or the limited divergence of the evolved electric field. For more details on the numerical aspects we refer to \cite{BHAC}.

\section{Relativistic reconnection in merging flux tubes}
\label{sect:application}

We conduct simulations for a range of resistivities and resolutions. We adopt a setup of two modified Lundquist tubes from \cite{SironiPorth} where a flux rope is described by the magnetic field 
\begin{equation}
\mathbf{B}(r \leq r_{\rm j}) = \alpha_{\rm t} c_{\rm t}  J_1(\alpha_{\rm t} r) \mathbf{e_{\phi}} + \alpha_{\rm t} c_{\rm t} \sqrt{J_0(\alpha_{\rm t} r)^2 + \frac{C}{(\alpha_{\rm t} c_{\rm t})^2}} \mathbf{e_z}.
\label{eq:lundquist}
\end{equation}
The toroidal magnetic field $B_{\phi}$ is initialised from a vector potential
\begin{equation}
A_z (r \leq r_{\rm j}) = c_{\rm t} J_0(\alpha_{\rm t} r),
\label{eq:vecpot}
\end{equation}
to keep $\nabla \cdot \mathbf{B} = 0$ using a staggered constrained transport scheme. The boundary conditions are periodic in all directions. All scales are given in terms of the radius of the rope $r_{\rm j} = 1$. Furthermore, we set the constant $C=0.01$ such that the minimum $B_{\rm z}$ component remains positive. $J_0$ and $J_1$ are Bessel functions of the zeroth and first kind respectively and the constant $\alpha_{\rm t} \approx 3.8317$ is the first root of $J_0$. The parameter $c_{\rm t} = 0.262$ is found such that the maximum value of the $\max(B_{\rm z}) = B_0$ field is unity. The solution for each tube is terminated at $r = r_{\rm j} = 1$, corresponding to the first zero of $J_1$, and the field remains uniform at $B_{\rm z} (r > r_{\rm j}) = B_{\rm z}(r_{\rm j}) = \alpha_{\rm t} c_{\rm t} \sqrt{J_0(\alpha_{\rm t})^{2} + C/(\alpha_{\rm t}  c_{\rm t})^{2}}$. {The toroidal field vanishes at the boundary of each rope where $A_{z} (r > r_{\rm j}) =  A_{z}(r_{\rm j}) = c_{\rm t}  J_0(\alpha_{\rm t}) $, such that the total current in each flux tube is zero (and no surface currents are present in the initial setup).} We exploit the modified version of Lundquist's force-free cylinders (including the factor proportional to C) as suggested by \cite{SironiPorth}, such that the guide-field in the out-of-plane direction does not change sign inside the ropes.
{The equations are solved on a 2D Cartesian grid ($x,y$), with $x \in [-3,3]$ and $y \in [-3,3]$. Each flux tube has a profile that can be described in polar coordinates (centred in the centre of the tube) as $(r,\theta) = (\sqrt{(x\pm1)^2+y^2},\arctan(y/(x\pm1)))$, where $\pm$ corresponds to the right and left tube, respectively. Initially the two tubes are just touching each other and their centres are located at $y=0$ and $x \pm 1$.}
The tubes are perturbed by a small electric field that pushes them towards each other
\begin{equation}
\mathbf{E} = -\mathbf{v}_{\rm kick}\times\mathbf{B},
\label{eq:vpush}
\end{equation}
with $\mathbf{v}_{\rm kick} = (\pm 0.1,0,0)$ the kick velocity (normalised to $c$) with the $\pm$ corresponding to the left and right rope respectively. The dynamical evolution is independent of the kick velocity, such that the choice of $\mathbf{v}_{\rm kick}$ is not restricting the validity of the conclusions (\citealt{SironiPorth}).
{The pressure $p=p_0$ and density $\rho = \rho_0$ are initially uniform such that the equilibrium is force-free.
The fluid obeys an equation of state for a relativistic ideal gas with adiabatic index $\hat{\gamma} = 4/3$, such that the specific enthalpy $h(\rho,p) = 1 + \frac{\hat{\gamma}}{(\hat{\gamma}-1)}p/\rho = 1 + 4p/\rho$. For the reference cases the pressure is set at $p_0=0.05$ and the density at $\rho_0 = 0.1$. The maximum of the dimensionless initial magnetic field magnitude is set as $B_0 = \max(B_{\rm z}) = 1$ in all cases, such that plasma-$\beta_0= 2 p_0 / B_0^2 = 0.1$ and $\sigma_0= B_0^2/(h_0 \rho_0) = 3.33$ in the reference cases.} {We distinguish two different Lundquist numbers; {one} at initialisation of each case, where it is assumed that the Alfv\'{e}n speed approaches the speed of light during reconnection, such that $v_{\rm A} \approx c = 1$ and the half-length of the sheet corresponds to the radius of the flux tube $r_{\rm j} = 1$, resulting in $S_{\rm ini} = \eta^{-1}$; {and} an effective Lundquist number $S_{\rm eff} = L v_{\rm A} / \eta$, where the half-length $L$ of the current sheet is measured during the evolution, and the Alfv\'{e}n speed is determined for each case as $v_A = c\sqrt{\sigma_0 / (\sigma_0 + 1)}$ (where $c=1$). Time is measured in units of light crossing time of the flux tube radius $t_{\rm c} = r_{\rm j} / c = 1$.}

The 2D domain is divided in a varying number of cells to assess numerical convergence, both with and without AMR. We explore the effect of resolution on the threshold for plasmoid formation by varying the base resolution and the maximum AMR level. In the AMR cases, the mesh is refined when there are steep gradients in $B_{\rm z}$, $E_z$ and $\rho$. These variables are chosen because they show the largest variations during the nonlinear phase of the evolution. 
{Furthermore we vary the resistivity $\eta$ to determine a resistivity threshold for plasmoid formation. The pressure and density are then varied to determine the effect of magnetisation and plasma-$\beta$ on the plasmoid instability. Finally, non-uniform resistivity models are explored to determine how they affect the reconnection rate.}

%op: This description is very tiring here.  Can we move the description to the corresponding sections where there are more relevant (or remove altogether as we certainly give numbers again)?  Apart from the concrete values we already said all this.  I will remove it for now, bring it back if really needed though.
%{Furthermore, we explore the effect of magnetisation and plasma-$\beta_0$ on plasmoid formation for the fiducial values of $\eta = 10^{-4}$ and $\eta = 5\times 10^{-5}$. The background magnetisation, $\sigma_0$ and plasma-$\beta_0$ are varied as $\beta_0 = 2 p_0 / B_0^2 \in [0.01, 1]$ by changing the pressure $p_0$, and $\sigma_0 = B_0^2/(h_0 \rho_0) \in [0.476, 3.33]$, by changing the density $\rho_0$ and the pressure $p_0$. In these cases the assumption that $L = 1$ and $v_{\rm A} = 1$ becomes invalid and the length of the current sheet and the Alfv\'{e}n speed are determined per case.}
%The reference case with base resistivity $\eta_0 = 10^{-4}$ and $\sigma_0=3.33$, $\beta_0=0.1$ is then used to explore effects of spatiotemporally dependent resistivity set by equation (\ref{eq:anomalousres_lingam}). The parameter $\Delta_{\rm ei}$ indicates the importance of unresolved kinetic physics and we explore four cases: $\Delta_{\rm ei}^2 = 0.001$; $\Delta_{\rm ei}^2 = 0.01$; $\Delta_{\rm ei}^2 = 0.1$; and the limiting case $\Delta_{\rm ei}^2 = 1$.}

All runs are listed in Table \ref{tab:example_table}, where we catalog the applied numerical method to resolve the stiff source terms, the $x$-component of the kick velocity $v_{\rm kick,x}$, the plasma-$\beta_0$, magnetisation $\sigma_0$, the uniform resistivity $\eta_0$, the non-uniform resistivity enhancement factor $\Delta^2_{\rm ei}$, the maximum resistivity in the domain measured over the whole period of the evolution, whether or not the plasmoid instability is triggered, the base resolution $N_{\rm base}$ (total number of cells on the lowest AMR level), and the effective resolution $N_{\rm eff}$ (total number of cells if all AMR levels are fully utilised).

In cases with $\beta_0 \geq 0.5$ we find it beneficial to use an ImEx scheme, rather than a Strang split scheme, whereas for plasma-$\beta \lesssim 0.5$ the Strang split scheme is more robust, confirming the findings of \cite{Palenzuela2}. The advantage of the Strang split scheme lies in its ability to capture steep gradients more accurately than the ImEx scheme, due to the stability constraint on the time step that depends linearly on the resistivity. This however results in a much more expensive computation for low resistivity. The two schemes are compared for cases Am2S and Am2, with $\eta = 5\times 10^{-5}$, $\beta_0=0.5$ and $\sigma_0 = 0.476$, confirming that they provide the same results even after 20 light crossing times.

\begin{table*}
	\centering
	\caption{The simulated cases and several characteristic parameters.}
	\label{tab:example_table}
	%\scriptsize
	%\tiny
	%\small
	%\footnotesize
	\begin{tabular}{lcccccccccccr}
		\hline
		Run & method & $v_{\rm kick}$ & $\beta_0$ & $\sigma_0$ &  $\eta_0$ & $\Delta^2_{\rm ei}$   & $\max(\eta)$ & plasmoids & $N_{\rm base}$ & $N_{\rm eff}$  \\
		\hline
		Al & Strang & $0.1c$ & 0.1 & 3.33& $5 \times 10^{-5}$ & 0 & $5 \times 10^{-5}$ & yes  & $4096^2$ &  $4096^2$  \\
		A & Strang& $0.1c$ & 0.1 & 3.33& $5 \times 10^{-5}$ & 0 & $5 \times 10^{-5}$  & yes & $512^2$ &  $8192^2$   \\
		Ah   & Strang& $0.1c$ & 0.1 & 3.33& $5 \times 10^{-5}$ & 0 & $5 \times 10^{-5}$  & yes & $2048^2$ &  $16384^2$  \\
		Am1 & Strang & $0.1c$ & 0.01 & 3.33& $5 \times 10^{-5}$  & 0 &$5 \times 10^{-5}$ & yes & $512^2$&  $8192^2$ \\
             Am2S  &Strang  & $0.1c$ & 1 & 0.476& $5 \times 10^{-5}$  & 0 &$5 \times 10^{-5}$ & no & $512^2$&  $8192^2$ \\
		Am2 & ImEx & $0.1c$ & 1 & 0.476& $5 \times 10^{-5}$  & 0 &$5 \times 10^{-5}$ & no & $512^2$&  $8192^2$ \\
		Am3 & ImEx & $0.1c$ & 1 & 0.497& $5 \times 10^{-5}$  & 0 &$5 \times 10^{-5}$ & no & $512^2$&  $8192^2$ \\
		Am4 & ImEx & $0.1c$ & 0.1 & 0.833& $5 \times 10^{-5}$  & 0 &$5 \times 10^{-5}$ & no & $512^2$&  $8192^2$ \\
		Am5 & ImEx & $0.1c$ & 0.01 & 0.980& $5 \times 10^{-5}$  & 0 &$5 \times 10^{-5}$ & no & $512^2$&  $8192^2$ \\
		Am6 & ImEx & $0.1c$ & 0.5 & 0.909& $5 \times 10^{-5}$  & 0 &$5 \times 10^{-5}$ & no & $512^2$&  $8192^2$ \\
		Am7 & ImEx & $0.1c$ & 0.5 & 0.990& $5 \times 10^{-5}$  & 0 &$5 \times 10^{-5}$ & yes & $512^2$&  $8192^2$ \\
		Am8 & ImEx & $0.1c$ & 0.5 & 0.999& $5 \times 10^{-5}$  & 0 &$5 \times 10^{-5}$ & yes & $512^2$&  $8192^2$ \\
            Bll  & Strang & $0.1c$ & 0.1 & 3.33& $1 \times 10^{-4}$  & 0 &$1 \times 10^{-4}$ & yes  & $2048^2$&  $2048^2$  \\
      	Bl  & Strang & $0.1c$ & 0.1 & 3.33& $1 \times 10^{-4}$  & 0 &$1 \times 10^{-4}$  & no & $4096^2$ & $4096^2$  \\
		Bh  & Strang & $0.1c$ & 0.1 & 3.33& $1 \times 10^{-4}$  & 0 &$1 \times 10^{-4}$  & no & $8192^2$  &  $8192^2$ \\
 		B  & Strang & $0.1c$ & 0.1 & 3.33& $1 \times 10^{-4}$  & 0 &$1 \times 10^{-4}$  & no & $512^2$&  $8192^2$ \\
		Bhl  & Strang & $0.1c$ & 0.1 & 3.33& $1 \times 10^{-4}$  & 0 &$1 \times 10^{-4}$ & no & $1024^2$&  $8192^2$ \\
		Blh  & Strang & $0.1c$ & 0.1 & 3.33& $1 \times 10^{-4}$  & 0 &$1 \times 10^{-4}$ & no & $512^2$&  $16384^2$ \\
		Bhh  & Strang & $0.1c$ & 0.1 & 3.33& $1 \times 10^{-4}$  & 0 &$1 \times 10^{-4}$ & no & $2048^2$&  $16384^2$ \\
             Bnr1  & Strang & $0.1c$ & 0.1 & 3.33& $1\times 10^{-4}$ & 1 &$404.7\times10^{-4}$ & no & $512^2$& $8192^2$  \\
		Bnr2   & Strang& $0.1c$ & 0.1 & 3.33& $1\times10^{-4}$ & 0.1 & $264.6\times10^{-4}$ & no & $512^2$& $8192^2$  \\
		Bnr3   & Strang& $0.1c$ & 0.1 & 3.33& $1\times10^{-4}$ & 0.01 &$6.7\times10^{-4}$ & no & $512^2$& $8192^2$  \\
		Bnr4   & Strang& $0.1c$ & 0.1 & 3.33& $1\times 10^{-4}$ & 0.001 & $1.3\times10^{-4}$ & no & $512^2$& $8192^2$ \\
		Bm1 & Strang & $0.1c$ & 0.01 & 3.33& $1 \times 10^{-4}$  & 0 &$1 \times 10^{-4}$ & no & $512^2$&  $8192^2$ \\
		Bm2 & Strang & $0.1c$ & 0.01 & 10.0& $1 \times 10^{-4}$  & 0 &$1 \times 10^{-4}$ & yes & $512^2$&  $8192^2$ \\
             C  & Strang & $0$ & 0.1 & 3.33& $1 \times 10^{-4}$  & 0 &$1 \times 10^{-4}$ & no  & $512^2$&  $8192^2$  \\
            Ci  & Strang & $0$ & 0.1 & 3.33& $0$  & 0 &$0$ & no & $512^2$&  $8192^2$  \\
             D  & Strang & $0.1c$ & 0.1 & 3.33& $1 \times 10^{-3}$  & 0 &$1 \times 10^{-3}$ & no & $512^2$&  $8192^2$ \\
		E & Strang  & $0.1c$ & 0.1 & 3.33& $1 \times 10^{-2}$  & 0 &$1 \times 10^{-2}$ & no & $512^2$&  $8192^2$ \\

		\hline
			\end{tabular}
\end{table*}

{\subsection{Dependence on resolution}}
%\op{Sorry to come across a bit negative, but from here onwards its a back-and-forth between convergence and physics which seems just sprinkled in for color.  Hence I'm not sure if the physical arguments are just there to add sense to the resolution study.  So I end up not paying much attention to either of the discussion points.  
%The section title is 'dependence on resistivity'.  The resolution study is admirable, however in a non-technical paper like this, its not in the focus (and might be demoted to the appendix by some, especially since we  achieve converged results -- then the unconverged results are ultimately uninteresting).  Suggestion would be to write a brief discussion on the convergence in its own section.  Then paste together the physical discussion.  This will shorten the paper a bit.  }

In this section we determine the resolution for which the results converge numerically in the nonlinear plasmoid regime. The effective maximum resolution is repeatedly doubled up to a point where the physics is independent of numerical resistivity. A higher resolution allows for the current sheet to become thinner, and for the current density and the electric energy density to become higher. We claim converged results if the thickness of the current sheet and the peaks of electric energy density no longer depend on resolution. For too low resolutions, numerical resistivity can artificially trigger the plasmoid instability, such that the electric energy density and the reconnection rate are overestimated. 
\begin{figure*}
\begin{center}
\subfloat{\includegraphics[width=0.75\columnwidth]{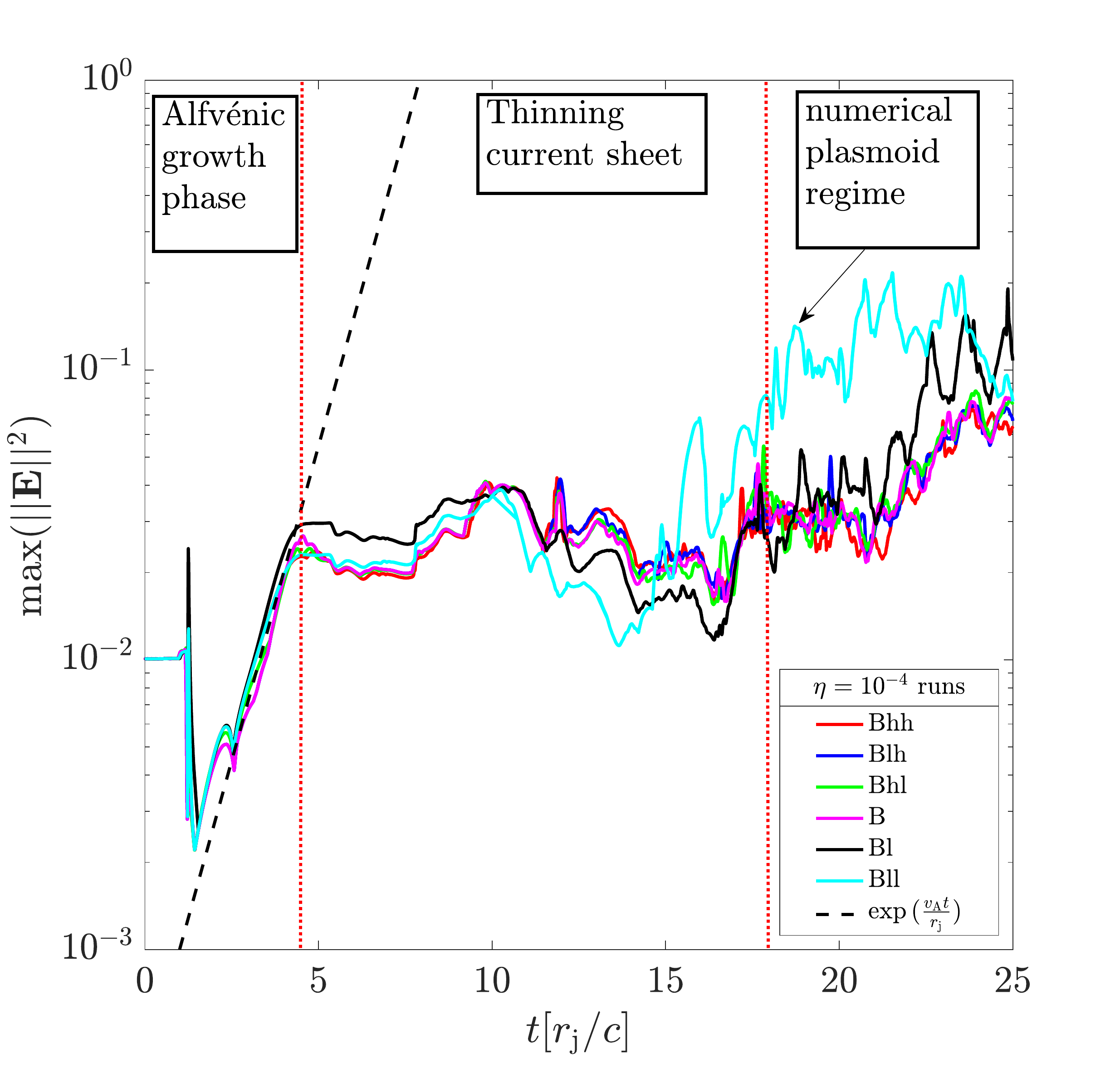}}
\subfloat{\includegraphics[width=0.75\columnwidth]{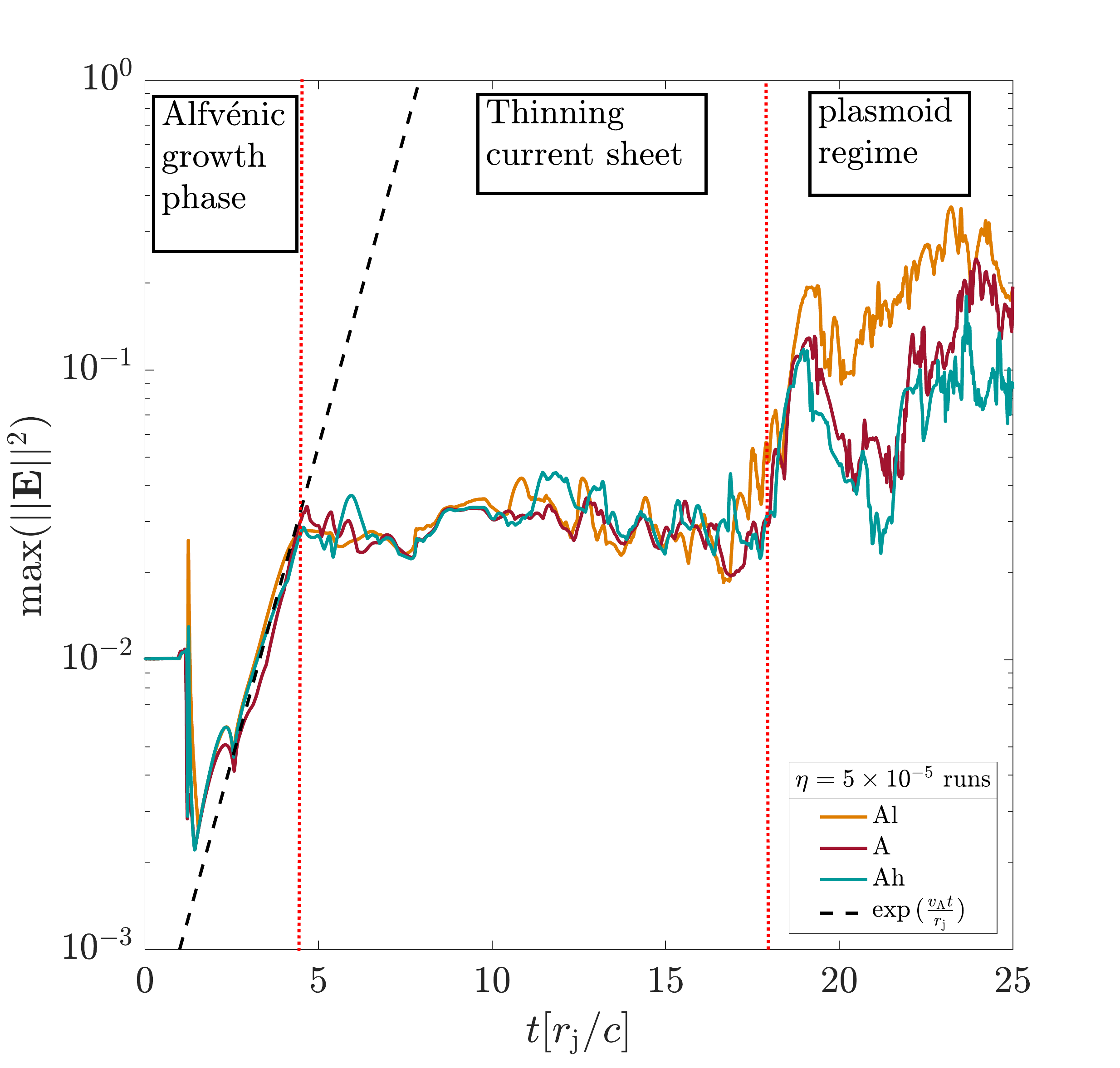}}
\caption{Peak electric energy for all runs with uniform resistivity $\eta = 10^{-4}$ (left-hand panel) and $\eta = 5 \times 10^{-5}$ (right-hand panel). The dashed black line shows the Alfv\'{e}nic growth rate $\propto \exp{(v_{\rm A} t / r_{\rm j})} \approx \exp{(c t/ r_{\rm j})}$ of the electric energy density. The vertical red dotted lines separate the three different evolutionary stages: The Alfv\'{e}nic growth phase of the flux tube coalescence, the thinning of the current sheet, and the phase where the plasmoid instability is triggered for $\eta = 5 \times 10^{-5}$. For $\eta = 10^{-4}$, the plasmoid instability is only triggered by numerical resistivity if the resolution is not high enough, in run Bll.}
\label{fig:peakfieldsdmin4}
\end{center}
\end{figure*}
In Figure \ref{fig:peakfieldsdmin4} the temporal evolution of the maximum energy density in the electric field is shown for all runs with $\eta = 10^{-4}$ in the left-hand panel and for all runs with $\eta = 5\times 10^{-5}$ in the right-hand panel, both with $\sigma_0 = 3.33$ and $\beta_0 = 0.1$. The maximum of the energy density in the electric field, $\max(||\mathbf{E}||^2)$ is determined over the whole domain, at all timesteps, and it is a very strict indicator for convergence of the MHD results at varying resolutions. Low resistivity results in the most restrictive resolution requirements, because the thickness of the forming current sheet is proportional to the square root of the resistivity. 

The initial exponential growth phase (left of the first vertical red dotted line, $t \lesssim 5t_{\rm c}$) where the energy density grows as $\propto \exp{(v_{\rm A} t/ r_{\rm j})} \approx \exp{(c t/ r_{\rm j})}$ (indicated by the dashed black lines), is accurately described by all considered resolutions. From $t \gtrsim 5 t_{\rm c}$, the coalescence instability has saturated and the current sheet has fully formed in between the flux tubes. The nonlinear regime (in between the two vertical red dotted lines), in which the current sheet gets thinner, is reached at $t \approx 5 t_{\rm c}$ for both resistivities, and for all resolutions. {In this period, the current sheet remains laminar and becomes thinner and the resolution needs to be sufficiently fine to avoid numerical resistivity to affect the evolution. During this stage no plasmoids are formed.} 
At $t \approx 17 t_{\rm c}$ the simulation reaches a secondary nonlinear regime (right of the second vertical red dotted line), where a higher variability of the electric field energy density and the current density is observed, that is related to the onset of the plasmoid instability. 
\begin{figure}
\begin{center}
\includegraphics[width=8cm]{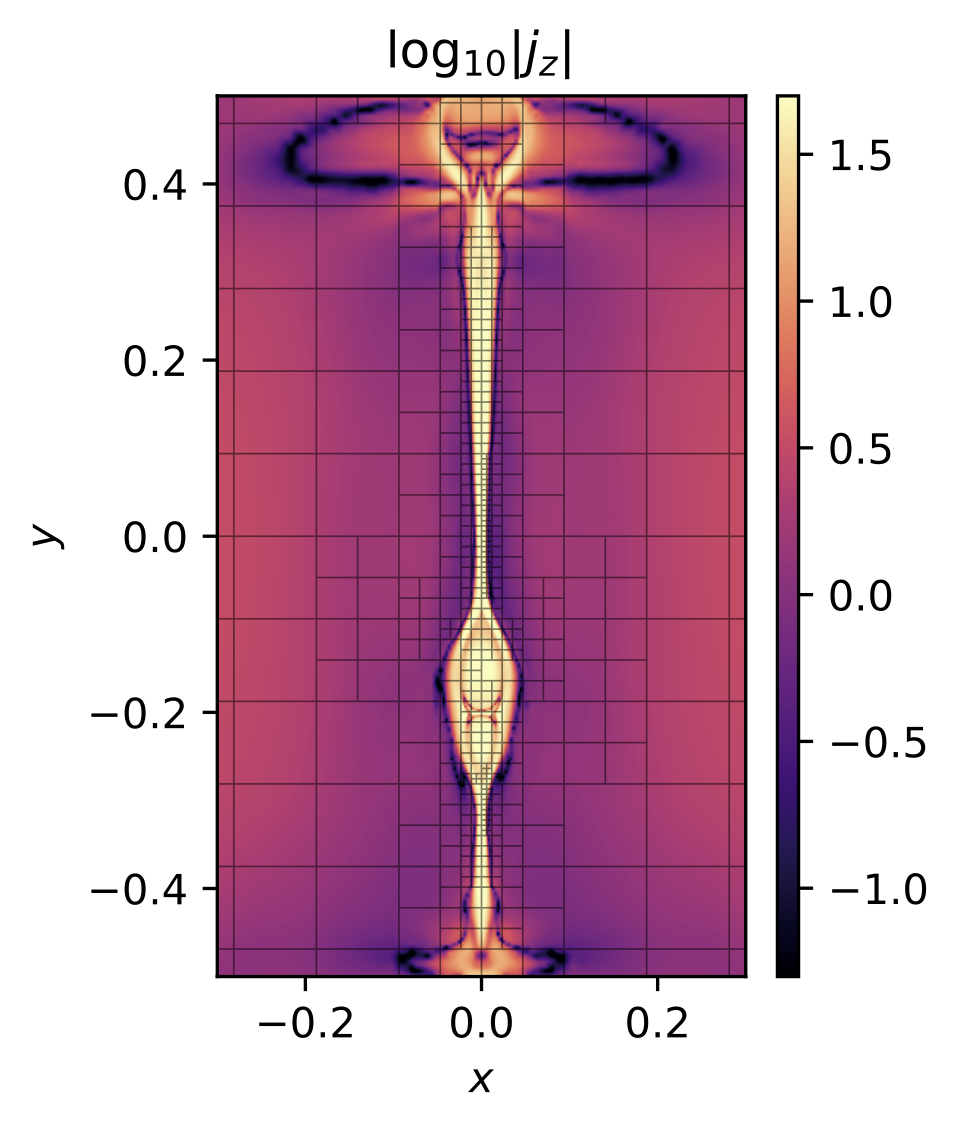}
\caption{Zoom of the out-of-plane current density magnitude $|J_{\rm z}|$ in the current sheet for run Am8 ($\eta = 5\times10^{-5}$; $\sigma_0=0.999$; $\beta_0=0.5$) at  $t=20 t_{\rm c}$. The AMR grid blocks containing 8x8 cells are plotted on top and the base resolution $512^2$ is visible at the left and right image boundary. The grid refines around the current sheet, where the maximum resolution $8192^2$ is visible at the plasmoids. The merger of two plasmoids and the formation of a secondary (horizontal) current sheet is accurately captured by the refined mesh. Previously formed plasmoids are expelled from the top and bottom outflow regions. The colour scale is constrained to range from 0.05 and 50.}
\label{fig:AMRzoom}
\end{center}
\end{figure}

For $\eta=10^{-4}$ the plasmoid regime is reached for run Bll with resolution $2048^2$, but not for higher resolution runs. In run Bll the electric energy density also increases significantly compared to the higher resolution runs. This shows that by choosing a resolution that is too low, reconnection and plasmoid formation is induced via locally prevailing numerical resistivity. Resolution $2048^2$ accurately resolves the growth phase, where resistivity has little effect, but fails to reproduce the higher resolution results in the nonlinear phase from $t \approx 10 t_{\rm c}$ onwards. In all runs with $N \geq 4096^2$ artificial plasmoid formation is avoided and the results converge even in the far nonlinear phase at $t \gtrsim 17 t_{\rm c}$ (i.e, our chosen resistivity is safely larger than the numerical resistivity). We find that the base resolution is unimportant for convergence, since for a (too) low base resolution, the AMR level increases in a large part of the domain from the start of the simulation, as confirmed in runs Ah, Bhl and Bhh. The AMR results have been confirmed by uniform resolutions up to $4096^2$ in runs Al, Bll and Bl. The nonlinear phase has been evolved until $t = 9 t_{\rm c}$ by run Bh with uniform resolution $8192^2$ after which it quickly became too expensive to continue without mesh refinement.

For $\eta = 5\times10^{-5}$ cases, $4096^2$ is the lowest effective resolution considered (corresponding to the highest refinement level). The orange line for run Al, in the right-hand panel, shows that even for this resolution, the maximum electric field is overestimated in the far nonlinear regime, even if it matches the higher resolution runs before $t \approx 17 t_{\rm c}$. For resolutions $N \gtrsim 8192^2$ convergence is obtained up to the far nonlinear regime. These resolutions are only attainable by applying multiple AMR levels.

Figure \ref{fig:AMRzoom} presents a zoom of the magnitude of the out-of-plane current density $|J_{\rm z}|$ in the current sheet of run Am8 ($\eta = 5 \times 10^{-5}$, $\sigma_0 = 0.999$, $\beta_0 = 0.5$), showing interacting plasmoids at $t = 20 t_{\rm c}$. The refined grid is plotted on top, showing all five AMR levels, with the finest level ($8192^2$) accurately capturing the plasmoids and the coarsest level ($512^2$) in between the initial flux ropes. Two plasmoids are expelled from the top and bottom outflow regions, and in the middle of the current sheet two plasmoids have merged and formed a secondary (horizontal) current sheet that is only captured with the highest refinement level. Applying AMR results in a speed-up between 10 and 100 times compared to uniform resolution runs at the high refinement level, and a data reduction of close to 16 times.

\subsection{Dependence on resistivity}
\label{sect:lundquistcomparison}

In this section we explore the dependence of the plasmoid instability on the (uniform) resistivity $\eta$ and hence on the Lundquist number $S_{\rm ini} \sim 1/\eta$. We determine the resistivity threshold for which the plasmoid instability is triggered and the reconnection rate is enhanced. The resistivity is varied in a range of $\eta = 10^{-2}$ (run E), $\eta = 10^{-3}$ (run D), $\eta = 10^{-4}$ (run B) and $\eta = 5\times10^{-5}$ (run A) to evaluate when the plasmoid instability occurs, while $\sigma_0=3.33$ and $\beta_0=0.1$ are kept constant and the resolution is set at $N_{\rm eff} = 8192^2$. The Alfv\'{e}n speed is determined as $v_{\rm A} = \sqrt{3.33 / 4.33} \approx 1$ and the typical length scale $L \approx r_{\rm j} = 1$, such that the Lundquist numbers become $S_{\rm ini} \approx 1/\eta = 10^2$, $S_{\rm ini} \approx 1/\eta = 10^3$, $S_{\rm ini} \approx 1/\eta = 10^4$ and $S_{\rm ini} \approx 1/\eta = 2 \times 10^4$ respectively.

\begin{figure*}
\begin{center}
%\subfloat{\includegraphics[width=0.2\columnwidth,  clip=true]{jz_etadmin2_8192x8192_t5}}
%\subfloat{\includegraphics[width=0.2\columnwidth,  clip=true]{jz_etadmin3_8192x8192_t5}}
\subfloat{\includegraphics[width=0.661\columnwidth,  clip=true]{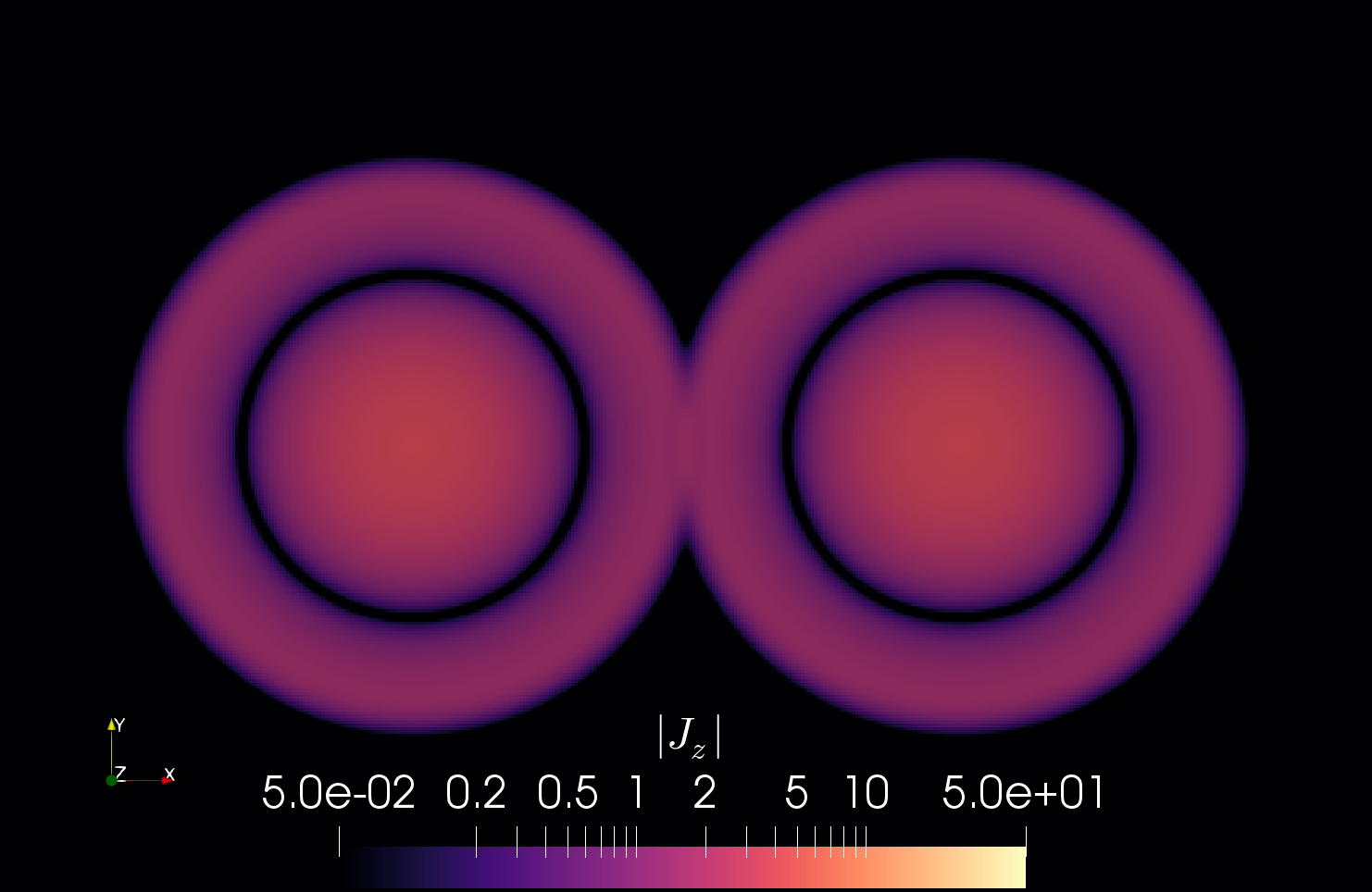}}
\subfloat{\includegraphics[width=0.66\columnwidth,  clip=true]{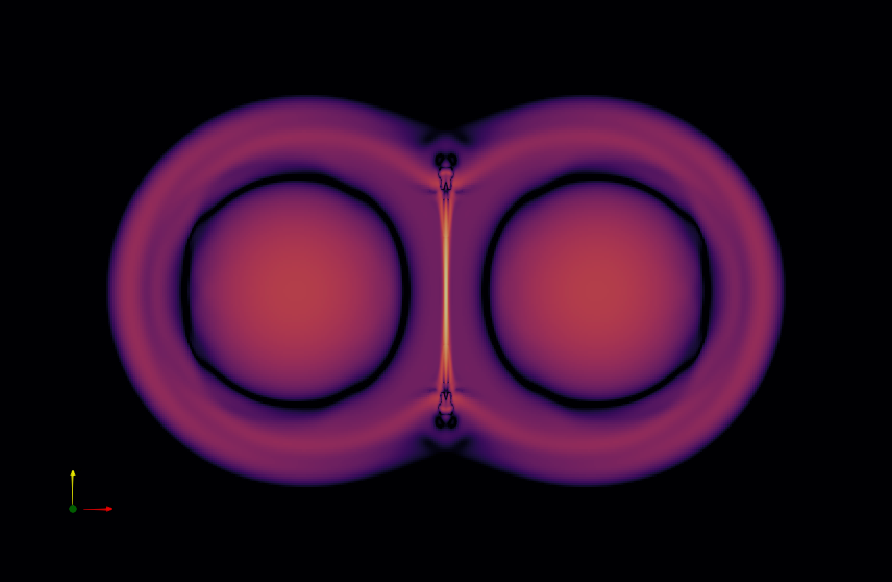}}
\subfloat{\includegraphics[width=0.66\columnwidth,  clip=true]{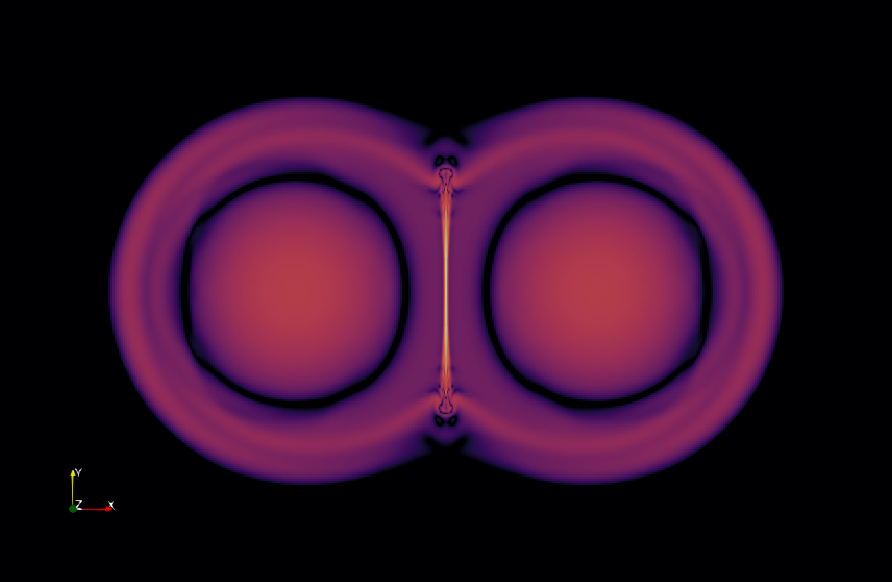}}

%\subfloat{\includegraphics[width=0.2\columnwidth,  clip=true]{jz_etadmin2_8192x8192_t10}}
%\subfloat{\includegraphics[width=0.2\columnwidth,  clip=true]{jz_etadmin3_8192x8192_t10}}
\subfloat{\includegraphics[width=0.66\columnwidth,  clip=true]{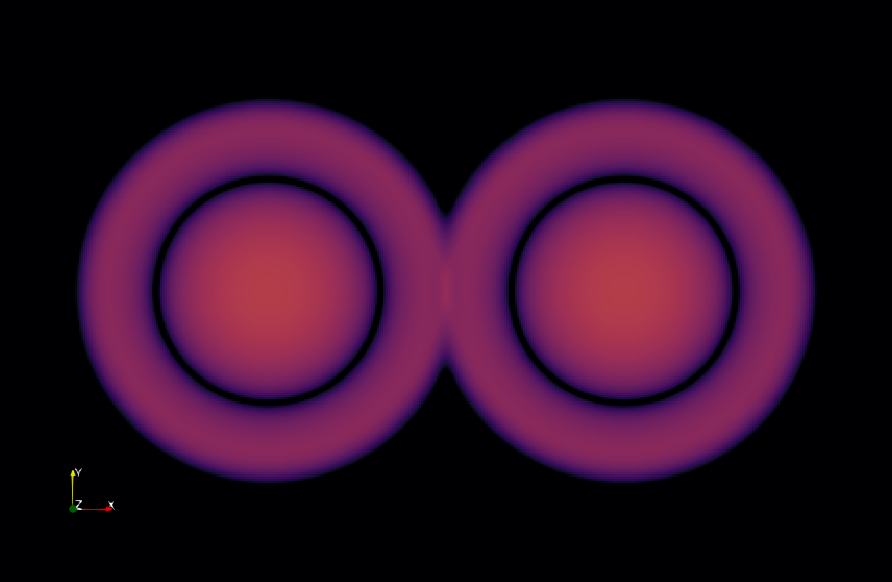}}
\subfloat{\includegraphics[width=0.66\columnwidth,  clip=true]{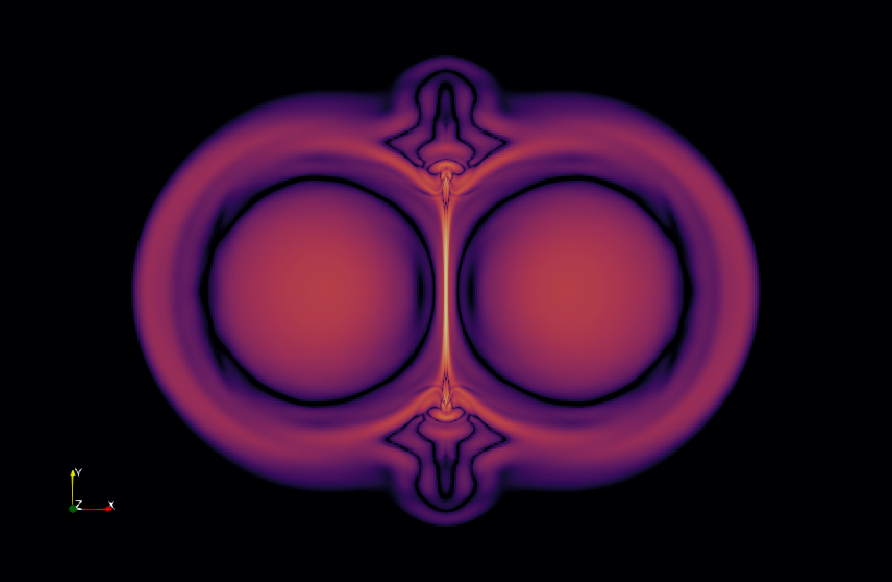}}
\subfloat{\includegraphics[width=0.66\columnwidth,  clip=true]{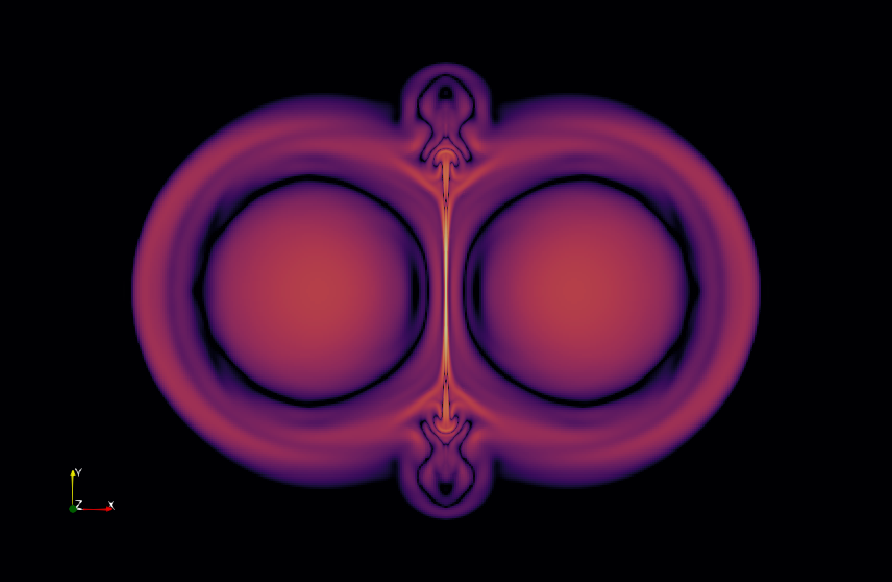}}

%\subfloat{\includegraphics[width=0.2\columnwidth,  clip=true]{jz_etadmin2_8192x8192_t18}}
%\subfloat{\includegraphics[width=0.2\columnwidth,  clip=true]{jz_etadmin3_8192x8192_t18}}
\subfloat{\includegraphics[width=0.66\columnwidth,  clip=true]{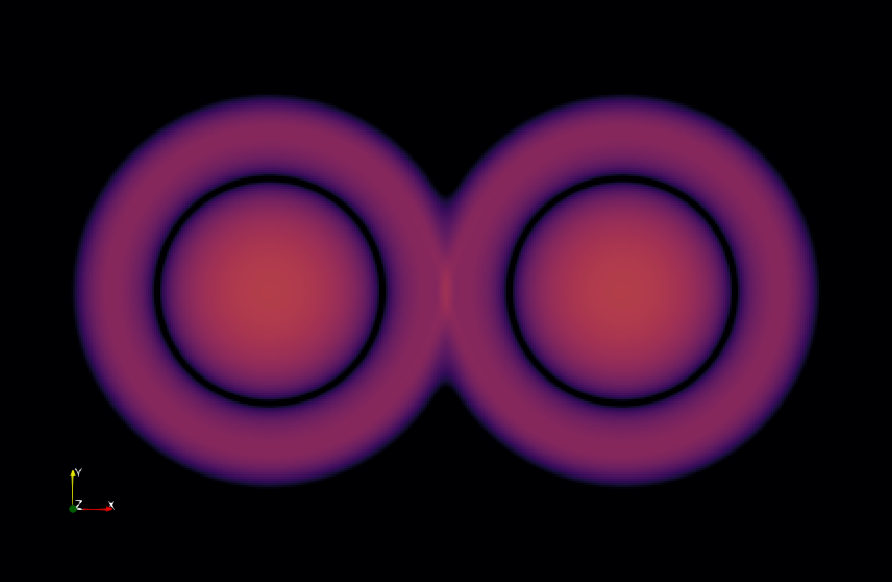}}
\subfloat{\includegraphics[width=0.66\columnwidth,  clip=true]{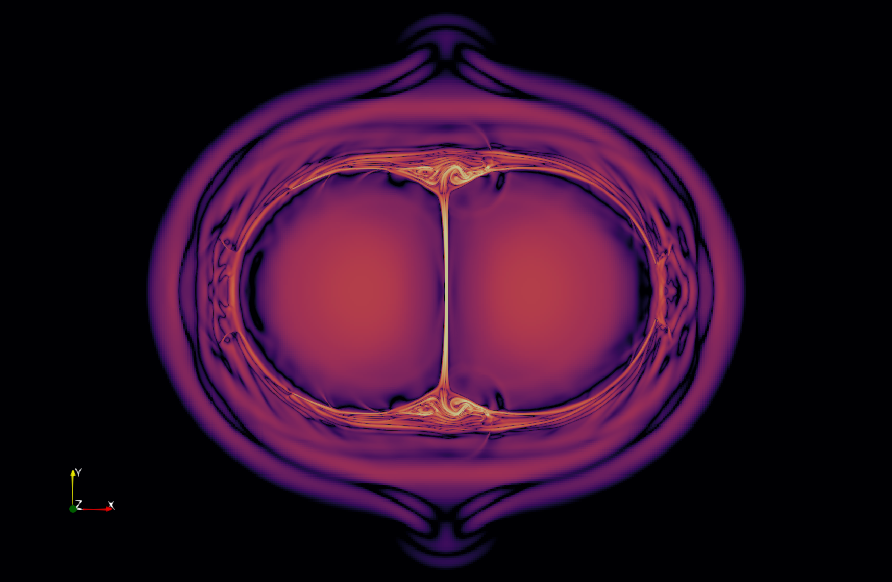}}
\subfloat{\includegraphics[width=0.66\columnwidth,  clip=true]{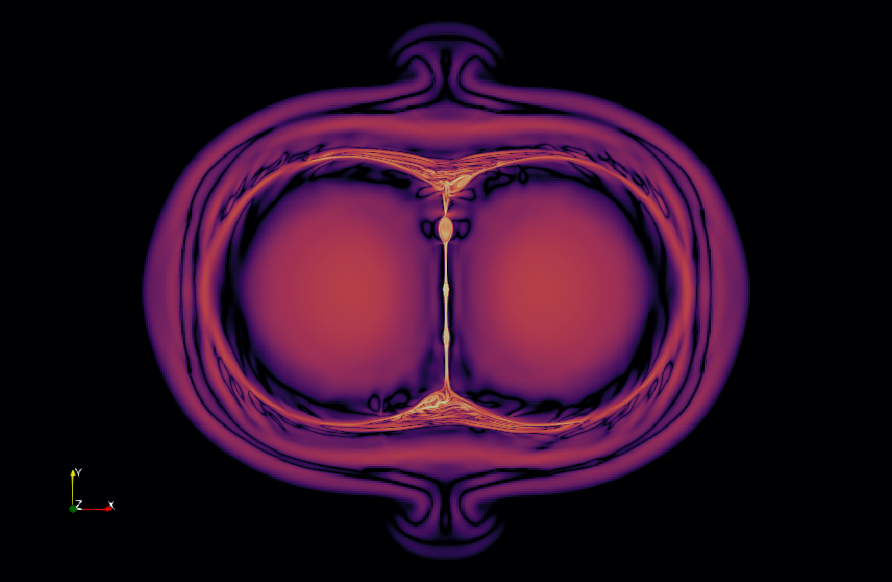}}

%\subfloat{\includegraphics[width=0.2\columnwidth,  clip=true]{jz_etadmin2_8192x8192_t24}}
%\subfloat{\includegraphics[width=0.2\columnwidth,  clip=true]{jz_etadmin3_8192x8192_t24}}
\subfloat{\includegraphics[width=0.66\columnwidth,  clip=true]{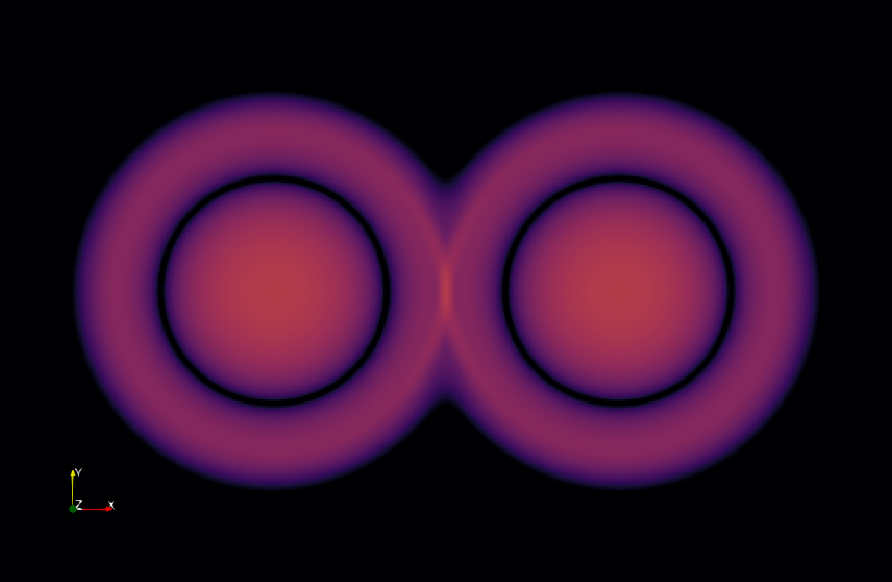}}
\subfloat{\includegraphics[width=0.66\columnwidth,  clip=true]{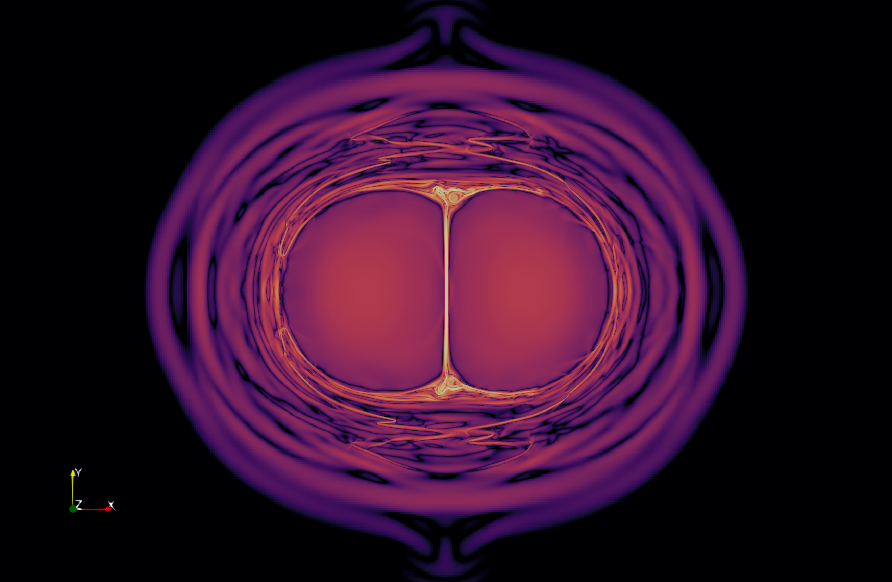}}
\subfloat{\includegraphics[width=0.66\columnwidth,  clip=true]{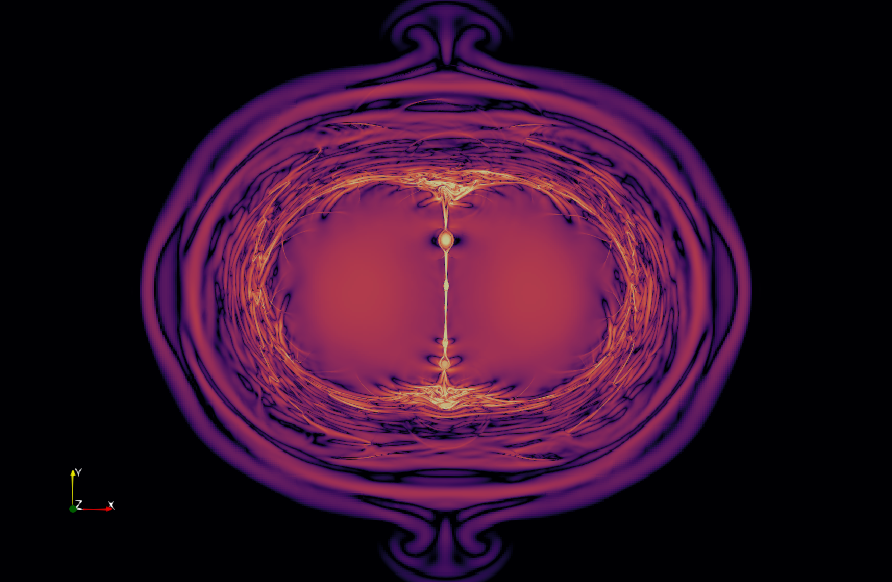}}

\caption{Out-of-plane current density magnitude $|J_{\rm z}|$ in runs (from left to right) C, B and A at, (from top to bottom) $t=5 t_{\rm c}$ (at the end of the exponential growth of the peak current density due to the coalescence instability), $t=10 t_{\rm c}$, $t=18 t_{\rm c}$ (at the start of the growth of the plasmoid instability in case A), $t=24 t_{\rm c}$ (at the far-nonlinear regime for all cases except the unperturbed case C). The logarithmic colour scale is shown in the top-left panel and is constrained to range from 0.05 and 50. The figures are cut to exclude the ambient plasma where $|J_{\rm z}| \ll 0.05$. Plasmoids are visible for case A at times $t=18t_{\rm c}$ and $t=24t_{\rm c}$, indicated by a strong localised $|J_{\rm z}|$.}
\label{fig:currentdensity}
\end{center}
\end{figure*}

In Figure \ref{fig:currentdensity} we compare the current density magnitude in the $(x,y)$-plane at times $t = 5 t_{\rm c}$, $t = 10 t_{\rm c}$, $t = 18 t_{\rm c}$ and $t = 24 t_{\rm c}$ (increasing from top to bottom panels) for runs A ($\eta = 5 \times 10^{-5}$), B ($\eta = 10^{-4}$), and the unperturbed (i.e. $\mathbf{v}_{\rm kick} = 0$) run C ($\eta = 10^{-4}$). The first column shows that the flux ropes do not (visually) coalesce on the time scales considered for unperturbed run C. In run B (middle column) a thin current sheet forms but the resistivity is too large for the plasmoid instability to be triggered. Run A (right-hand column) shows an even thinner current sheet and plasmoids are observed at $t = 18 t_{\rm c}$ and $t = 24 t_{\rm c}$. Runs A and B serve as fiducial runs for Sections \ref{sect:sigmacomparison} and \ref{sect:anomalousrescomparison}, where we compare the effects of magnetisation $\sigma$, plasma-$\beta$ and non-uniform resistivity. 
%op: removing as the statement hangs in the air.
%Regions of large current density correspond to regions of large $\mathbf{E}\cdot\mathbf{B} \neq 0$. Regions with a strong current and parallel, resistive electric fields are prone to be efficient particle acceleration sites. 
High resistivity runs D ($\eta = 10^{-3}$) and E ($\eta = 10^{-2}$) are not shown, since the magnetic energy diffuses away before $t = 5 t_{\rm c}$ due to the high resistivity, destroying the current sheet before it can stably form.

\begin{figure*}
\begin{center}
\subfloat{\includegraphics[width=0.75\columnwidth]{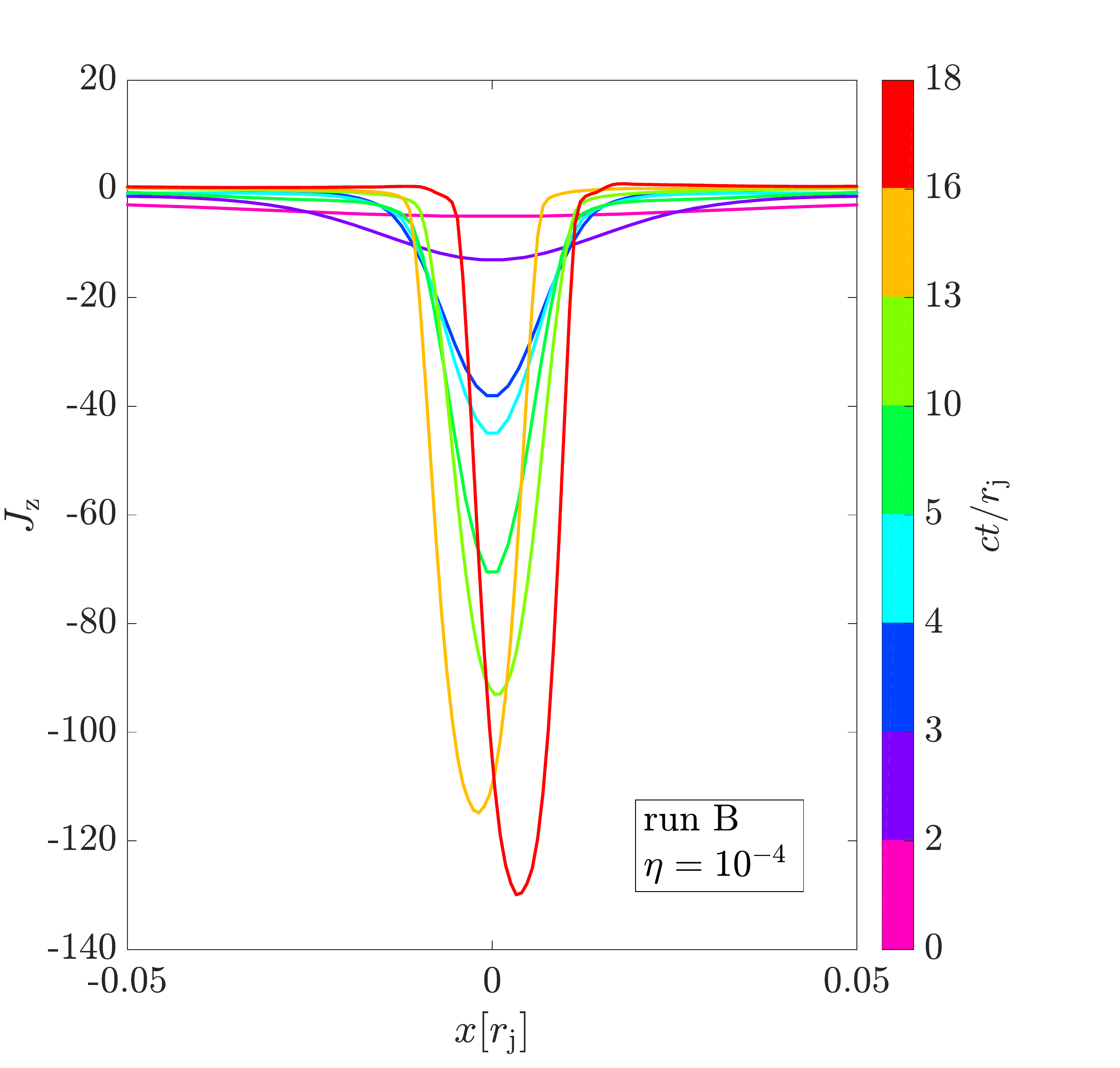}}
\subfloat{\includegraphics[width=0.75\columnwidth]{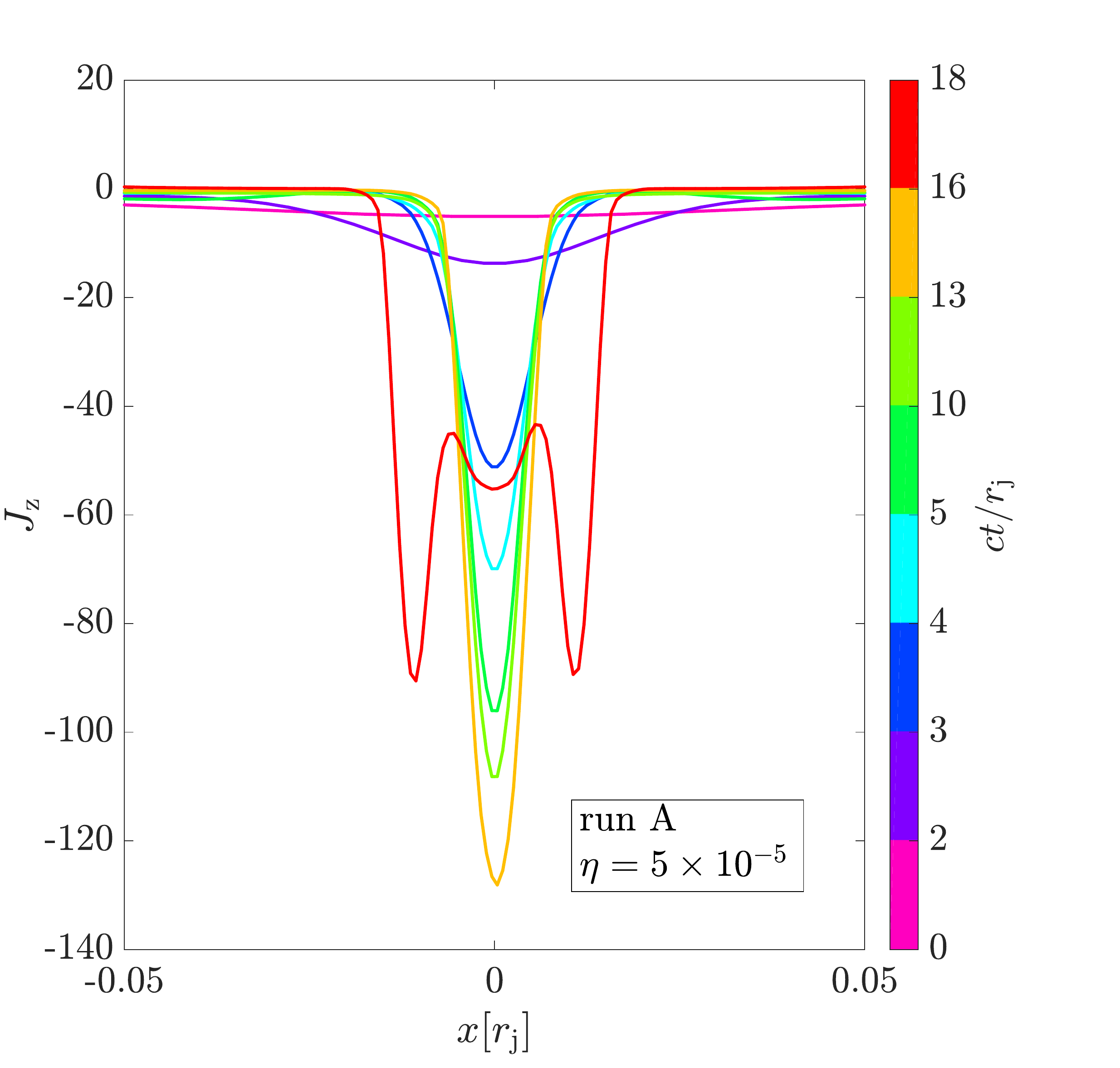}}

\subfloat{\includegraphics[width=0.75\columnwidth]{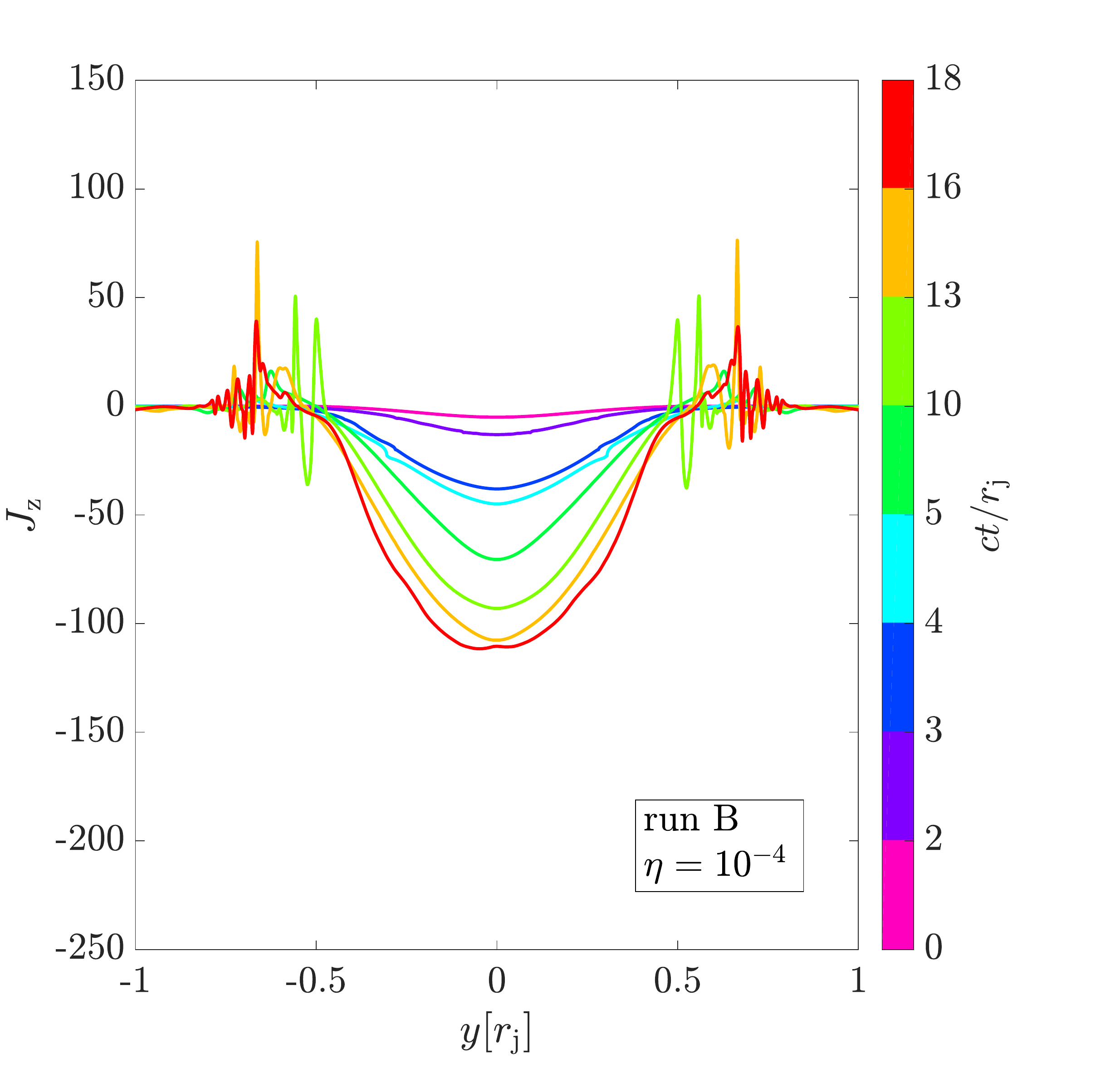}}
\subfloat{\includegraphics[width=0.75\columnwidth]{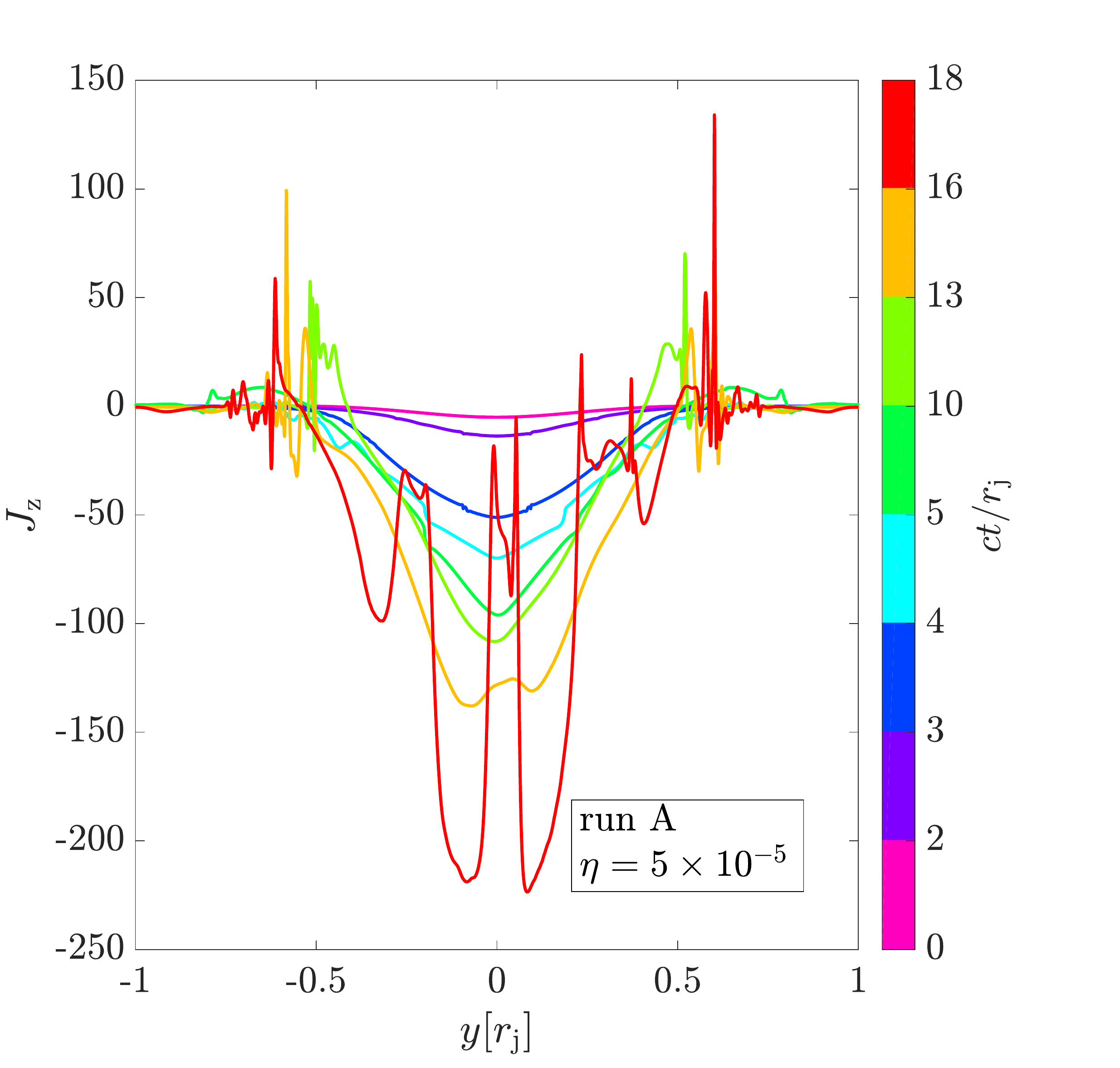}}
\caption{Cuts along $y = 0$ (top panels) and $x = 0$ (bottom panels) for run B with $\eta=10^{-4}$ (left-hand panels) and run A with $\eta=5\times10^{-5}$ (right-hand panels). The evolution of the $z$-component of the current density $J_{\rm z}$ is shown at selected times. The colour scale indicates the time. For run B ($\eta=10^{-4}$), no plasmoids are detected and the current sheet gets thinner in time (see the top-left panel), while the current density increases (see the bottom-left panel). For run A ($\eta=5\times10^{-5}$), the current sheet becomes even smaller and the current density is higher until $t = 18 t_{\rm c}$. From that moment onwards, the current sheet breaks up due to the ideal tearing mode, and plasmoids can be detected. At plasmoid locations, the current sheet broadens and the current density decreases (see the top-right panel). The plasmoids are then advected and expelled in the outflow regions at $y = \pm 1 r_{\rm j}$ (see the bottom-right panel).}
\label{fig:currentslices}
\end{center}
\end{figure*}

{Plasmoids are detected by taking cuts along the axes $x=0$ and $y=0$ in Figure \ref{fig:currentslices}. For runs B ($\eta = 10^{-4}$, left-hand panels in Figure \ref{fig:currentslices}) and A ($\eta = 5\times 10^{-5}$, right-hand panels in Figure \ref{fig:currentslices}) the current sheet becomes notably thinner and the current density keeps growing on the time scales considered here. In run A (the right-hand panels), plasmoids are detected at $t = 18 t_{\rm c}$ both in the cut along $x$ (top panel), indicated by the split of the peak in the out-of-plane current density into two peaks, corresponding to the edges of the plasmoid that have a higher current density than the plasmoid-centre. These plasmoids move either up or down along $y$ and end up in the outflow regions at $y \approx \pm 0.5$.}

\begin{figure}
\begin{center}
\includegraphics[width=0.75\columnwidth]{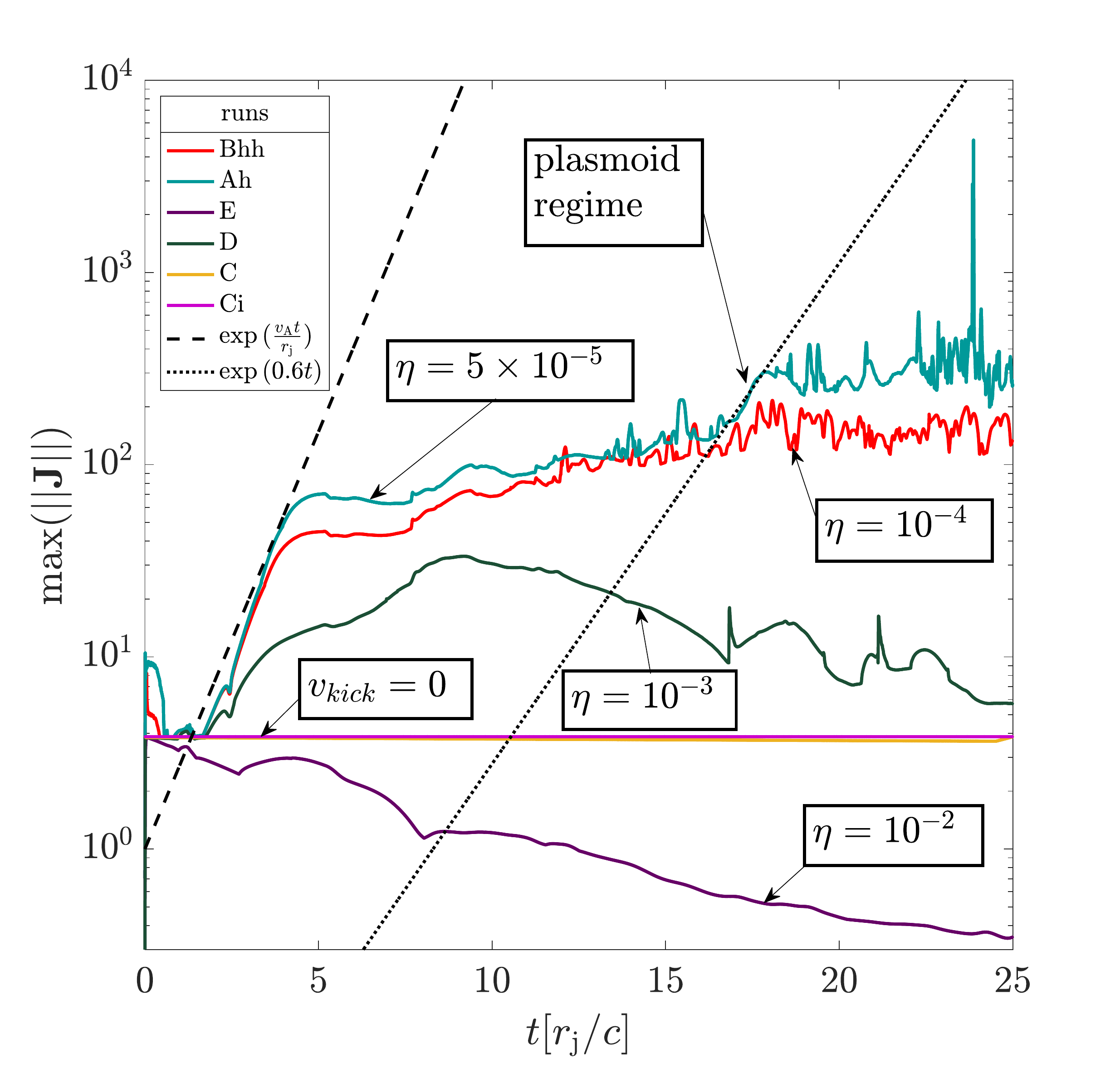}
\caption{Peak current density for high-resolution runs Ah, Bhh, E, D with resistivity $\eta = 5 \times 10^{-5}$, $\eta = 10^{-4}$, $\eta = 10^{-3}$, $\eta = 10^{-2}$, run C without an initial perturbation and $\eta = 10^{-4}$ and run Ci without initial perturbation and in ideal SRMHD ($\eta = 0$). Run Ah, with lower resistivity, is liable to the plasmoid instability and the other runs, with higher resistivity, are not. Runs Ah and Bhh have equal effective resolution $16384^2$, whereas runs Ci, C, D and E have effective resolution $8192^2$. Runs C and Ci are unperturbed ($\mathbf{v}_{\rm kick} = 0$) such that the flux tubes do not coalesce on the time scale considered. This results in a constant peak current density, since no current sheet forms in between the tubes. The dashed line shows the Alfv\'{e}nic growth rate $\propto \exp{(v_{\rm A} t/ r_{\rm j})} \approx \exp{(c t/ r_{\rm j})}$ of the peak current density for runs Ah and Bhh until $t \approx 5 t_{\rm c}$. The dotted line shows the exponential growth rate $\gamma_{\rm max} \simeq 0.6 \tau_{\rm A}^{-1}$ for run Ah at $t \approx 17 t_{\rm c}$.}
\label{fig:peakcurrent}
\end{center}
\end{figure}
In Figure \ref{fig:peakcurrent} we show the peak current density $\max(||\mathbf{J}||)$, again taken over the whole domain, for the highest resolution runs Ah ($\eta = 5 \times 10^{-5}$), Bhh ($\eta = 10^{-4}$), D ($\eta = 10^{-3}$), E ($\eta = 10^{-2}$) and unperturbed runs C ($\eta = 10^{-4}$) and Ci ($\eta = 0$). An Alfv\'{e}nic growth phase is visible at $t \lesssim 5 t_{\rm c}$ showing that the peak current density grows as $\propto \exp{(v_{\rm A} t/ r_{\rm j})} \approx \exp{(c t/ r_{\rm j})}$ (indicated by the dashed-line) for runs Ah and Bhh, where the resistivity is small enough not to affect the ideal MHD behaviour of the coalescence instability. {In this stage of runs Ah and Bhh, the current density reduces to the non-relativistic result $\mathbf{J} = \nabla \times \mathbf{B}$, that is also valid for ideal relativistic MHD in a co-moving frame where the displacement current vanishes.} 
The trend for low resistivity runs Ah and Bhh follows the evolution of the maximum electric field energy in Figure \ref{fig:peakfieldsdmin4}. The maximum current density depends inversely on the resistivity and a stable current sheet forms after the initial growth of the coalescence instability (at $t \gtrsim 5 t_{\rm c}$). {Between $5t_{\rm c} \lesssim t \lesssim 17 t_{\rm c}$ the semi-stationary current sheet becomes thinner in runs Ah and Bhh and this stage corresponds to the laminar Sweet-Parker reconnection regime.} For run Ah, with resistivity $\eta = 5\times10^{-5}$, the plasmoid instability is then triggered at $t \approx 17 t_{\rm c}$ as can be seen by the second exponential growth phase of the current density. {The second exponential growth phase grows as $\gamma_{\rm max} \sim 0.6 \tau_{\rm A}^{-1}$ (as indicated by the black dotted line), resulting in the tearing instability and the formation of multiple secondary plasmoids. Note however that the non-relativistic scaling for the ideal tearing instability $\gamma_{\rm max} \sim 0.6 \tau_{\rm A}^{-1}$ as found by \cite{pucci2014} considers a current sheet that is still in the process of formation and thinning, whereas in our simulations the plasmoid instability is developing in a thin and long current sheet that has reached a semi-stationary state in the interval $5 t_{\rm c} \leq t \leq 17 t_{\rm c}$.}
For run Bhh the secondary plasmoid regime is not attained after the formation of a stable current sheet, due to a too large resistivity. Runs E ($\eta = 10^{-2}$) and D ($\eta = 10^{-3}$) show that even higher resistivity induces the diffusion of the current density before the coalescence growth phase can finish and no stable current sheet can form. Run C shows that for an unperturbed case (i.e. $\mathbf{v}_{\rm kick} = 0$) the current density remains constant and no current sheet forms. Run Ci overlaps with run C, showing that for an unperturbed case in ideal SRMHD the numerical resistivity does not affect the evolution on the considered timescales and the current remains constant. In case Ci the current density is calculated as $\mathbf{J} = \nabla \times \mathbf{B}$.

\begin{figure*}
\begin{center}
\subfloat{\includegraphics[width=0.75\columnwidth]{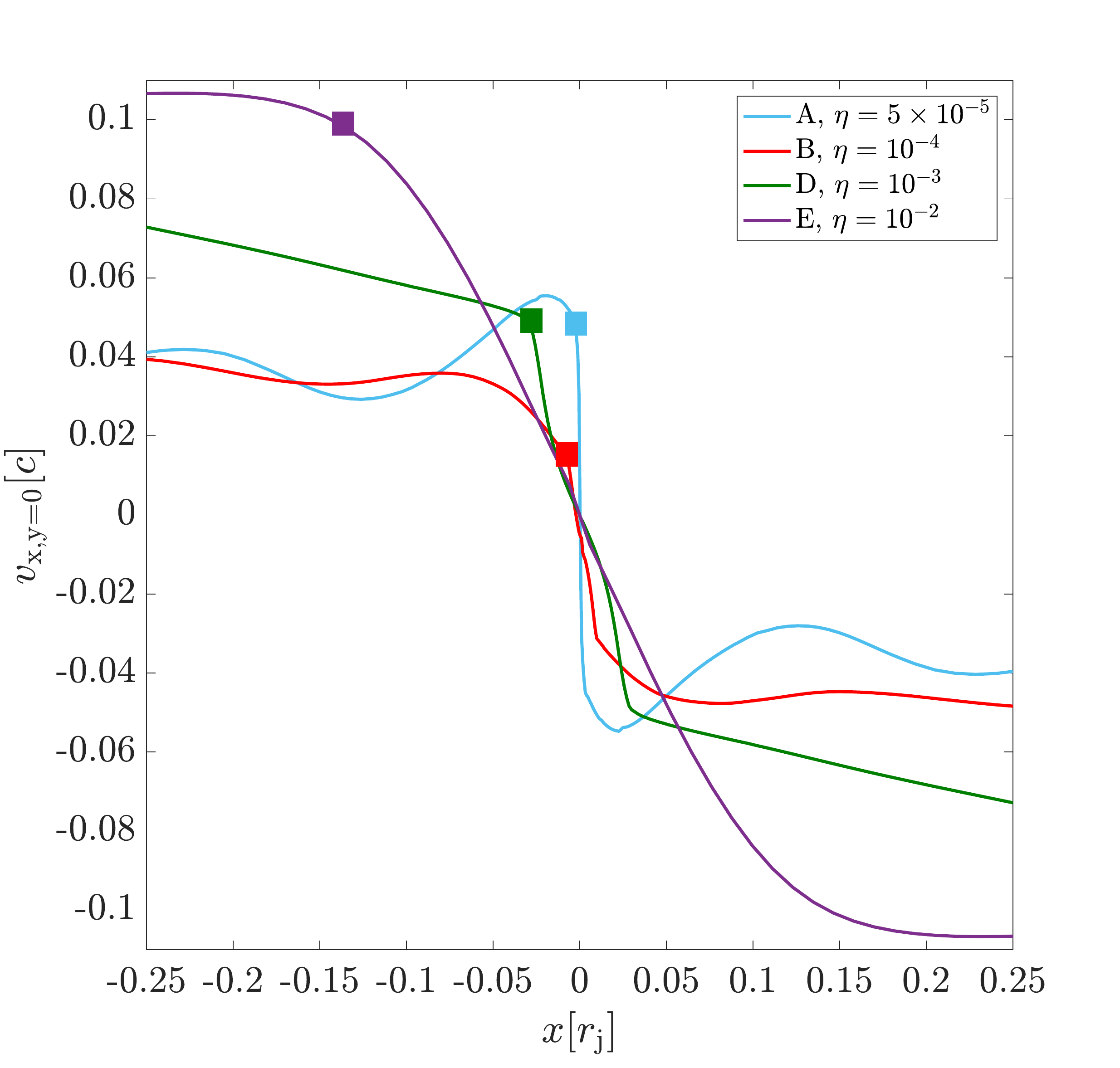}}
\subfloat{\includegraphics[width=0.75\columnwidth]{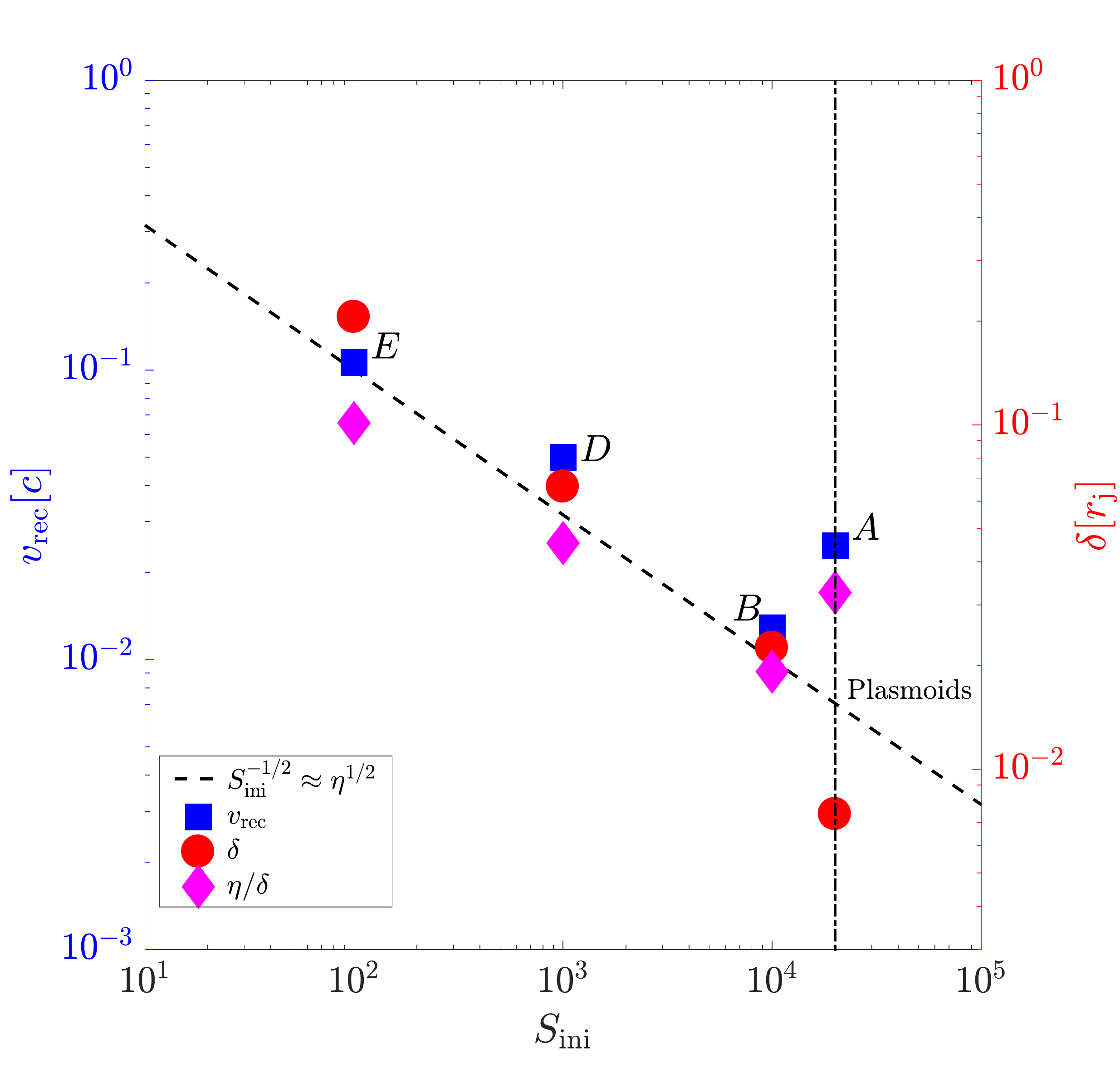}}
\caption{Left-hand panel shows the flow velocity into the current sheet $v_{\rm x,y=0}$ versus $x$-coordinate at $y=0$, for high-resolution runs A, B, E and D with uniform resistivity $\eta = 5 \times 10^{-5}$, $\eta = 10^{-4}$, $\eta = 10^{-3}$, $\eta = 10^{-2}$ respectively.
All runs have equal effective resolution $8192^2$. For case A, the reconnection rate $v_{\rm rec} = v_{\rm in} = v_{\rm x,y=0}$ is taken at time $t = 18.5 t_{\rm c}$, after the first plasmoids have been formed (see Figures \ref{fig:currentdensity} and \ref{fig:peakcurrent}). This specific time has been selected such that there are no plasmoids present in the current sheet, although the plasmoid instability has been triggered. In all other cases, the reconnection rate is taken at time where the peak current density has a maximum ($t = 18 t_{\rm c}$ for case B, $t = 9t_{\rm c}$ for case D, and $t = 4t_{\rm c}$ for case E). The reconnection rate is determined at the point where $v_{\rm x,y=0}$ has an inflection point, i.e. where the function changes from being convex to concave (as indicated by the coloured squares in the left-hand panel). The right-hand panel shows the reconnection rate $v_{\rm rec} := v_{\rm in}$ (blue squares), $v_{\rm rec} \propto \eta/\delta$ (magenta diamonds) and the thickness of the current sheet $\delta$ (red circles), versus the Lundquist number $S_{\rm ini} = \eta^{-1}$. The thickness $\delta$ is determined as the full-width at half-maximum of the out-of-plane current density $J_{\rm z}$ for each case (see e.g. Figure \ref{fig:currentslices} for cases A and B) at the same time as the reconnection rate.
The dashed line in the right-hand panel represents the Sweet-Parker scaling $v_{\rm rec} \sim S^{-1/2}$. The vertical dash-dotted line indicates the critical resistivity $\eta = 5 \times 10^{-5}$ for which the plasmoid instability is triggered.}
\label{fig:recrate}
\end{center}
\end{figure*}
{In the left-hand panel of Figure \ref{fig:recrate} we show the inflow velocity $v_{\rm in} \simeq v_{\rm x,y=0}$ (in units of $c$) into the current sheet. For case A ($\eta = 5 \times 10^{-5}$), the inflow velocity is taken at time $t = 18.5 t_{\rm c}$ such that no plasmoids are present in the current sheet, yet the plasmoid instability has been triggered already (see Figures see Figures \ref{fig:currentdensity}, \ref{fig:currentslices} and \ref{fig:peakcurrent} where plasmoids are formed at $t = 18 t_{\rm c}$ for example). In all other cases, where the plasmoid instability is not triggered, the inflow velocity is taken at time where the peak current density has a maximum, i.e. $t = 18 t_{\rm c}$ for case B ($\eta = 10^{-4}$), $t = 9t_{\rm c}$ for case D ($\eta = 10^{-3}$), and $t = 4t_{\rm c}$ for case E ($\eta = 10^{-2}$).} To calculate the reconnection rate $v_{\rm rec} := v_{\rm in} \simeq v_{\rm x,y=0}$ (for $v_{\rm out} = c=1$), the inflow speed $v_{\rm in}$ is taken at a cut along the $x$-coordinate, at the point where $v_{\rm x,y=0}$ has an inflection point, i.e. where the function changes from being convex to concave, as exemplified by the coloured squares in the left-hand panel. $v_{\rm rec}$ is then determined by averaging the inflow speed $v_{\rm in}$, over a vertical line $y \in [-0.1,0.1]$. 
In the Sweet-Parker regime, the reconnection speed is proportional to the ratio between the resistivity and the thickness of the sheet $v_{\rm in} \propto \eta/\delta$. The thickness $\delta$ is determined as the full-width at half-maximum of the out-of-plane current density $J_{\rm z}$ at $y = 0$  (see e.g. the top panels of Figure \ref{fig:currentslices} for cases A and B) at the same time as the reconnection rate for each case. In the right-hand panel of Figure \ref{fig:recrate} we show the scaling of the reconnection rate $v_{\rm in}/c$, the thickness $\delta$ and the ratio $\eta/\delta$ in units of $c$ with the Lundquist number $S_{\rm ini} \simeq 1/\eta$  for cases A, B, D and E. 
The dashed line depicts the Sweet-Parker scaling $v_{\rm rec} \sim S_{\rm ini}^{-1/2}$. The dash-dotted vertical line shows the critical resistivity $\eta = 5\times10^{-5}$ for the plasmoid instability to occur. 
The three high-resistivity cases follow the slow Sweet-Parker scaling, but the lowest resistivity run A, is liable to the plasmoid instability and diverges from the scaling for all three quantifiers, confirming the occurrence of the ideal tearing mode as observed in the peak current density in Figure \ref{fig:peakcurrent}. The rightmost blue square and magenta diamond indicate higher reconnection rates than expected for a Sweet-Parker scaling and the rightmost red circle is lower than the expected thickness due to the plasmoid instability. Note that indeed approximately $v_{\rm rec} \propto \eta/\delta$ for the runs without plasmoids. 
%\begin{figure*}
%\begin{center}
%\subfloat{\includegraphics[width=\columnwidth]{etadmin2_current_sliceX.eps}}
%\subfloat{\includegraphics[width=\columnwidth]{etadmin3_current_sliceX.eps}}
%
%\subfloat{\includegraphics[width=\columnwidth]{etadmin2_current_sliceY.eps}}
%\subfloat{\includegraphics[width=\columnwidth]{etadmin3_current_sliceY.eps}}
%\caption{Cuts along $y = 0$ (top panels) and $x = 0$ (bottom panels) for run Q with $\eta=10^{-2}$ (left-hand panels) and run P with $\eta=10^{-3}$ (right-hand panels). The evolution of the $z$-component of the current density $J_{\rm z}$ is shown at selected times. The colour scale indicates the time. The thickness of the current sheet can be read from the top panels, and the length of the current sheet can be read from the bottom panels. Note the larger $x$-range for run Q to capture the thicker current sheet length.}
%\label{fig:currentsliceshigheta}
%\end{center}
%\end{figure*}
%
%

\subsection{Dependence on magnetisation}
\label{sect:sigmacomparison}
In this section we explore the dependence of the plasmoid instability on plasma-$\beta$ and magnetisation $\sigma$. The range of $0.01 \leq \beta_0 \leq 1$, $0.4 \leq \sigma_0 \leq 3.33$ considered in this section is particularly relevant for black hole accretion disks where flux tubes continuously emerge out of the disk due to magnetic buoyancy (\citealt{Tout1992}; \citealt{uzdensky2008}; \citealt{Goodman}; \citealt{Yuan2009}), which is conjectured to result in magnetic reconnection and flaring (\citealt{melzani2014}; \citealt{ball2016}; \citealt{rowan2017}; \citealt{werner2017}; \citealt{ball2017}; \citealt{werner2018}; \citealt{ball2018}). Regimes with $\sigma_0 \geq 1$ and $\beta_0 < 1$ also have a direct relevance for outflows from black holes and neutron stars, with a typically much larger magnetisation. 

\begin{figure*}
\begin{center}
\subfloat{\includegraphics[width=0.66\columnwidth]{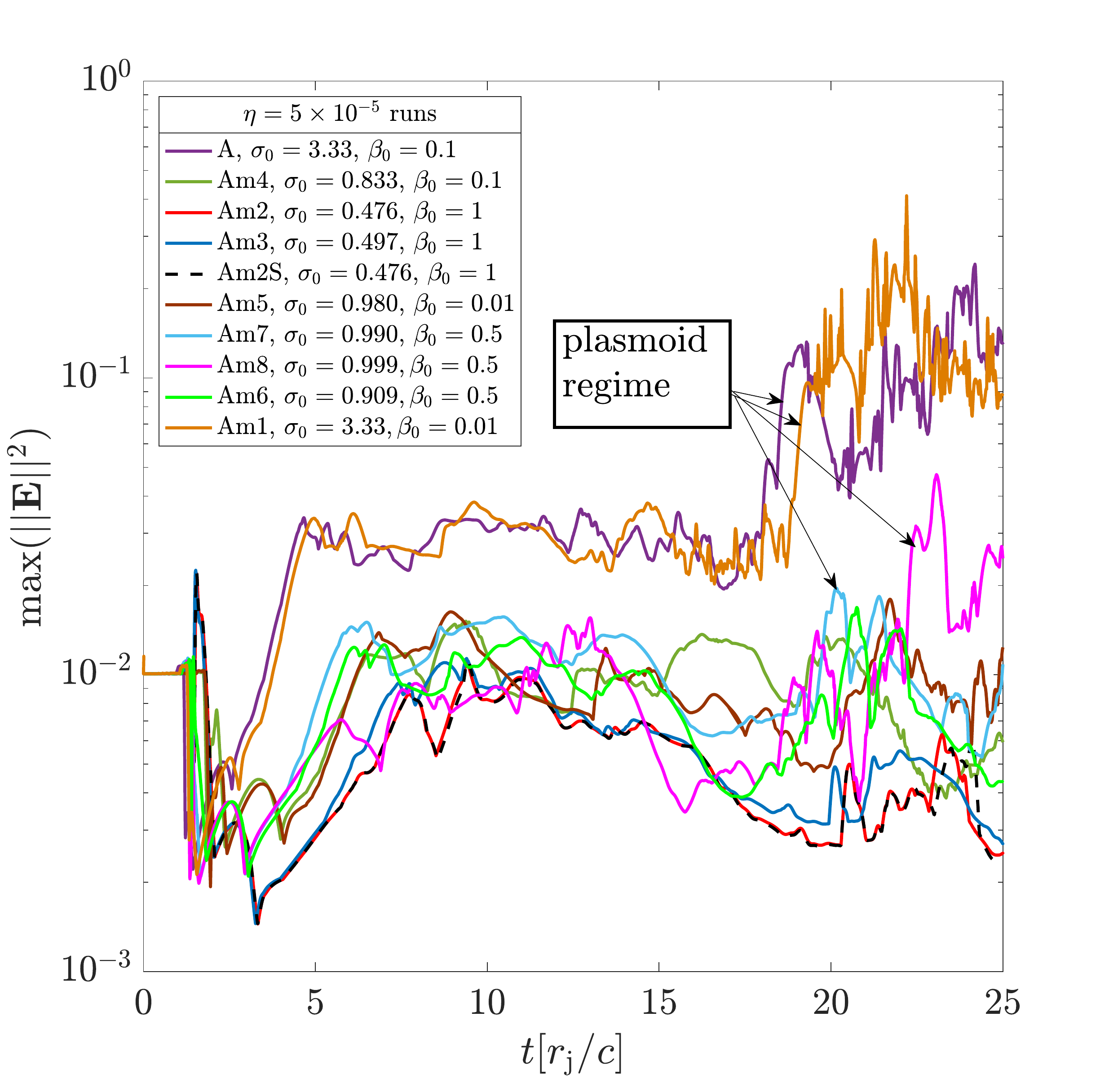}}
\subfloat{\includegraphics[width=0.66\columnwidth]{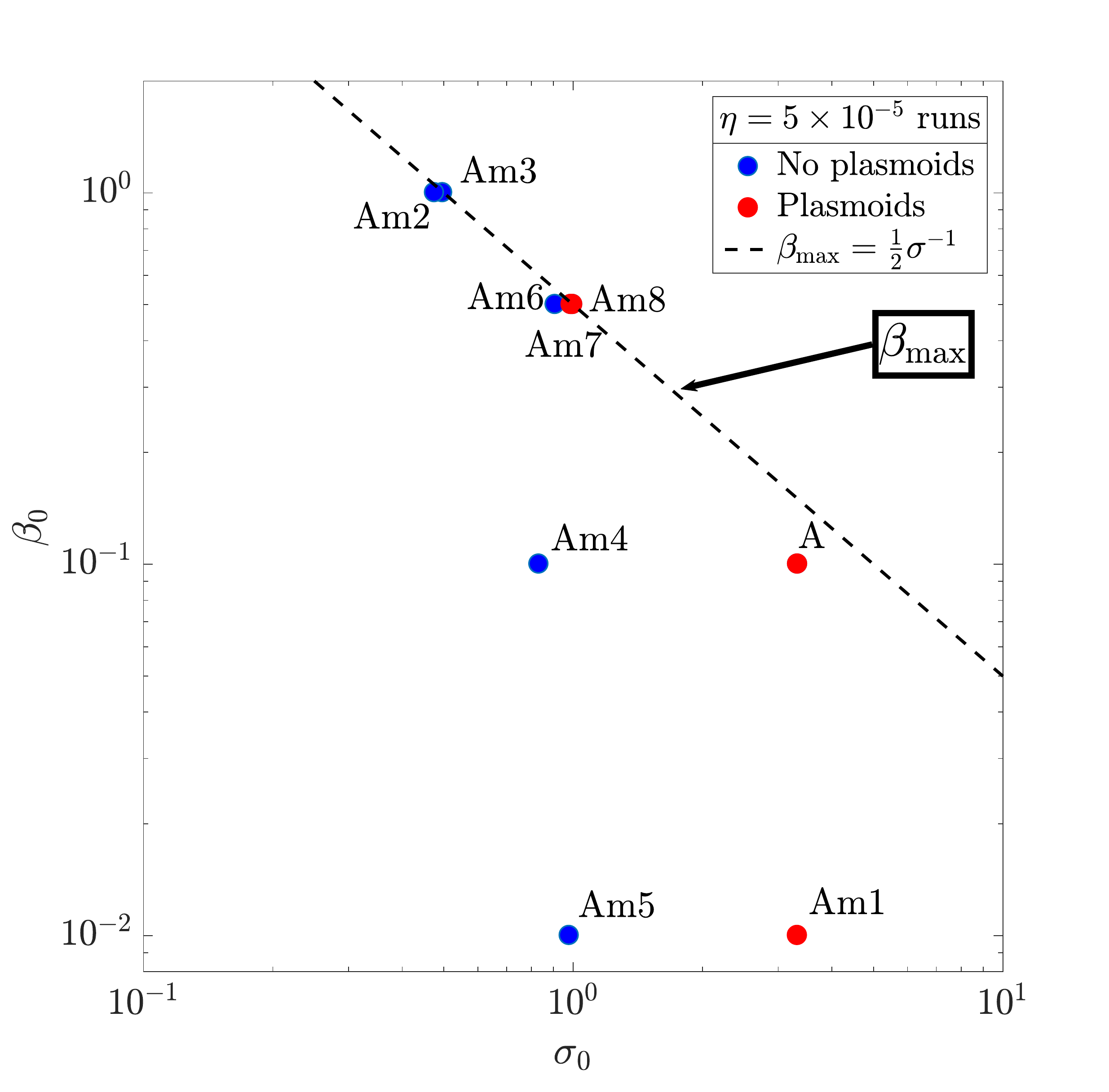}}
\subfloat{\includegraphics[width=0.66\columnwidth]{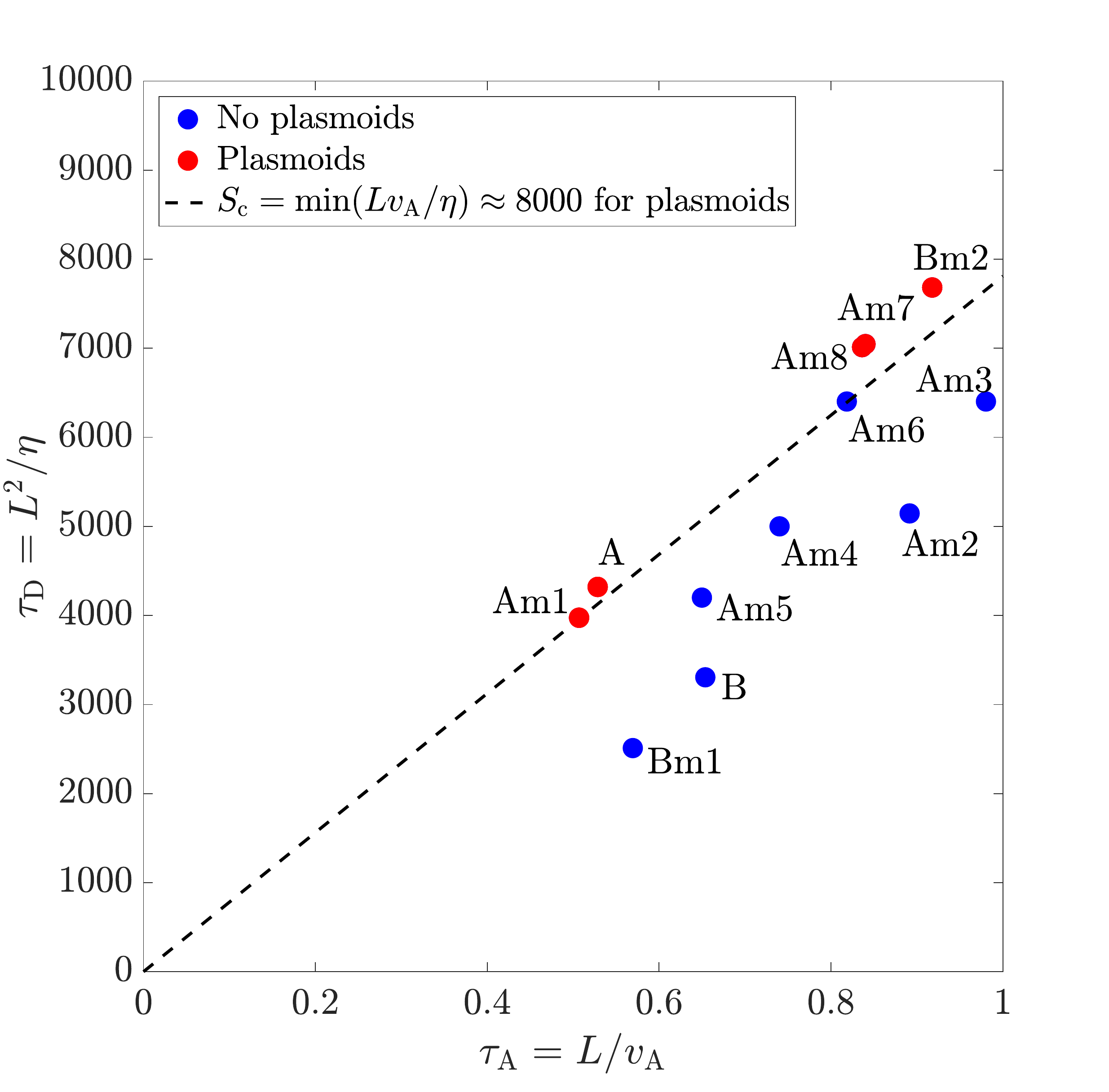}}
\caption{Left-hand panel: Peak electric energy for all runs with uniform resistivity $\eta = 5\times10^{-5}$ and resolution $8192^2$. Plasma-$\beta$ and $\sigma$ are varied. The exponential growth phase of the electric field energy depends on the Alfv\'{e}n speed as $\propto \exp{(v_{\rm A} t / r_{\rm j})}$, such that a smaller $\sigma_0$, results in a slower growth rate. The plasmoid instability is triggered for runs Am1 ($\sigma=3.33$, $\beta=0.01$), Am7 ($\sigma_0 = 0.990$, $\beta_0=0.5$), Am8  ($\sigma_0 = 0.999$, $\beta_0=0.5$) and fiducial run A. The Strang split scheme and the ImEx scheme show strong visual agreement for run Am2 (red solid line) and Am2S (black dashed line), whose curves overlap.
Middle panel: The values of $\beta_0$ and $\sigma_0$ that are explored in this work. The runs with $\sigma_0 \lesssim 0.99$ for $\eta = 5 \times 10^{-5}$ do not show the formation of plasmoids (blue markers), whereas the red markers indicate runs A, Am1, Am7 and Am8 where plasmoids are observed. The maximum attainable plasma-$\beta_0 = \sigma^{-1}/2$ is indicated by the black dashed line.
Right-hand panel: The diffusion time scale $\tau_{\rm D} = L^2/\eta$ versus the Alfv\'{e}n time scale $\tau_{\rm A} = L / v_{\rm A}$, where $L$ is the full-width half-maximum of the out-of-plane current density $J_{\rm z}$ at $t = 11 t_{\rm c}$ for all cases (see e.g. Figure \ref{fig:currentslices}). At this point a stable current sheet has formed in all cases and no plasmoids have formed yet. Cases where the ratio between the time scales $\tau_{\rm D} / \tau_{\rm A}$ is large enough for plasmoids to form are indicated by red circles, where cases $\tau_{\rm D} / \tau_{\rm A}$ below the threshold for plasmoid formation are indicated by blue circles. By taking the minimum of this ratio for all cases with plasmoids (A, Am1, Am7, Am8, Bm2), we determine a critical Lundquist number for plasmoid formation $S_{\rm c} = \min(\tau_{\rm D}/\tau_{\rm A}) = \min(L v_{\rm A} / \eta) \approx 8000$ in case Am7 (as indicated by the black dashed line).}
\label{fig:EmaxImEx}
\end{center}
\end{figure*}

The left-hand panel of Figure \ref{fig:EmaxImEx} shows the evolution in time of the maximum electric field energy for several runs. The resolution is kept at $8192^2$ and $\eta = 5\times10^{-5}$ for all cases. 
We compare all runs to the fiducial run A, with $\sigma_0 = 3.33$ and $\beta_0 = 0.1$ (the purple line in the left-hand panel of Figure \ref{fig:EmaxImEx}), where we observed plasmoid formation around $t \gtrsim 17 t_{\rm c}$ (see e.g. the third column in Figure \ref{fig:currentdensity}). The dependence on $\beta_0$ and  $\sigma_0$ is analysed by varying the pressure $p_0$ and density $\rho_0$. Note that by adapting the density $\rho_0$, only $\sigma_0 = B^2/(h \rho_0)$ is changed, whereas varying the pressure $p_0$, affects both $\beta_0 = 2 p_0/B^2$ and $\sigma_0 = B^2/(h \rho_0) = B^2 / (\rho_0 + 4 p_0)$ due to the change in enthalpy density  $h = 1 + 4 p_0/\rho_0$. {The adiabatic index is set as $\hat{\gamma} = 4/3$ as appropriate for a relativistically hot ideal gas. For cases with $p_0 \ll \rho_0$ the plasma is initially not ultra-relativistically hot. Assuming $\hat{\gamma} = 4/3$ is clearly a simplification on the thermodynamics and preferably the effect of different equations of state should be considered. Here however, we choose to only vary the pressure and mass density to stay consistent between cases. In cases that are initially not relativistically hot (i.e. case Am4 with $p_0 / \rho_0 = 0.05$ and case Am5 with $p_0 / \rho_0 = 0.005$), we observe an increase to $p / \rho \approx 0.2$ in the current sheet and outflow region at the final stage of the evolution at $t = 25 t_{\rm c}$.}

{When varying magnetisation $\sigma_0$, the Alfv\'{e}n speed $v_{\rm A} = c\sqrt{\sigma_0 / (\sigma_0 + 1)}$ and half-length of the sheet $L$ can significantly alter the Lundquist number $S_{\rm eff} = L v_{\rm A} \eta^{-1}$ between cases with equal resistivity. For low magnetisation the approximation $v_{\rm A} \sim c$ becomes invalid. The half-length of the current sheet enters the Lundquist number as well, and can differ significantly between cases. The Lundquist number can effectively drop below the threshold for plasmoid formation when $v_{\rm A} < c$ or $L \neq 1$, compared to fiducial case A. Therefore the approximation $S_{\rm ini} \approx \eta^{-1}$ can be inaccurate and we determine the half-length $L$ and the Alfv\'{e}n speed $v_{\rm A}$ in each case specifically to obtain the effective Lundquist number $S_{\rm eff} = L v_{\rm A} \eta^{-1}$.}

All runs in the left-hand panel of Figure \ref{fig:EmaxImEx} show an exponential evolution of the electric energy density during the coalescence of the flux tubes, for all values of $\sigma_0$ and $\beta_0$. In particular, the exponential growth of the coalescence instability is proportional to the Alfv\'{e}n speed as $\propto \exp{(v_{\rm A} t / r_{\rm j})}$, such that for runs with smaller $\sigma_0$ and hence smaller Alfv\'{e}n speed, exponential growth is slower. After the initial growth phase, all runs reach a plateau in the electric energy density. The value of the plateau depends on $\beta_0$, $\sigma_0$ and whether the plasmoid instability is triggered or not. The ImEx and the Strang schemes are compared in runs Am2 and Am2S for $\eta = 5 \times 10^{-5}$, $\beta_0 = 1$ and $\sigma_0 = 0.476$. Very good agreement is observed up to the far nonlinear regime (see the red solid line for run Am2 and black dashed line for run Am2S in the left-hand panel of Figure \ref{fig:EmaxImEx}). Only minor differences are observed in the maximum electric field energy density after $t \gtrsim 23 t_{\rm c}$ in the far nonlinear phase. For such low resistivity the Strang scheme is extremely computationally expensive, whereas the ImEx scheme is not. We find a factor $\sim 9$ speedup for the ImEx scheme as compared to the Strang split scheme.

From the middle panel of Figure \ref{fig:EmaxImEx} we can conclude that plasmoids can only form when $\sigma_0 \gtrsim 1$ for runs with resistivity $\eta = 5\times10^{-5}$ (indicated by the red circles). In all other cases the effective Lundquist number $S_{\rm eff} = L v_{\rm A} / \eta$ is too low to trigger the plasmoid instability (indicated by the blue circles). Plasma-$\beta_0$ has no directly observable effect on the plasmoid formation threshold in the cases explored here. In the non-relativistic resistive magnetohydrodynamics regime, \cite{ni2012} and \cite{baty2014} numerically studied the dependence of the onset of the plasmoid instability on plasma-$\beta$. They find that the critical Lundquist number for the plasmoid instability slightly decreases for higher plasma-$\beta$, with lower reconnection rates in lower plasma-$\beta$ systems. This can be explained by the fact that magnetisation $\sigma$ and plasma-$\beta$ are coupled (\citealt{ball2018}). By taking the enthalpy density into account in the definition of the magnetisation as $\sigma_0 = B^2 / (\rho_0 h_0) = B^2 / (\rho_0 + 4 p_0) = B^2 / (\rho_0 + 2 \beta_0 B^2)$, we obtain a maximum value for plasma-$\beta_0$ of $\beta_{\rm max} = \frac{1}{2} \sigma^{-1}$ (as indicated by the black dashed line in the middle panel of Figure \ref{fig:EmaxImEx}). By raising $\beta_0$, the magnetisation effectively decreases, resulting in a lower effective Lundquist number.

Comparing runs with $\eta=10^{-4}$ to runs with $\eta=5\times10^{-5}$, the Lundquist number $S_{\rm eff} = L v_{\rm A} / \eta$ changes due to differences in resistivity, magnetisation (and thus Alfv\'{e}n speed), and also the typical length of the current sheet $L$. 
In the right-hand panel of Figure \ref{fig:EmaxImEx}, we show the typical Alfv\'{e}n time $\tau_{\rm A} = L / v_{\rm A}$ versus the diffusion time $\tau_{\rm D} = L^2 / \eta$ for cases B ($\sigma_0 = 3.33$, $\beta_0 = 0.1$), Bm1 ($\sigma_0 = 3.33$, $\beta_0 = 0.01$) and Bm2 ($\sigma_0 = 10$, $\beta_0 = 0.01$) with resistivity $\eta = 10^{-4}$, and compare to cases A ($\sigma_0=3.33$, $\beta_0=0.1$), Am1 ($\sigma_0=3.33$, $\beta_0=0.01$), Am2 ($\sigma_0=0.476$, $\beta_0=1.0$), Am3 ($\sigma_0=0.497$, $\beta_0=1.0$), Am4 ($\sigma_0=0.833$, $\beta_0=0.1$), Am5 ($\sigma_0=0.98$, $\beta_0=0.01$), Am6 ($\sigma_0=0.909$, $\beta_0=0.5$), Am7 ($\sigma_0=0.99$, $\beta_0=0.5$) and Am8 ($\sigma_0=0.999$, $\beta_0=0.01$) with resistivity $\eta = 5\times 10^{-5}$.
The full-width half-maximum of the out-of-plane current density $J_{\rm z}$ is used to determine the typical length $L$ of the current sheet at $t = 11 t_{\rm c}$ when a stable current sheet without plasmoids has formed in all cases. In this way, the growth phase of the coalescence instability has a negligible effect on the length of the sheet, and no plasmoids have formed yet in cases liable to the tearing instability. 
We find plasmoids for runs A, Am1, Am7, Am8 and Bm2 (as indicated by the red dots), showing that for a high magnetisation $\sigma_0 \gtrsim 10$, and hence Alfv\'{e}n speed $v_{\rm A} \approx 0.95c$, the Lundquist number can increase above the critical threshold for lower resistivity of $\eta=10^{-4}$ as well. Cases Am2, Am3, Am4, Am5, Am6, and cases B and Bm1 have a too small Lundquist number for the plasmoid instability to be triggered (as indicated by the blue dots).
Cases where plasmoids can form have a typical ratio between diffusion time and Alfv\'{e}n time $S_{\rm eff} = \tau_{\rm D} / \tau_{\rm A} = L v_{\rm A} /\eta$ that is large enough for the secondary tearing instability to be triggered. By comparing this ratio for these cases, we can determine a critical Lundquist number $S_{\rm c} = \min(L v_{\rm A} / \eta) = 8000$ (as indicated by the black dashed line in the right-hand panel), corresponding to case Am7. By increasing the Alfv\'{e}n speed, for example in the force-free magnetodynamics simulations of \cite{SironiPorth} where $v_{\rm A} = c$, plasmoids can form already for a resistivity $\eta \sim 10^{-3}$. Vice versa, in non-relativistic magnetohydrodynamics plasmoids can form, but a smaller resistivity is necessary to counterbalance the lower typical speeds $v_{\rm A} \ll c$. 

\begin{figure} 
\begin{center}
\subfloat{\includegraphics[width=0.75\columnwidth]{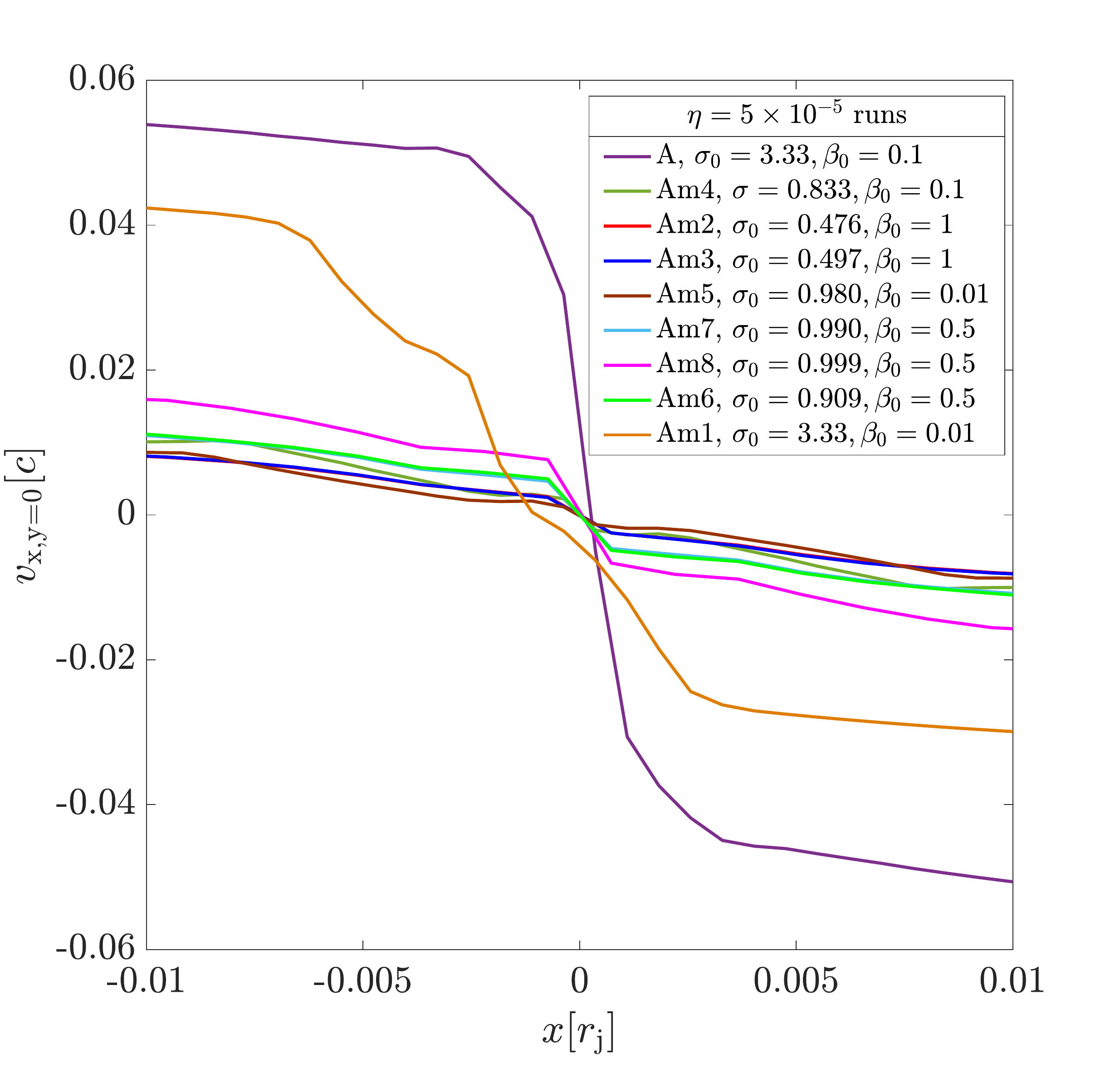}}
\caption{The flow velocity into the current sheet $v_{\rm x,y=0}$ versus $x$-coordinate at $y=0$, for runs with uniform resistivity $\eta = 5\times10^{-5}$, A ($\beta_0 = 0.1$, $\sigma_0=3.33$), Am1 ($\beta_0 = 0.01$, $\sigma_0=3.33$), Am2 ($\beta_0 = 1$, $\sigma_0=0.476$), Am3 ($\beta_0 = 1$, $\sigma_0=0.497$), Am4 ($\beta_0 = 0.1$, $\sigma_0=0.833$), Am5 ($\beta_0 = 0.01$, $\sigma_0=0.980$), Am6 ($\beta_0 = 0.5$, $\sigma_0=0.909$), Am7 ($\beta_0 = 0.5$, $\sigma_0=0.990$) and Am8 ($\beta_0 = 0.5$, $\sigma_0=0.999$). All runs have equal effective resolution $8192^2$. The reconnection rate  $v_{\rm rec} = v_{\rm in} = v_{\rm x,y=0}$ is taken at time $t = 18 t_{\rm c}$ where the the plasmoid instability is triggered for fiducial run A (see the left-hand panel of Figure \ref{fig:EmaxImEx}). When this time coincides with the formation of a plasmoid at $y = 0$, we take the reconnection rate shortly before the plasmoid formation.}
\label{fig:recrate_sigmabeta}
\end{center}
\end{figure}

In Figure \ref{fig:recrate_sigmabeta} we show the inflow velocity into the current sheet, giving an indication of the reconnection rate, for all runs with uniform resistivity $\eta = 5\times 10^{-5}$ and varying plasma-$\beta_0$ and $\sigma_0$ at $t = 18 t_{\rm c}$. This time corresponds to the onset point of the plasmoid instability for run A, Am1 and Am8. For these cases, the reconnection rate is clearly enhanced by the plasmoid instability. For case Am7 plasmoids only form very late in the evolution, after $t = 23 t_{\rm c}$, indicating that it is at the threshold for plasmoid formation. Case Am7 therefore shows minimal differences with case Am6, where no plasmoids form. Generally, a higher magnetisation $\sigma_0$ results in thinner and longer current sheets, higher Lundquist numbers and a higher reconnection rate.

%recrate: 
%delta2 0 t18 0.006958 till 0.009988, 0.0129, max eta dmin4
%delta2 0.001 t18 from -0.00476074 to 0.012085, 0.0114, maxeta  1.3dmin4
%delta2 0.01 t18 from -0.00219727 to 0.0274658, 0.0215, maxeta 6.7dmin4
%delta2 0.1 t10 from -0.0465088 to 0.0406496, 0.0927, maxeta 264.6dmin4
%delta2 1 t2 from -0.0644531 to 0.06544531, 0.0088, maxeta 404.7dmin4 

\subsection{Dependence on non-uniform resistivity}
\label{sect:anomalousrescomparison}

In this section we implement the nonlinear, current-dependent resistivity $\eta(\mathbf{x},t) = \eta_0 (1+\Delta_{\rm ei}^2 J)$ of Section \ref{sect:anomalousresistivity}, while $\sigma_0 = 3.33$ and $\beta_0 = 0.1$ are kept constant and all runs are conducted with a resolution of $8192^2$. This description results in an enhanced resistivity in reconnection regions, like the current sheet, and a low base resistivity in the ambient. A non-uniform resistivity, which is enhanced in the current sheet, may broaden the reconnection layer and therewith enhance the reconnection rate.
 
The base resistivity is set to $\eta_0 = 10^{-4}$, for which no sign of plasmoid formation or fast reconnection is observed in case of uniform resistivity (see run B for $\Delta_{\rm ei}=0$, $\eta = 10^{-4}$). The asymptotically small parameter that represents the importance of sub-grid non-uniform effects is varied between $\Delta_{\rm ei}^2 \in [0.001, 1]$. The maximum resistivity in each run is given in Table \ref{tab:example_table} and always occurs in the current sheet forming in between the merging flux ropes. 

For run Bnr4, with small parameter $\Delta_{\rm ei}^2 = 0.001$, the resistivity is only mildly enhanced $\eta_{\rm max} \approx 1.3 \times \eta_0$. For run Bnr3 ($\Delta_{\rm ei}^2 = 0.01$), the increase is still small with $\eta_{\rm max} \approx  6.7 \times \eta_0$ and for runs Bnr2 ($\Delta_{\rm ei}^2 = 0.1$) and Bnr1 ($\Delta_{\rm ei}^2 = 1$), the non-uniform resistivity is two orders of magnitude larger, resulting in $\eta_{\rm max} \approx 264.5 \times \eta_0$ and $\eta_{\rm max} \approx 404.7 \times \eta_0$, respectively.

\begin{figure*} 
\begin{center}
%\subfloat{\includegraphics[width=0.2\columnwidth,  clip=true]{jz_etadmin2_8192x8192_t5}}
%\subfloat{\includegraphics[width=0.2\columnwidth,  clip=true]{jz_etadmin3_8192x8192_t5}}
\subfloat{\includegraphics[width=0.66\columnwidth,  clip=true]{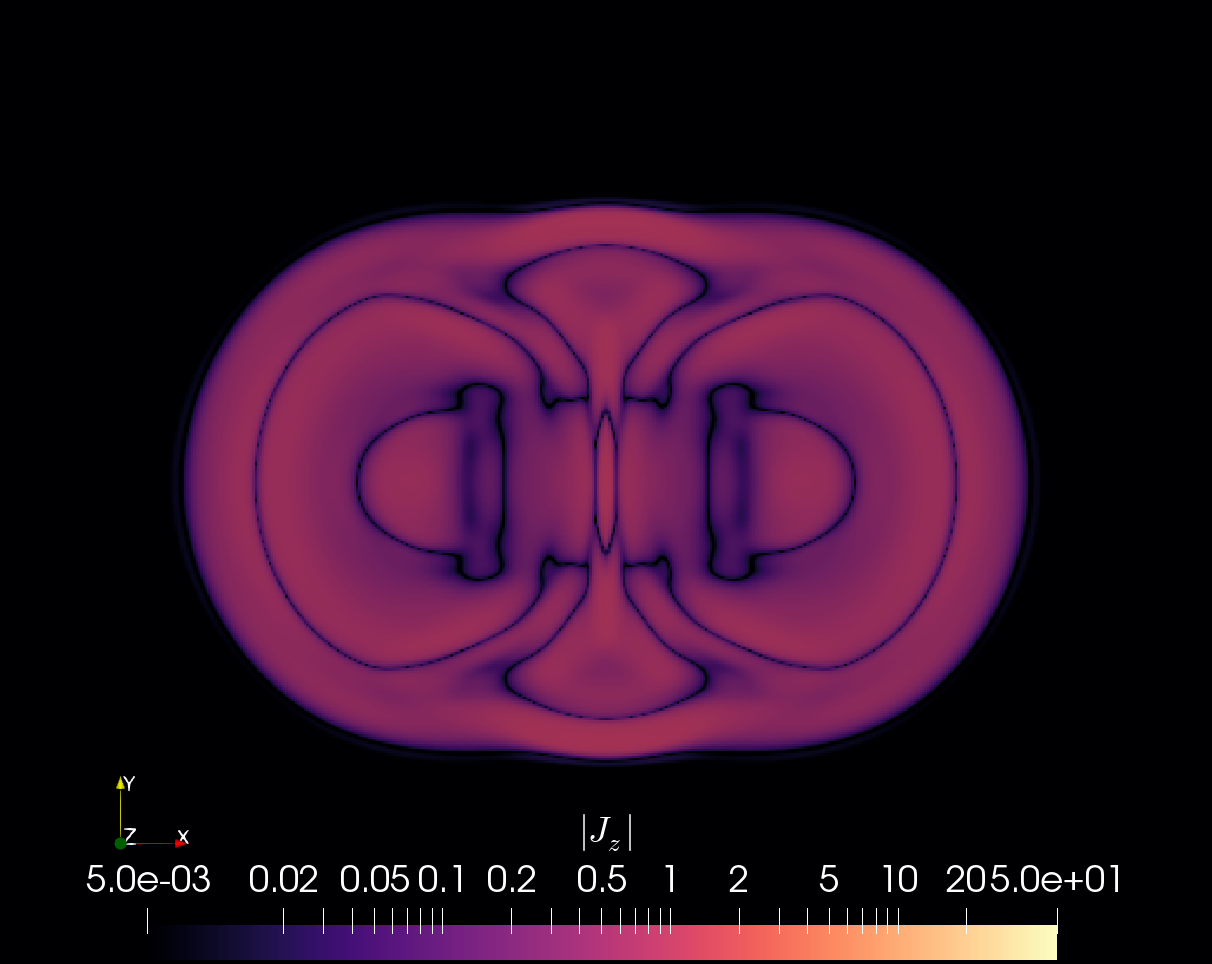}}
\subfloat{\includegraphics[width=0.66\columnwidth,  clip=true]{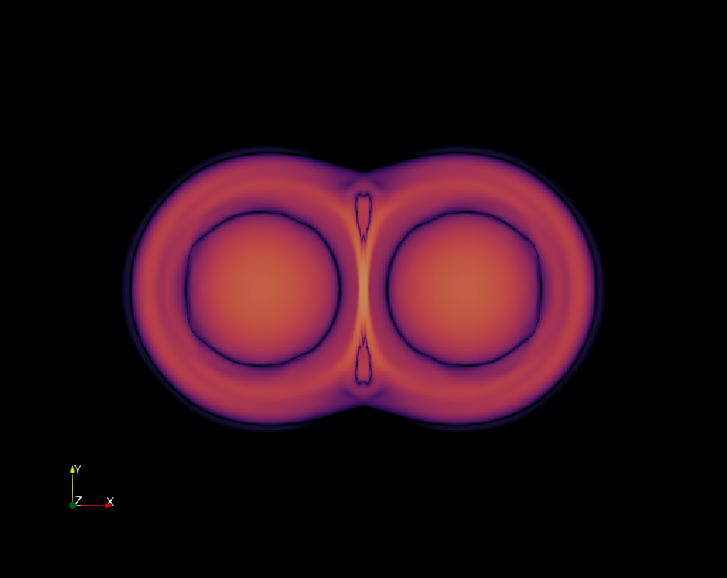}}
%\subfloat{\includegraphics[width=0.25\columnwidth,  clip=true]{jz_etaAR_8192x8192_t5}}
\subfloat{\includegraphics[width=0.66\columnwidth,  clip=true]{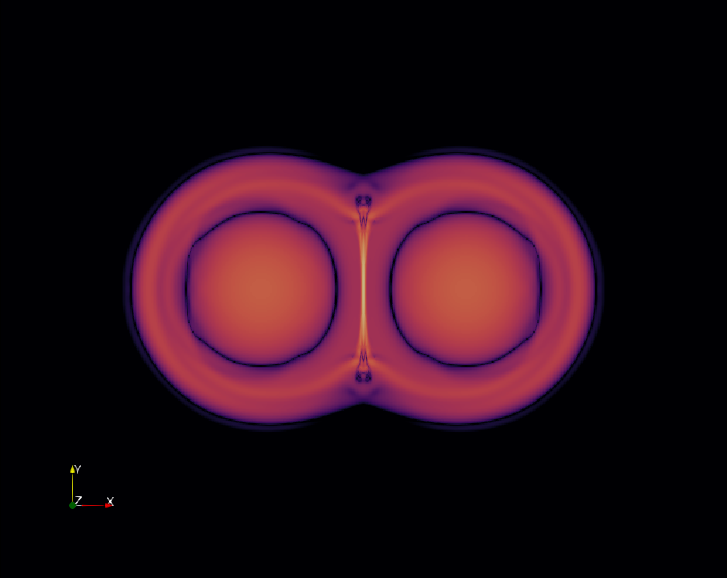}}

\subfloat{\includegraphics[width=0.66\columnwidth,  clip=true]{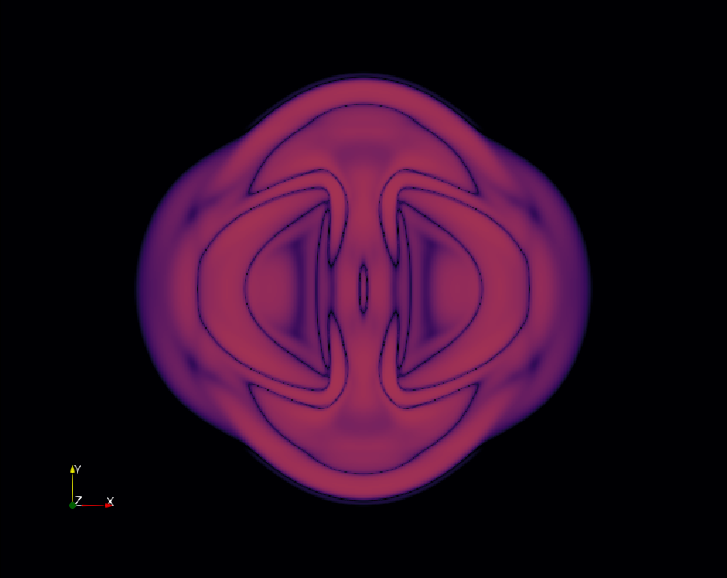}}
\subfloat{\includegraphics[width=0.66\columnwidth,  clip=true]{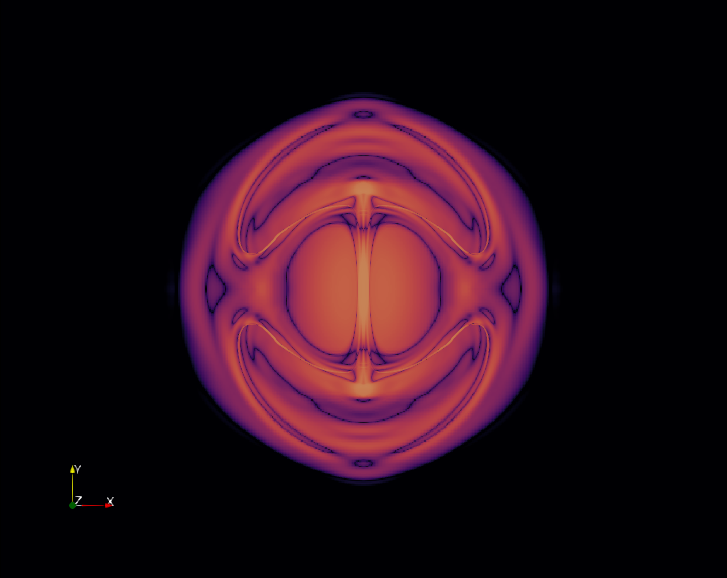}}
%\subfloat{\includegraphics[width=0.25\columnwidth,  clip=true]{jz_etaAR_8192x8192_t5}}
\subfloat{\includegraphics[width=0.66\columnwidth,  clip=true]{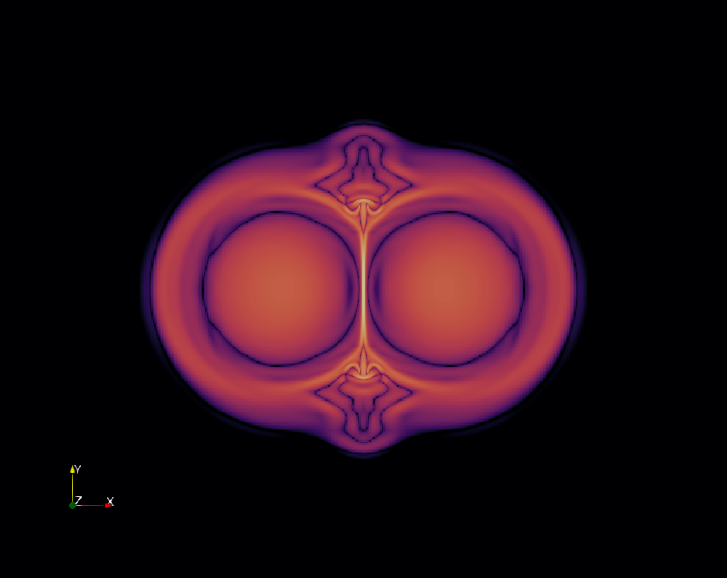}}

\subfloat{\includegraphics[width=0.66\columnwidth,  clip=true]{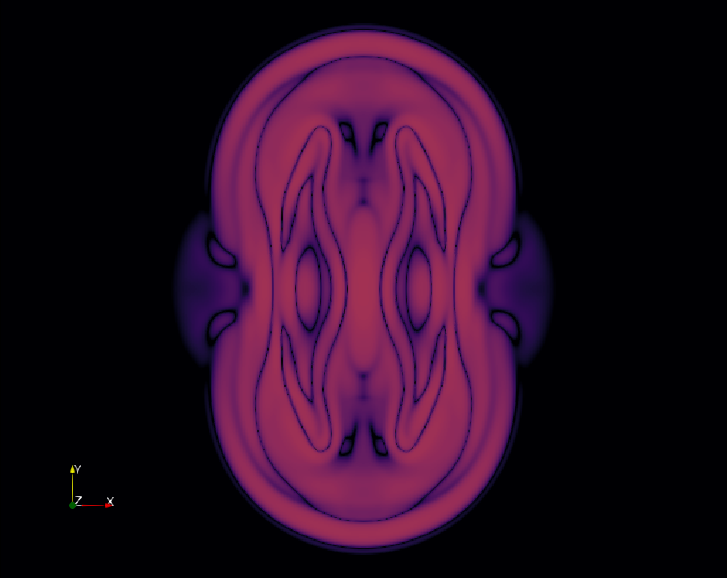}}
\subfloat{\includegraphics[width=0.66\columnwidth,  clip=true]{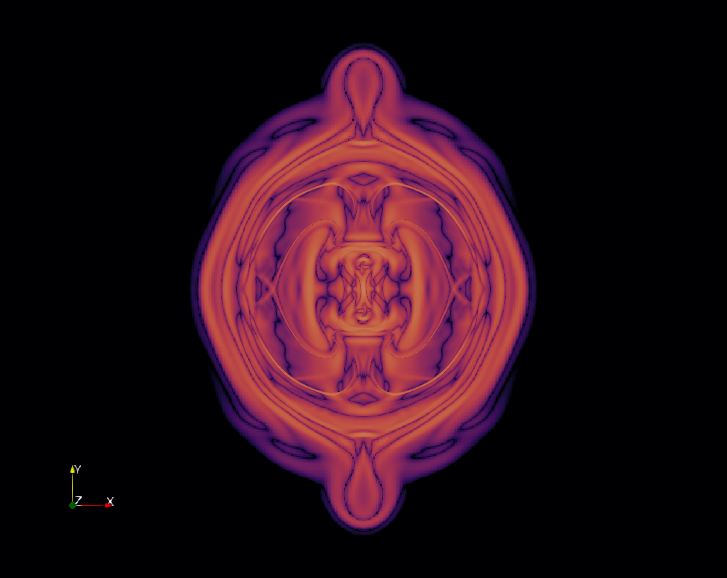}}
%\subfloat{\includegraphics[width=0.25\columnwidth,  clip=true]{jz_etaAR_8192x8192_t5}}
\subfloat{\includegraphics[width=0.66\columnwidth,  clip=true]{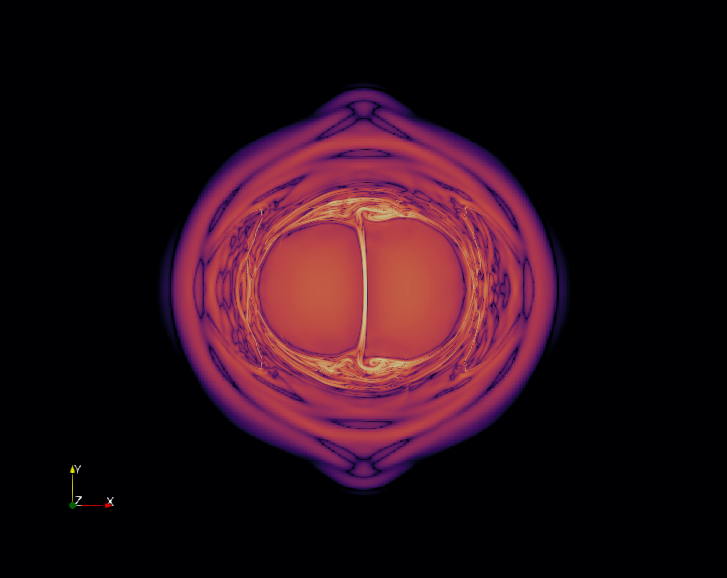}}

\subfloat{\includegraphics[width=0.66\columnwidth,  clip=true]{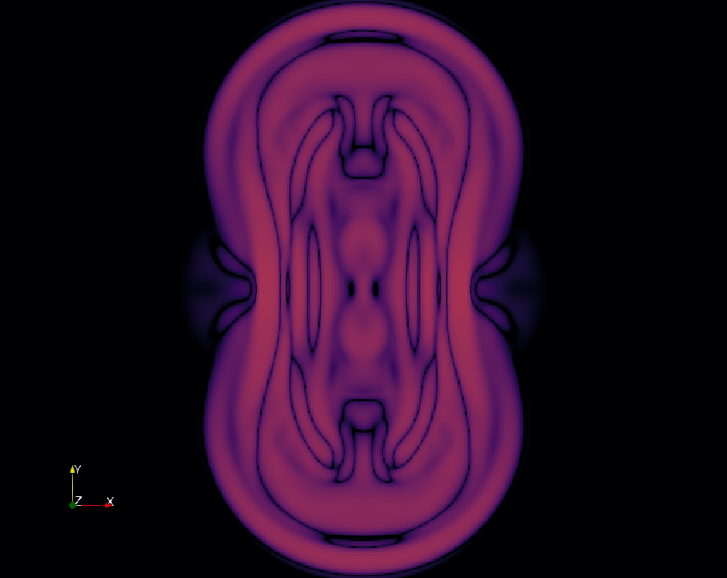}}
\subfloat{\includegraphics[width=0.66\columnwidth,  clip=true]{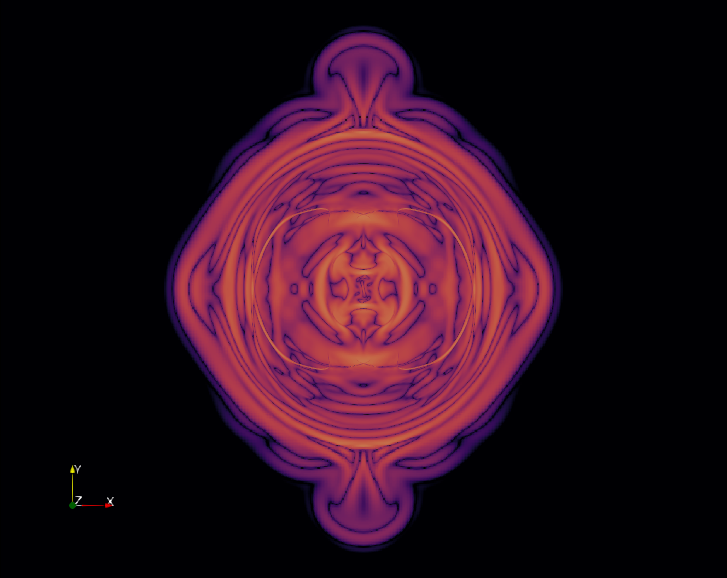}}
%\subfloat{\includegraphics[width=0.25\columnwidth,  clip=true]{jz_etaAR_8192x8192_t5}}
\subfloat{\includegraphics[width=0.66\columnwidth,  clip=true]{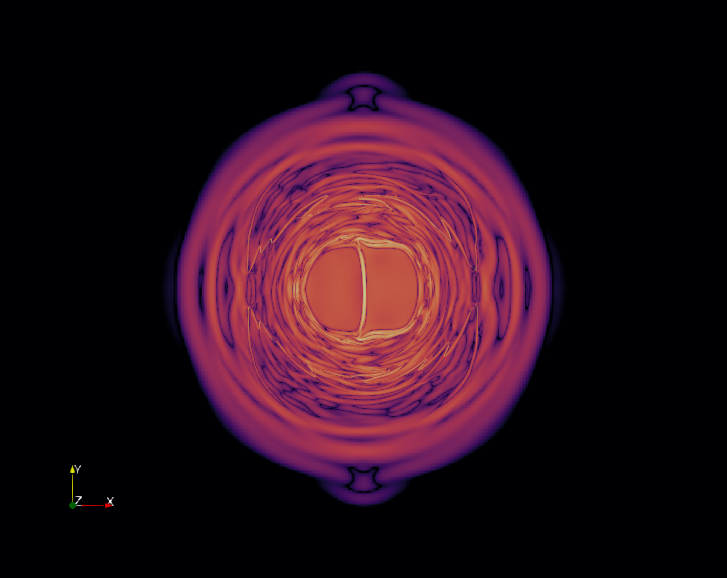}}
\caption{Out-of-plane current density magnitude $|J_{\rm z}|$ for runs with non-uniform resistivity $10^{-4}(1+\Delta_{\rm ei}^2J)$ (from left to right) Bnr1 ($\Delta^2_{\rm ei} = 1$), Bnr2 ($\Delta^2_{\rm ei} = 0.1$), and Bnr3 ($\Delta^2_{\rm ei} = 0.01$) at, (from top to bottom) $t=5 t_{\rm c}$, $t=10 t_{\rm c}$, $t=18 t_{\rm c}$ and $t=24 t_{\rm c}$. The logarithmic colour scale is shown in the top-left panel and is constrained to range from 0.005 and 50. Compare to run B with uniform resistivity $10^{-4}$ (middle column Figure \ref{fig:currentdensity}).}
\label{fig:currentdensityAR}
\end{center}
\end{figure*}
In Figure \ref{fig:currentdensityAR} the effect of the non-uniform resistivity on the evolution (time increases from top to bottom panels) of the spatial distribution of the current density magnitude is shown for runs Bnr1 (left-hand panels, $\Delta^2_{\rm ei} = 1$), Bnr2 (middle panels, $\Delta^2_{\rm ei} = 0.1$) and Bnr3 (right-hand panels, $\Delta^2_{\rm ei} = 0.01$). For increasing $\Delta_{\rm ei}$, the current sheet thickness $\delta$ increases. For run Bnr3 (right-hand panels), the broadening of the current sheet remains minimal, whereas for run Bnr2 (middle panels) there is a clear broadening and a faster merger of the flux tubes. The broadening is caused by the locally enhanced resistivity in the reconnection layer where the current density is very high. The thickness of the current sheet is limited by the lower resistivity in the ambient, where the current density is less high. This avoids the diffusion of the current sheet as can be seen in run Bnr1. In run Bnr1 (left-hand panels), the evolution occurs on such a fast time scale, due to strong enhancement of the resistivity, that no clear current sheet is observed and the flux tubes quickly merge and diffuse. In run Bnr4, with $\Delta^2_{\rm ei} = 0.001$, there is no observable difference with uniform resistivity run B and it is therefore not shown. 

\begin{figure} 
\begin{center}
\subfloat{\includegraphics[width=0.75\columnwidth]{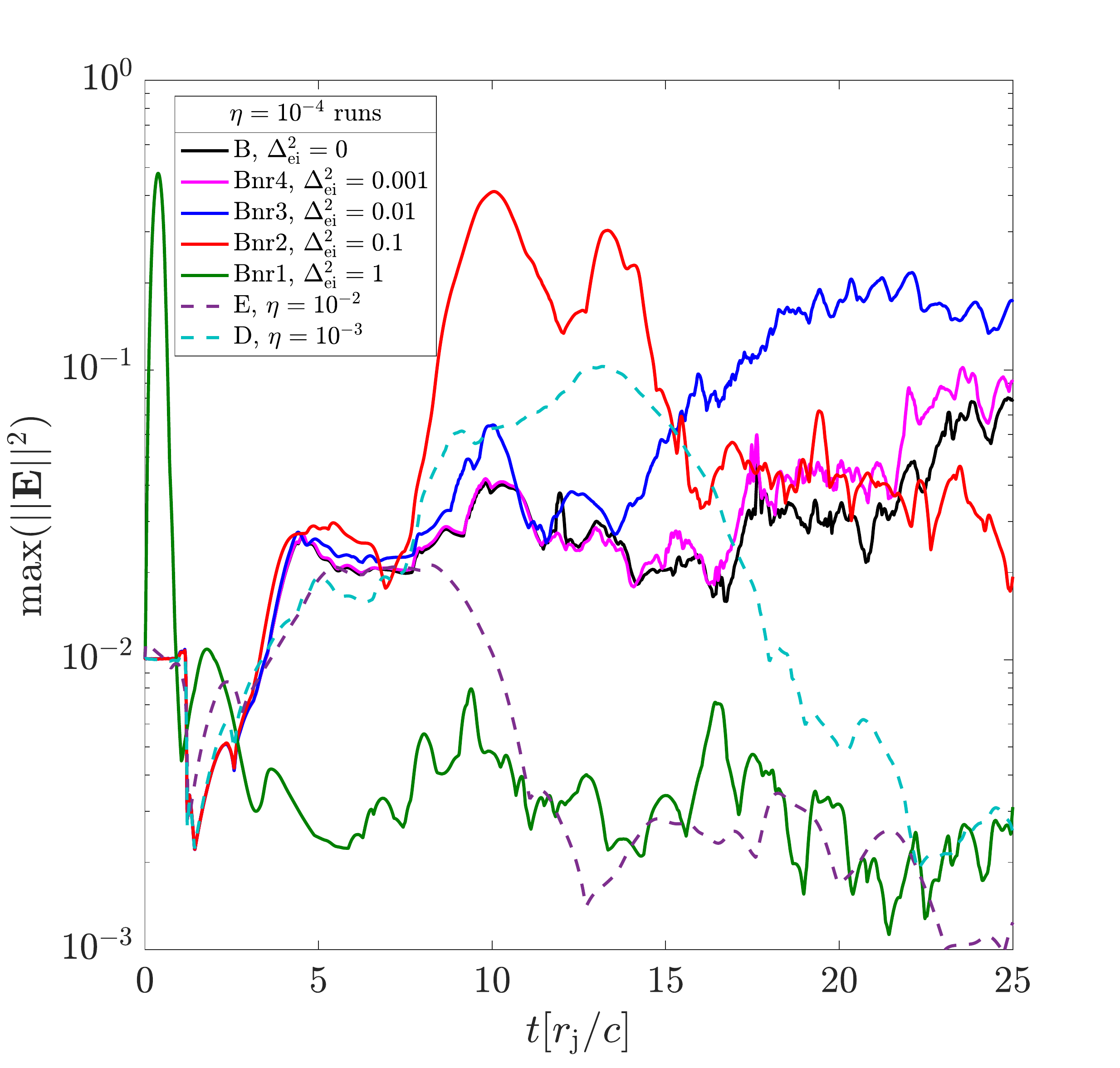}}
\caption{Peak electric energy for all runs with non-uniform resistivity $\eta = 10^{-4}(1+\Delta_{\rm ei}^2J)$ and resolution $8192^2$. $\Delta^2_{\rm ei}$ is varied between 1 and 0. For run Bnr4 with small  $\Delta_{\rm ei}^2 = 0.001$, the behaviour of run B with uniform resistivity $\eta = 10^{-4}$ and $\Delta_{\rm ei}=0$ is retrieved. Non-uniform resistivity clearly enhances the electric energy density in cases Bnr3 ($\Delta_{\rm ei}^2 = 0.01$) and Bnr2 ($\Delta_{\rm ei}^2 = 0.1$), compared with uniform resistivity run B. For run Bnr1 ($\Delta_{\rm ei}^2 = 1$), the resistivity becomes too high and the magnetic energy diffuses before a stable current sheet can form. Runs with uniform resistivity, E ($\eta = 10^{-2}$) and D ($\eta = 10^{-3}$) are shown, to compare to runs Bnr1 and Bnr2 with a maximum resistivity $\eta_{\rm max} \approx  10^{-2}$ and to run Bnr3 with $\eta_{\rm max} \approx  10^{-3}$.}
\label{fig:EmaxAR}
\end{center}
\end{figure}
Figure \ref{fig:EmaxAR} depicts the evolution of the maximum electric energy density in the domain over time for all non-uniform resistivity runs. 
Run B (black line), with uniform (i.e. $\Delta_{\rm ei}=0$) resistivity $\eta = 10^{-4}$ equivalent to the chosen base resistivity $\eta_0$, run D (light-blue dashed line), with uniform resistivity $\eta = 10^{-3}$ and run E (purple dashed line), with uniform resistivity $\eta = 10^{-2}$ are also shown for comparison. The current-dependent resistivity varies in time with current density, hence it induces fluctuations of the current density itself, resulting in a larger variability in the nonlinear phase. 
The evolution of $\max (||\mathbf{E}||^2)$ quantifies the results in Figure \ref{fig:currentdensityAR}, showing that run Bnr4 (pink line, $\Delta_{\rm ei}^2 = 0.001$) closely resembles uniform resistivity run B ($\Delta_{\rm ei}^2 = 0$, $\eta = 10^{-4}$) up to the far nonlinear stage. This confirms that for smaller $\Delta_{\rm ei}$, the uniform resistivity limit is retrieved. 

Run Bnr3 (blue line, $\Delta_{\rm ei}^2 = 0.01$) shows a similar evolution up to $t \approx 13 t_{\rm c}$, after which the electric field energy density dramatically increases. For comparison, we show run D (dashed light-blue line in Figure \ref{fig:EmaxAR}), that has a uniform resistivity $\eta = 10^{-3}$ comparable to the maximum resistivity in run Bnr3, $\eta_{\rm max} \approx 0.67 \times 10^{-3}$. The non-uniform resistivity enhancement only starts to grow from $t \approx 14 t_{\rm c}$ onwards. At this time, the electric energy density in run D already decreases.

In run Bnr2 (red line, $\Delta_{\rm ei}^2 = 0.1$), the increase of $\max (||\mathbf{E}||^2)$ due to non-uniform resistivity occurs shortly after the Alfv\'{e}nic growth, at $t \approx 7 t_{\rm c}$. For comparison we show run E (dashed purple line), with uniform resistivity $\eta = 10^{-2}$, that is comparable to the maximum resistivity attained in run Bnr2, $\eta_{\rm max} \approx 2.6 \times 10^{-2}$. During the secondary growth phase at $t \approx 7 t_{\rm c}$ due to non-uniform resistivity in run Bnr2, the electric energy density decreases strongly in run E due to the diffusion of the current sheet.

In run Bnr1 (green line, $\Delta_{\rm ei}^2 = 1$), the diffusion occurs almost instantaneously at $t \approx 1 t_{\rm c}$, due to the strongly enhanced resistivity, such that no stable current sheet can form. Plasmoids are not observed in any of the non-uniform resistivity cases.

\begin{figure*} 
\begin{center}
\subfloat{\includegraphics[width=0.75\columnwidth]{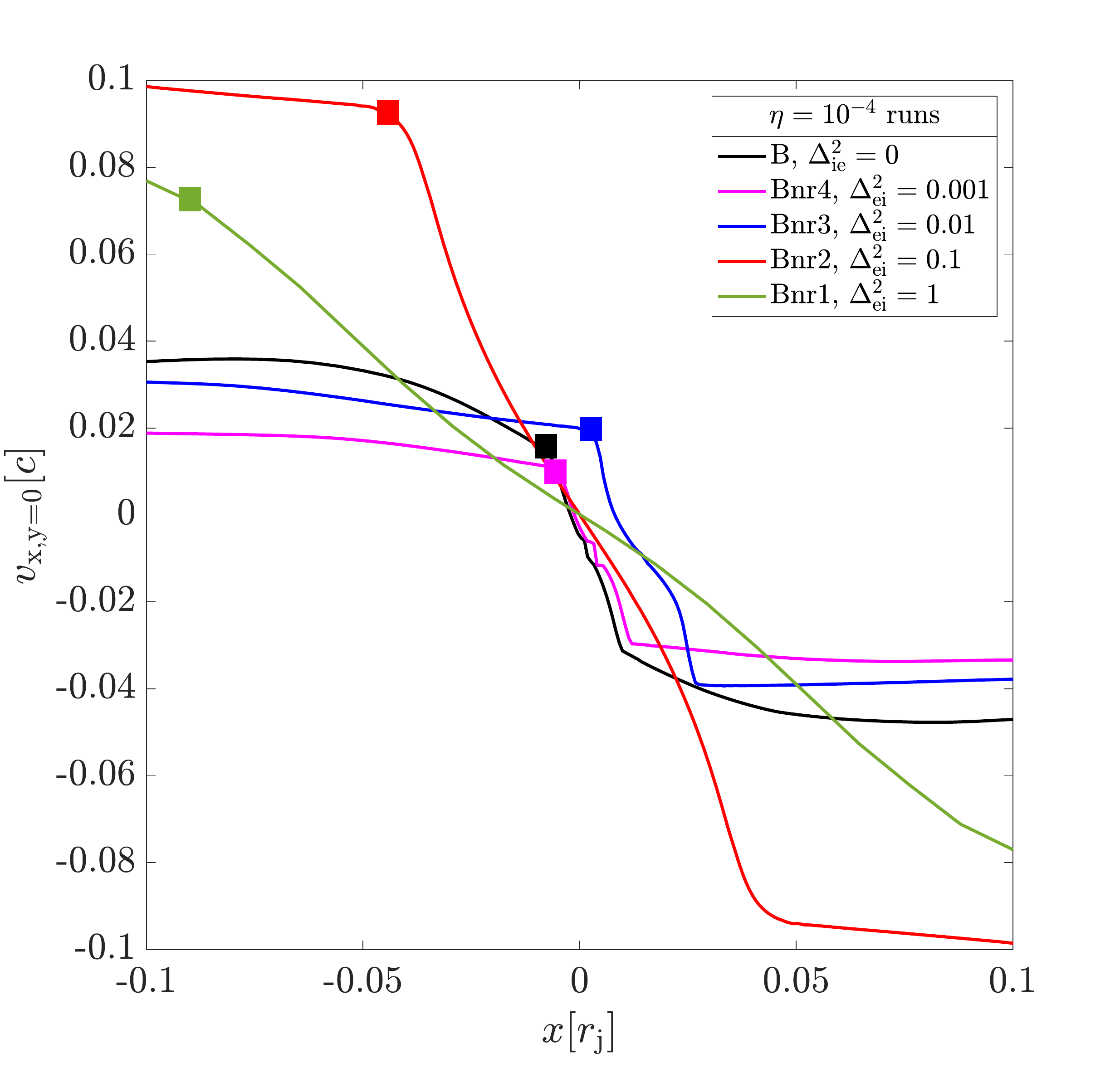}}
\subfloat{\includegraphics[width=0.75\columnwidth]{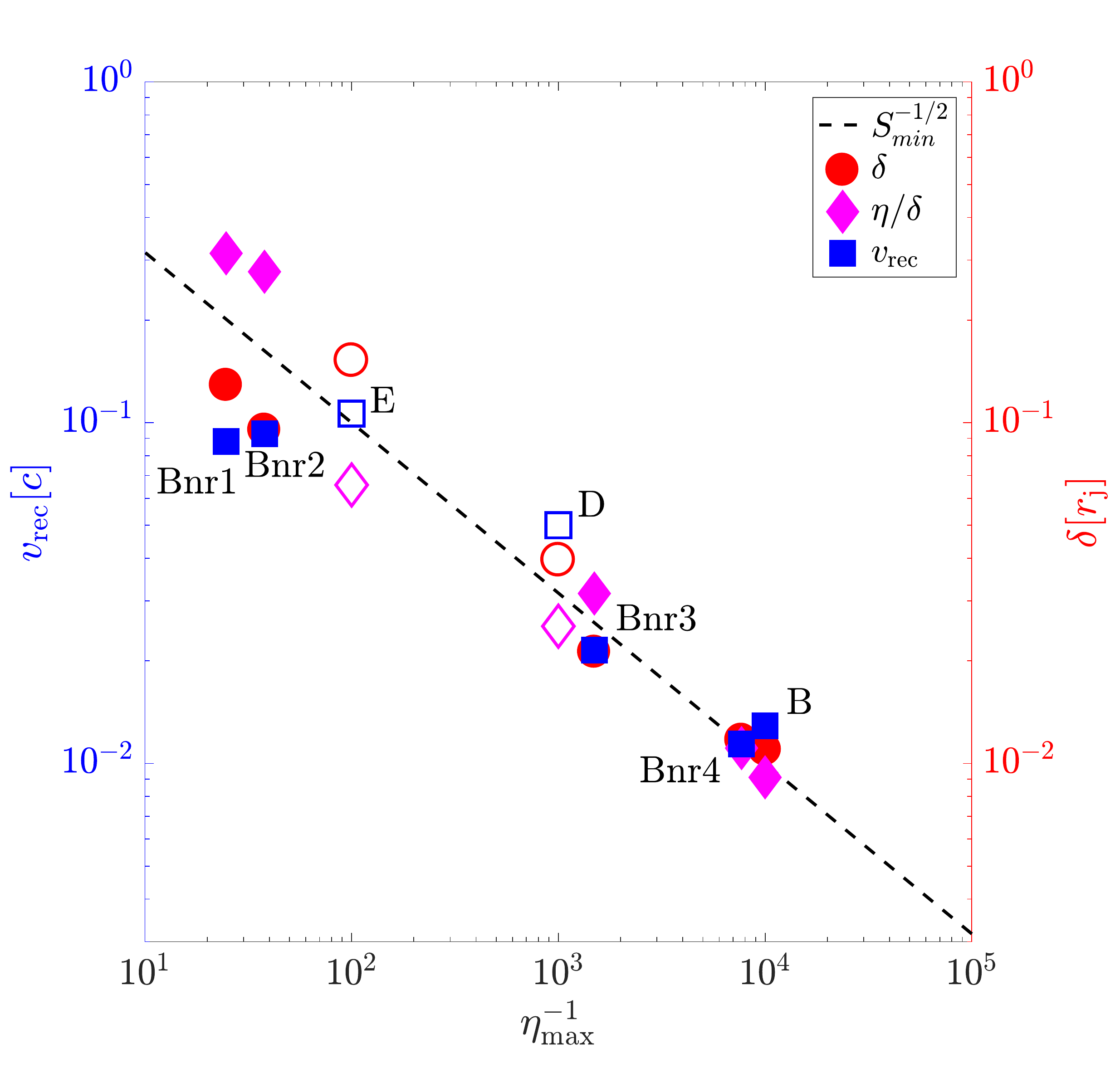}}
\caption{Left-hand panel shows the flow velocity into the current sheet $v_{\rm x,y=0}$ versus $x$-coordinate at $y=0$, for non-uniform resistivity runs B, Bnr4, Bnr3, Bnr2 and Bnr1 with base resistivity $\eta_0= 10^{-4}$, and $\Delta^2_{\rm ei} = 0$, $\Delta^2_{\rm ei} = 0.001$, $\Delta^2_{\rm ei} = 0.01$, $\Delta^2_{\rm ei} = 0.1$ and $\Delta^2_{\rm ei} = 1$ respectively. All runs have equal effective resolution $8192^2$. The reconnection rate $v_{\rm rec} = v_{\rm in} = v_{\rm x,y=0}$ is taken at times $t = 2 t_{\rm c}$ (run Bnr1), $t = 10 t_{\rm c}$ (run Bnr2), $t = 18 t_{\rm c}$ (runs Bnr3, Bnr4 and B) at the point where $v_{\rm x,y=0}$ has an inflection point, i.e. where the function changes from being convex to concave, e.g. where the dotted line crosses $v_{\rm x,y=0}$ for run Bnr3. 
The right-hand panel shows the reconnection rate $v_{\rm rec} := v_{\rm in}$ (blue squares), $v_{\rm rec} \sim \eta/\delta$ (magenta diamonds) and the thickness of the current sheet $\delta$ (red circles), versus the inverse of the maximum of the non-uniform resistivity $(\max(\eta))^{-1}$ taken over the whole domain and over the whole simulation time, for all non-uniform resistivity runs. The thickness $\delta$ is determined as the full-width at half-maximum of the out-of-plane current density $J_{\rm z}$ for each case at the same point in time as the reconnection rate. For comparison, runs with uniform resistivity B ($\eta = 10^{-4}$) D ($\eta = 10^{-3}$) and E ($\eta = 10^{-2}$) are shown (indicated by open symbols), where $\eta_{\rm max}^{-1} = \eta^{-1}$ in this case. The dashed line shows a Sweet-Parker scaling $v_{\rm rec} \sim S_{\rm min}^{-1/2}$, where $S_{\rm min} = \eta^{-1}_{\rm max}$. For clarity we only label the reconnection rate (blue squares), and magenta diamonds and red circles correspond to the blue square at the same $\eta^{-1}_{\rm max}$.}
\label{fig:recrate_AR}
\end{center}
\end{figure*}
The nonlinear, current-dependent resistivity is mainly enhanced within the current sheet, where the strongest current density occurs (see Figure \ref{fig:currentdensityAR}). Due to the locally larger resistivity, the current sheet broadens and the reconnection rate $v_{\rm rec} \propto \delta$, is expected to increase. 
In Figure \ref{fig:recrate_AR} the reconnection rate $v_{\rm rec} := v_{\rm in}  \simeq v_{\rm x,y=0} $ is determined at times $t = 2 t_{\rm c}$ (run Bnr1), $t = 10 t_{\rm c}$ (run Bnr2), $t = 18 t_{\rm c}$ (runs Bnr3, Bnr4 and B), where the current density has its peak for all non-uniform resistivity cases. In the left-hand panel the inflow speed $v_{\rm x,y=0}$ is depicted in units of $c$ for runs Bnr4, Bnr3, Bnr2, Bnr1 and B. Compared to uniform resistivity run B (black line,  $\eta= 10^{-4}$, $\Delta^2_{\rm ei} = 0$), the inflow velocity is higher in the non-uniform resistivity run Bnr2 (red line, $\eta_0 = 10^{-4}$, $\Delta^2_{\rm ei} = 0.1$), due to the strongly enhanced resistivity in the current sheet $\eta_{\rm max} \approx 264.6 \times 10^{-4}$. For runs Bnr4 (magenta line, $\eta_0 = 10^{-4}$, $\Delta^2_{\rm ei} = 0.001$) and Bnr3 (blue line, $\eta_0 = 10^{-4}$, $\Delta^2_{\rm ei} = 0.01$), the resistivity enhancement is only minor and the inflow velocity does not significantly increase compared to run B. In run Bnr1 (green line, $\eta_0 = 10^{-4}$, $\Delta^2_{\rm ei} = 1$), the resistivity is so large that a stable current sheet can never form due to diffusion of the magnetic energy density. 

%The inflow speed $v_{\rm in}$ is again determined by averaging over a vertical line $y \in [-0.1,0.1]$ and taken at a cut along the $x$-coordinate, at the point where $v_{x, y=0}$ has an inflection point, i.e. where the function changes from being convex to concave, as exemplified by the coloured squares in the left-hand panel of Figure \ref{fig:recrate_AR}. 
%{The thickness $\delta$ of the sheet is again determined as the full-width at half-maximum of the out-of-plane current density $J_{\rm z}$ for each case at the same point in time as the reconnection rate.} 
%OP: not needed since we are consistent.  
In the right-hand panel of Figure \ref{fig:recrate_AR} the scaling of the reconnection rate $v_{\rm in}/c$, the thickness $\delta$ and the ratio $\eta/\delta$ in units of $c$ is plotted versus the minimum Lundquist number $S_{\rm min} \simeq 1/\eta_{\rm max}$ for all non-uniform resistivity runs. The dashed line depicts the Sweet-Parker scaling $v_{\rm rec} \sim S_{\rm min}^{-1/2}$. Note that indeed the reconnection rate increases for a larger resistivity $\eta_{\rm max}$, and hence for larger $\Delta_{\rm ei}$, and the sheet thickness grows accordingly. 

Uniform resistivity runs B ($\eta = 10^{-4}$), D ($\eta = 10^{-3}$) and E ($\eta = 10^{-2}$) are shown for comparison in the right-hand panel of Figure \ref{fig:recrate_AR}. Comparison to run B (with equivalent base resistivity $\eta_0 = 10^{-4}$ as all non-uniform runs) shows that the non-uniform resistivity has the effect of increasing the reconnection rate and the current sheet thickness $\delta$. Comparing run E ($\eta = 10^{-2}$) to run Bnr2 (with similar maximum resistivity $\eta_{\rm max} \approx 2.6 \times 10^{-2}$), shows that a non-uniform resistivity results in a thinner current sheet ($0.1 r_{\rm j}$ versus $0.5 r_{\rm j}$ for run E) and a comparable reconnection rate of 0.1c. Comparing run D ($\eta = 10^{-3}$) to run Bnr3 (with similar maximum resistivity $\eta_{\rm max} \approx 0.67 \times 10^{-3}$), shows that for small values of $\Delta_{\rm ei}$, the non-uniform resistivity has small, albeit non-negligible effects, as is confirmed by the electric energy density evolution in Figure \ref{fig:EmaxAR}.

\section{Conclusions}
\label{sect:conclusions}

The GRRMHD equations are solved with the new resistive {\tt BHAC} code, to study relativistic magnetic reconnection in Minkowski spacetime. Reconnection is triggered by the coalescence instability in an initially force-free 2D equilibrium of two Lundquist tubes (\citealt{SironiPorth}). A reconnection layer forms in between the merging flux tubes, such that the details of reconnection and plasmoid formation are not affected by the assumption of a pre-existing current sheet.

All results are confirmed to converge up to the nonlinear plasmoid regime. The convergence is determined based on the evolution of the maximum electric energy density and the maximum current density in the system while progressively doubling the resolution. This strict convergence criterion is met for resolutions of $N \geq 8192^2$ for all runs and is confirmed for different combinations of AMR levels and base resolutions. We find that a too low resolution of $N = 2048^2$ causes the plasmoid instability to be artificially triggered due to numerical artifacts. This effect disappears for higher resolutions of $N \geq 4096^2$.

We explored a range of Lundquist numbers $S$ by changing $\eta$ and by varying plasma-$\beta$ and $\sigma$ in astrophysically relevant regimes of the parameter space. {We find that the plasmoid instability is triggered for a Lundquist number larger than the critical value of $S_{\rm c} \approx 8000$.} In these cases, secondary plasmoids form efficiently and the reconnection rate is increased up to a maximum of $v_{\rm rec} \approx 0.03c$, compared to a slow Sweet-Parker scaling $v_{\rm rec} \sim S^{-1/2} \approx 0.01c$. For smaller Lundquist numbers $S_{\rm eff} \lesssim S_{\rm c}$, i.e. higher resistivities and lower magnetisation, we find a slow Sweet-Parker scaling. Current-dependent non-uniform resistivity is implemented and compared to cases with uniform resistivity. We find that non-uniform resistivity can increase the reconnection rate up to $v_{\rm rec} \sim 0.1c$ versus $v_{\rm rec} \sim 0.01c$ for the fiducial case with uniform resistivity $\eta = 10^{-4}$. We show that the resistivity model and magnitude have a large impact on the reconnection rate and on plasmoid formation. Therefore, realistic models for the resistivity in astrophysical systems are an absolute necessity. 

Our 2D findings for merging flux ropes in SRRMHD provide a model for consistently forming current sheets and plasmoids and we provide the necessary resolutions required to resolve relativistic reconnection and the tearing instability in high-Lundquist number plasmas. Regimes with small and spatiotemporally dependent resistivity are extremely demanding for GRRMHD codes and both schemes implemented in {\tt BHAC} appear to be able to handle these conditions well. With a combination of an ImEx scheme and AMR, the required accuracy can also be obtained in large 3D domains, relevant for high-energy astrophysics.

Recently, the first advances to analytically describe reconnection in Kerr spacetime have been by made \cite{asenjo2017} and \cite{comisso2018}. In \cite{ball2017} the properties of reconnecting current sheets are determined with ideal GRMHD simulations, purely based on numerical resistivity. They find reconnection sites in black hole accretion disks with magnetised plasma ranging from $\beta = 10^{-2}$ to $10^3$ and magnetisations of $\sigma = 10^{-3}$ to 10, comparable to the range of $\beta$ and $\sigma$ used for the interacting flux tubes considered in our work. Resolving magnetic reconnection within GRRMHD simulations and accounting for a physically-motivated resistivity will provide a more realistic model for plasmoid formation and subsequent flaring variability. With the GRRMHD module in {\tt BHAC}, it will soon be possible to explore general relativistic reconnection based on physical resistivity up to a nonlinear regime that is currently inaccessible both analytically and with ideal GRMHD simulations.

\section*{Acknowledgements}
This research was supported by projects GOA/2015-014 (2014-2018 KU Leuven) and the Interuniversity Attraction Poles Programme by the Belgian Science Policy Office (IAP P7/08 CHARM). OP is supported by the ERC synergy grant `BlackHoleCam: Imaging the Event Horizon of Black Holes' (Grant No. 610058). LS acknowledges support from DoE DE-SC0016542, NASA Fermi NNX16AR75G, NASA ATP NNX-17AG21G, NSF ACI-1657507, and NSF AST- 1716567.
The computational resources and services used in this work were provided by the VSC (Flemish Supercomputer Center), funded by the Research Foundation Flanders (FWO) and the Flemish Government - department EWI.
BR would like to thank Luca Comisso for useful discussions on resistivity. The authors would like to thank the anonymous reviewer for significantly improving the manuscript.

%%%%%%%%%%%%%%%%%%%%%%%%%%%%%%%%%%%%%%%%%%%%%%%%%%



\bibliographystyle{mnras}
\bibliography{mylib3,mylib4,astro} % if your bibtex file is called example.bib

%\begin{thebibliography}{99}


%\end{thebibliography}

%%%%%%%%%%%%%%%%%%%%%%%%%%%%%%%%%%%%%%%%%%%%%%%%%%

%%%%%%%%%%%%%%%%% APPENDICES %%%%%%%%%%%%%%%%%%%%%

%\appendix



%%%%%%%%%%%%%%%%%%%%%%%%%%%%%%%%%%%%%%%%%%%%%%%%%%

% Don't change these lines
\bsp	% typesetting comment
\label{lastpage}
\end{document}